\documentclass[a4paper,11pt]{article}

\usepackage[a4paper,
  left=2.5cm, right=2.5cm,
  top= 3cm, bottom=4cm]{geometry}

\usepackage{lmodern}
\usepackage{amsmath}
\numberwithin{equation}{section}
\usepackage{mathtools}
\usepackage{oldgerm}
\usepackage{amssymb}
\usepackage{bbm}
\usepackage{graphicx}
\usepackage{subcaption}
\usepackage{tikz-cd} 
\usetikzlibrary{matrix,arrows,decorations.pathmorphing}
\usepgfmodule{shapes}
\usetikzlibrary{backgrounds, arrows,calc,shapes,decorations.pathreplacing, decorations.markings, automata,positioning}
\usepackage{cite}

\usepackage{youngtab}
\usepackage{ytableau}
\newcommand{\todo}[1]{}
\renewcommand{\todo}[1]{{\color{red} TODO: {#1}}}
\newcommand{\by}{\begin{ytableau}}
\newcommand{\ey}{\end{ytableau}}

\usetikzlibrary{arrows,decorations.markings}
\usepackage{color}
\usepackage{pgf}
\usepackage{pgffor}

\usepackage{epic}
\usepackage{hyperref}
\usepackage{bbold}

\newcommand{\be}{\begin{equation}}  
\newcommand{\ee}{\end{equation}}  
\newcommand{\bp}{\begin{pmatrix*}[r]}  
\newcommand{\ep}{\end{pmatrix*}}  
\newcommand{\bpp}{\begin{pmatrix}}  
\newcommand{\epp}{\end{pmatrix}}  
\newcommand{\bcd}{\begin{equation}
\begin{tikzcd}}
\newcommand{\ecd}{\end{tikzcd} \end{equation}}

\def\C{\mathbb{C}}

\def\cS{\mathcal{S}}
\def\1{\mathbb{1}}

\def\cN{\mathcal{N}}
\def\cX{\mathcal{X}}
\def\cY{\mathcal{Y}}
\def\cZ{\mathcal{Z}}

\newcommand{\tr}{\text{tr}\,}

\begin{document}
\begin{titlepage}
 
 
\begin{flushright}

\end{flushright}
 
\vskip 1cm
\begin{center}
 
{\huge \bf \boldmath T-branes, monopoles and S-duality} 
 
 \vskip 2cm
 
Andr\'es Collinucci$^\blacklozenge$, Simone Giacomelli$^\spadesuit$, Roberto Valandro$^{\clubsuit,\bigstar,\spadesuit}$

 \vskip 0.4cm
 
 {\it  $^\blacklozenge$Physique Th\'eorique et Math\'ematique and International Solvay Institutes,\\ Universit\'e Libre de Bruxelles, C.P. 231, 1050
Bruxelles, Belgium \\[2mm]
$^\spadesuit$ICTP, Strada Costiera 11, 34151 Trieste, Italy \\[2mm]
$^\clubsuit$Dipartimento di Fisica, Universit\`a di Trieste,\\ Strada Costiera 11, 34151 Trieste, Italy\\[2mm]
$^\bigstar$INFN, Sezione di Trieste, Via Valerio 2, 34127 Trieste, Italy 
 }
 \vskip 2cm
 
\abstract{\noindent M2 branes probing T-brane backgrounds in M-theory with ADE surface singularities perceive deformations on their worldvolume superpotentials by monopole operators. The dynamics and moduli spaces of the resulting theories can be studied using a dual description involving conventional superpotential terms and (the dimensional reduction of) class S trinion theories. By using the S-dual description of N=2 SU(N) SQCD with 2N flavors in four dimensions, we are able to study T-branes corresponding to all minimal nilpotent orbits for the whole ADE series. 
Our proposed dualities are supported by the match in squashed sphere partition functions.
} 

\end{center}
\end{titlepage}

\tableofcontents

\newpage
\section{Introduction} 

The duality between M-theory and IIA string theory establishes a striking connection between gauge theories and geometry. For instance, gauge groups and charged matter on D6-branes in IIA find an incarnation as geometric singularities in M-theory. However, this picture is lacking, not only fundamentally due to the absence of a microscopic formulation of the former, but even at the level of supersymmetric vacua. 

For example, a stack of parallel N D6-branes, which hosts a $U(N)$ gauge theory, uplifts in M-theory to an $A_{N-1}$ singularity given by the hypersurface
$uv = \det(z \mathbb{1}_N-\Phi_{\rm D6})$, where, say $\Phi_{\rm D6} = \phi_1+i\phi_2$ are two of the three transverse worldvolume scalars.
Giving a vev to these scalars \emph{may} correspond to separating the branes of the stack, which in M-theory corresponds to deforming the singularity, as the r.h.s. of the hypersurface equation starts seeing non-trivial Casimirs of the $\Phi_{\rm D6}$ turned on.

However, if we switch on worldvolume flux, the D-term equation $F \sim [\Phi_{\rm D6}, \Phi_{\rm D6}^\dagger]$ allows for non-diagonalizable scalars. For instance, for a stack of two branes, we could have
\be
\Phi_{\rm D6} = \begin{pmatrix} 0 & 1\\0 & 0 \end{pmatrix}\,.
\ee
This clearly breaks the $U(2)$ gauge group to the overall $U(1)$, but leaves the singularity in M-theory intact, since all its Casimirs vanish. Such configurations were studied originally in \cite{Gomez:2000zm,Donagi:2003hh}, and more recently in \cite{Donagi:2011jy} and in \cite{Cecotti:2010bp}, where they were dubbed \emph{T-branes}, where the `T' highlights the upper \emph{triangular} form of the vev. More generally, a T-brane can be characterized as a non-Abelian bound state between D-branes given by condensing stretched strings. In gauge theory, this corresponds to a nilpotent vev for the complex adjoint scalar on a stack of D6 or D7-branes.

This clash between what we expect from gauge theory, i.e. a broken gauge group, and what the M-theory geometry is displaying, i.e. an undeformed and unresolved singularity, is vexing. In hindsight, though, we should not be surprised that 11d supergravity data does not correctly capture all degrees of freedom when the spacetime curvature runs high. T-branes remind us that geometry alone is insufficient to understand even supersymmetric vacua.  We need a definition of what a T-brane means in M-theory. 
Direct attempts at defining T-brane data in M-theory have been made, see \cite{Anderson:2013rka,Collinucci:2014taa,Anderson:2017rpr}. They all involve heavy mathematical machinery leaving the microscopic origin of the T-brane in M-theory obscured.\footnote{For further recent work on T-branes and their applications see \cite{Collinucci:2014qfa,Marchesano:2015dfa,Cicoli:2015ylx,Bena:2016oqr,Marchesano:2016cqg,Ashfaque:2017iog,Bena:2017jhm}.}

In \cite{Collinucci:2016hpz}, we initiated the study of D2-branes probing T-branes. Specifically, we studied deformations of three-dimensional $\mathcal{N}=4$ gauge theories that correspond to D2-branes probing T-brane configurations with parallel D6-branes, and parallel D6-branes on top of an O6-plane. From the vantage point of the D2-brane, two of the three D6 adjoint scalars appear as complex mass parameters in the 3d theory:
\be
\Delta W = \langle {\Phi_{\rm D6}}^j_i \rangle Q_j \tilde Q^i\,.
\ee
From this perspective, a T-brane corresponds to a deformation of the 3d theory by a nilpotent complex mass term.\footnote{This possibility has been pointed out in \cite{Benini:2009qs} for the case of two intersecting D6-branes.}

The 3d mirror symmetry of \cite{Intriligator:1996ex} (and further developed in \cite{Hanany:1996ie,deBoer:1996mp,Porrati:1996xi,deBoer:1996ck,deBoer:1997ka,deBoer:1997kr,Aharony:1997bx}) states that these theories are infrared dual to 3d quiver gauge theories, whose quiver graph is the affine Dynkin diagram corresponding to the flavor group on the D2-brane, which is the gauge group on the D6-stack for the $A$ and $D$ series. The mirror symmetry exchanges the Higgs and Coulomb branches of the theories, and sends nilpotent mass terms into superpotential terms involving \emph{monopole operators}. E.g.
\be
m Q_{i-1} \tilde Q^{i} \quad \longrightarrow m W_{i, +}\,,
\ee
where $W_{i, +} \sim \exp(\sigma_i + i \gamma_i)$ is roughly the exponential of the dual $i$-th photon $\gamma_i$. More precisely, a monopole operator is a local disorder operator which can be defined directly in the infrared CFT \cite{Kapustin:1999ha,Borokhov:2002ib,Borokhov:2002cg}.
Monopole superpotentials arise also in the context of D3 branes suspended between pq-webs in Type IIB  \cite{Benvenuti:2016wet}.

In \cite{Collinucci:2016hpz} we studied the effective theories of $A$ and $D$ type quiver gauge theories deformed by such monopole operators, restricting to cases where the quiver node associated to the deformation was Abelian. We learned a few valuable lessons: 
\begin{enumerate}
\item The resulting effective theory is a quiver theory with one node missing, but a new `fundamental meson' with a particular superpotential.

\item $\mathcal{N}=4$ is broken to $\mathcal{N}=2$, and the Coulomb branch is reduced in dimension, usually to a complex but not quaternionic variety.
\item Strikingly, the Higgs branch remains intact as a complex manifold, still displaying the original ADE singularity of the parent $\mathcal{N}=4$ theory.

\item The monopole deformation can be shown to correspond to an insertion of a coherent state of membranes wrapping a vanishing $\mathbb{CP}^1$ of the singular geometry. This matches IIA expectation, since a T-brane is a coherent state of strings stretched between different D6-branes, which under mirror symmetry map to D2-branes wrapping the spheres. This point, in our view, elucidates the physical meaning of a T-brane in M-theory.
\end{enumerate}

This treatment of T-branes as monopole deformations, and their interpretation as coherent states of vanishing membranes gives us a definition of what a T-brane is in M-theory that can be generalized to cases where mirror symmetry is not straightforward, such as the exceptional singularities.

In this paper, we will actually tackle the T-brane problem for any minimal nilpotent orbit of any simple Lie algebra. In other words, we will study monopole deformations also on non-Abelian nodes of the $D$ and $E$ type quiver theories. For the latter, we will discover that the resulting effective theory contains a non-Lagrangian block connected to the rest of the quiver.

The purpose of this investigation is twofold: 
\begin{itemize}
\item On the one hand, we want to deepen our understanding of M-theory, and its connection to string theory. The T-brane is a perfect example of something that should be captured by holomorphic data, but is not naively encoded in the geometry of the 11d space. 
\item On the other hand, we are uncovering a new class of three-dimensional $\mathcal{N}=2$ theories that behave partly as if they enjoyed $\mathcal{N}=4$ supersymmetry: Their Coulomb branches are complex varieties that are not hyper-K\"ahler, yet their Higgs\footnote{Strictly speaking, this terminology is reserved for $\mathcal{N}=4$. We will give a more precise definition later of what we mean by `Higgs branch'.} branches are the well-known affine surfaces with ADE singularities.
\end{itemize}

We will use a novel approach to study the effect of monopole operator deformations on non-Abelian gauge theories. Four-dimensional SCFT's enjoy S-dualities that reduce straightforwardly to three-dimensional dualities\footnote{The caveat of \cite{Aharony:2013dha} is evaded if the UV theory is $\cN=2$ superconformal because a dynamically generated monopole superpotential would break extended supersymmetry.}. This will allow us to turn any non-Abelian node with gauge group $U(N)$ into a system with $G = U(1) \times SU(2)$ plus some non-Lagrangian theory. The original monopole deformation will always translate into a monopole with respect to  the new $U(1)$ factor, which we already learned to handle in our previous work \cite{Collinucci:2016hpz}.

Since we will be deriving our effective theories via rather indirect means, we will supplement our claims by comparing the squashed three-sphere partition functions of the monopole deformed theories and their proposed effective theories. We will find a perfect match.

What is novel in our approach, is the counterintuitive move of voluntarily replacing perfectly Lagrangian theories into something intrinsically strongly coupled in order to \emph{gain} control of the calculation. We will show that this approach lets us come to grips with mirror symmetry and treat any node of any Dynkin quiver in the same manner, thereby finding the effective theory for any T-brane corresponding to a minimal nilpotent orbit. 

This paper is organized as follows: In Section \ref{Sec:GoalStrategy} we state the goal of this work and explain our strategy. In Sections \ref{Sec:U2node},  \ref{Sec:U3node}, and \ref{Sec:trinions}, we implement our strategy on gauge nodes with $U(2), U(3)$ and $U(N\geq4)$ gauge groups, respectively. This allows us to handle monopole operators corresponding to nilpotent orbits in all Dynkin type quiver gauge theories.
In Section \ref{Sec:dualnew} we propose new field theory dualities for such quiver gauge theories which reveal the hidden but expected Weyl symmetries in the quantum enhanced ADE global symmetries. In section \ref{Sec:squasheds3}, we will introduce and compute the partition functions of all discussed theories on the squashed three-sphere. 
In particular we will study SQED with two flavors (\emph{aka} $T(SU(2))$) and  $U(N)$ SQCD with $2N$ flavors deformed by a monopole operator and show their equivalence with the proposed effective theories.
In Appendix \ref{app:higgsade} we review how to derive the Higgs branches of the undeformed $E$ and $D$ quivers. Finally, in Appendix \ref{app:DNU1extnode} we review our treatment of a monopole deformation along a $U(1)$ node of a $D_N$ quiver.

\section{Goal and general strategy}\label{Sec:GoalStrategy}

Our goal for this paper is to study what happens to a $d=3, \cN=4$ ADE quiver gauge theory when we deform it by a single monopole operator that is charged with respect to  the topological $U(1)$ of a single node. This is a restricted class of deformations, which nevertheless covers all minimal nilpotent orbits of the ADE flavor group of the quiver in question. 
In physical terms, we will study all deformations that correspond, on the mirror side, to nilpotent `mass matrix' deformations of vanishing degree two. By `mass matrix' we mean superpotential deformations that are linear in the moment map of the theory's flavor symmetry. In \cite{Heckman:2010qv}, such mass deformations were studied from the point of view of D3-branes probing F-theory. The three-dimensional point of view, however, has many advantages: It allows us to recover the ADE singularity as a branch of the theory, and we have mirror symmetry at our disposal.

\subsection{Reducing to the Abelian problem}\label{Sec:GoalStrategy1}
When the deformed node is a balanced $U(1)$ gauge theory (i.e. has two flavors), the effective theory is known \cite{Collinucci:2016hpz}. We will review the procedure to obtain such a deformation in Section \ref{Sec:U1monop}.
However, when the node is non-abelian, this procedure cannot be applied directly. The point of this paper is to tackle this case. The strategy is to use dual theories in which the relevant monopole operator will be charged under a $U(1)$ node. The dual theories are worked out in the following way:
\begin{itemize}
\item We start from $\mathcal{N}=2$ $SU(N)$ SQCD with $2N$ flavors in four dimensions. As was studied in \cite{Chacaltana:2010ks} 
following \cite{Argyres:2007cn} and \cite{Gaiotto:2009we}, this 
theory has a dual description involving an $SU(2)$ gauge group coupled to one doublet and to a strongly-coupled SCFT (called 
$R_{0,N}$ in \cite{Chacaltana:2010ks}) which is best described as a three-punctured sphere in the language of \cite{Gaiotto:2009we}. 
In this dual description the $SU(2N)$ 
global symmetry is carried by the SCFT whereas the baryon number is carried by the doublet of $SU(2)$ and its charge is fixed 
by anomaly matching, as noticed in \cite{Argyres:2007cn}. 
\item This duality is preserved when we compactify to three dimensions. In \cite{Aharony:2013dha}, it was argued that, in general theories that flow to the same IR point may not longer enjoy such a duality upon dimensional reduction, since dimensionally reducing and going to low energies are operations that do not commute. Instead, a duality might be salvaged at the cost of generating potentials with monopole operators.
In our case, however, this caveat does not apply since the (undeformed) dual theories are $\cN=2$ superconformal in four dimensions and need not flow. Furthermore, a monopole superpotential cannot be generated in the compactification simply because it is not compatible with enhanced supersymmetry. The duality is rather obvious in the 
light of \cite{Benini:2010uu}, since these two theories have the same mirror.

\item Since we are interested in $U(N)$ SQCD in three dimensions, we should gauge the baryon number on the Lagrangian side of the 
duality. This amounts on the other side of the duality to gauging the $U(1)$ global symmetry carried by 
the $SU(2)$ doublet. We then conclude that $U(N)$ SQCD with $2N$ flavors in 3d has a dual description involving a 
$U(1)\times SU(2)$ gauge theory coupled to the dimensional reduction of $R_{0,N}$. 

\item In the case $N=2$ the duality simplifies considerably since $R_{0,2}$ is just a free theory describing three hypermultiplets in the doublet of $SU(2)$. Notice that the surviving global symmetry is $SU(4)$ in both cases. This duality was actually already 
discussed in \cite{Gaiotto:2008ak}. 

\item On the $U(N)$ SQCD 
side the topological symmetry is known to enhance to $SU(2)$ due to the presence of monopole operators of R-charge one. We 
expect the same to happen on the dual side and indeed the $U(1)$ node is balanced in the sense of \cite{Gaiotto:2008ak}, implying the enhancement 
to $SU(2)$. We then conclude that the monopoles $V_{\pm}$ of R-charge one on the SQCD side are mapped to the monopole operators $W_\pm$
of the $U(1)$ gauge node on the dual side.
\end{itemize}
We can now use this map to construct theories that are dual to the ADE quiver gauge theories.
\begin{itemize}
\item[$\triangleright$] Choose a node of interest from the quiver and ungauge its neighboring nodes. This leaves us with the $d=3, \mathcal{N}=4$ SQCD theory with $U(N)$ gauge group and $2 N$ fundamental hypers coupled to the adjoint scalar of the vector multiplet via the superpotential
\be
W = Q_i^a \Phi^a_b \tilde Q^i_b
\ee
where $i=1, \ldots, 2 N$ are the flavor indices, and $a=1, \ldots, N$ the color indices. From this, we can form the gauge invariant mesons:
\be
{M^i}_j = \tilde Q^i_a Q_j^a 
\ee
which satisfy tr$M=0$, and $M^2=0$ by the F-term constraints. These mesons are only gauge invariant with respect to  the $U(N)$ node in question. When this node is attached back to the quiver, the indices $(i, j)$ become gauged, and we must couple the moment maps ${M^i}_j$ to the adjoint scalar in the vector multiplet of the neighboring nodes.
\item[$\triangleright$] We perform the 3d version of the S-duality map.
 This invariantly leads to a setup of the form described in Figure \ref{fig:sduality}.

 \begin{figure} 
\begin{center}
\begin{tikzpicture}[->, every node/.style={circle,draw},thick, scale=0.8]

 \node[inner sep=1](1) at (-1.5,0){$U(N)$};
 \node[rectangle, draw, minimum height=30pt, minimum width=30pt,
thick](2) at (.5,0){$ 2N$};

 \node[circle, draw, thick](3) at (6,0){$U(1)$};
 \node[circle, draw, thick](4) at (8,0){\small{$SU(2)$}};
 \node[rectangle, draw, minimum height=30pt, minimum width=30pt,
thick](5) at (10,0){$R_{0,N}$};

\draw[=>] (1.5,0)-- node[draw=none, above] {S-duality} (5,0) ;

\draw[-] (1) -- (2);
\draw[-] (3) -- (4);
\draw[-] (4) -- (5);

\end{tikzpicture}
\caption{S-dual theory of a single quiver node.} \label{fig:sduality}
\end{center}
 \end{figure}
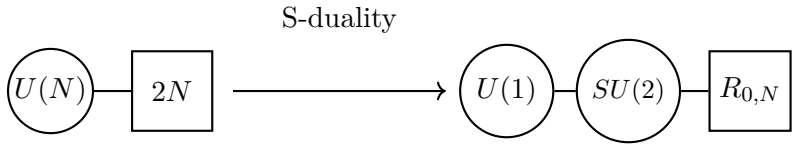
 
  The gain is that we now have isolated a $U(1)$ theory with two flavors.  
 
\item[$\triangleright$] We now recouple the dual theory to the rest of the quiver by gauging its global symmetry appropriately.
\end{itemize}
We obtain what we will call a `modified ADE quiver'.

We are now ready to discuss monopole deformation along any node in a 3d ADE quiver gauge theory.
Let us then describe the strategy we will apply throughout this paper. Although its implementation will vary drastically from one example to the next, the basic idea is exactly the same. 

In the original ADE theory we wish to deform the superpotential with a monopole operator $V_{-}$ relative to a $U(N)$ node, i.e.
\be
\Delta W = m V_{-}\:.
\ee
This operator has charge one with respect to  the topological $U(1)$ that shifts the scalar dual to the overall photon in $U(N)$. It is defined as the operator that creates a pointlike object with magnetic charges $(-1, \underbrace{0, \dots, 0}_{N-1})$ with respect to  to the $N$ $U(1)$ groups.
As was shown in \cite{Gaiotto:2008ak}, such operators have R-charge one and hence dimension one in the IR. (By the adjoint action of $U(N)$ we can always move the magnetic charge to the first entry.)

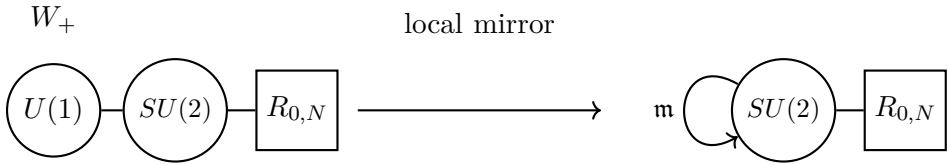
\begin{figure} 
 \begin{center}
\begin{tikzpicture}[->, every node/.style={circle,draw},thick, scale=0.8]
\node[draw=none] at (0,1.5){$W_+$};
 \node[circle, draw, thick](3) at (0,0){$U(1)$};
 \node[circle, draw, thick](4) at (2,0){\small{$SU(2)$}};
 \node[rectangle, draw, minimum height=30pt, minimum width=30pt,
thick](5) at (4,0){$R_{0,N}$};

\node[circle, draw, thick](1) at (12,0){\small{$SU(2)$}};
 \node[rectangle, draw, minimum height=30pt, minimum width=30pt,
thick](2) at (14,0){$R_{0,N}$};

\path[every node/.style={font=\sffamily\small,
 		fill=white,inner sep=1pt}]
(1) edge [loop, out=150, in=210, looseness=4] 
node[left=0.8mm]{{$\mathfrak{m}$}}  (1);

\draw[-] (1) -- (2);
\draw[-] (3) --(4);
\draw[-] (4) --(5);
\draw[=>] (5,0)-- node[draw=none, above] {local mirror} (9,0) ;
\end{tikzpicture}
\caption{Effective theory of a single quiver node after monopole deformation.} \label{fig:effective}
\end{center}
\end{figure}

Contrary to the original quiver gauge theory, in the dual theory the monopole operator that deforms the theory is charged under the topological symmetry of a $U(1)$ node. 
In fact, by matching the two topological $U(1)$'s, one concludes that turning on a superpotential term involving $V_{-}$ is equivalent to turning on the same type of superpotential involving the monopole operator $W_{+}$ relative to a $U(1)$ node on the dual side (see Section \ref{Sec:monu2}). Using now the result of \cite{Collinucci:2016hpz}, we can equivalently describe the resulting theory as an $SU(2)$ gauge theory coupled to the dimensional reduction of $R_{0,N}$ and to a chiral multiplet in the adjoint of $SU(2)$ despicted in Figure \ref{fig:effective}. In this way we can get a description of the theory without monopole operators. 

We can extract more information from this resulting theory. Specifically, we can study their Higgs branches. 
Before moving on, let us define what we mean by `Higgs branch', a term usually reserved for $\mathcal{N}=4$ theories (in 3d): Before breaking $\mathcal{N}=4 \rightarrow \mathcal{N}=2$, we can distinguish the Higgs and Coulomb branches thanks to the $SU(2)_H \times SU(2)_C$ R-symmetry. The fields of the gauge theory, collectively denoted $\Phi, W_+, {M^i}_j$ for adjoint scalar, monopole operator and meson, are charged under the Cartan of the R-symmetry and the topological $U(1)_T$ as follows:
\be
\begin{array}{c|c|c|c}
&U(1)_{H}&U(1)_{C}&U(1)_T\\ \hline
\Phi & 0 & 1  & 0 \\ \hline
W_+ & 0 & 1 & 1\\ \hline
{M^i}_j & 1 & 0 & 0 \\ \hline
d^2 \theta & -1 & -1 & 0 
\end{array}
\ee
For convenience, we included the charges of the half-superspace measure.
The R-symmetry of the $\cN=2$ subalgebra is $U(1)_{R}^{UV} = U(1)_{H+C}$. The independent $U(1)_{H-C}$ and $U(1)_T$ transformations are ordinary global symmetries from the $\cN=2$ point of view.  The deformation by a monopole operator breaks the supersymmetry in half, but preserves $U(1)_C$ and $U(1)_{H+T}$,  one combination of which will become the infrared R-symmetry. Henceforth, we refer to the branch along which  $U(1)_C$ is preserved, as the `Higgs branch'.

In this paper, we will see in all examples that, the original Higgs branch of the undeformed $\mathcal{N}=4$ theory, which is an algebraic surface with an ADE singularity, will be preserved as a branch of the effective $\mathcal{N}=2$ theory. First one shows that the Higgs branch of the `modified ADE quiver' is the same as for the dual ADE quiver. Second, we will easily see that the monopole deformation does not alter the coordinates of the Higgs branch nor their relations.

The general argument for proving the first statement is simple, and goes as follows: In all examples considered, the part of the `local' theory that we will recouple to the quiver will always contain an $su(2 N)$-valued moment map $X$ with the property that $X^2=0$. This will be true regardless of whether that theory is Lagrangian or not. We now want to show that this property ensures that we can treat $X$ as if it were a meson, satisfying standard $\mathcal{N}=4$ F-term conditions for the Higgs branch.

Now let us prove that such an $X$ can be written as a sum of $N$ bilinears, as is the case of the meson on the $U(N)$ Lagrangian side.
First let us show that rk$(X) \leq N$. The nilpotency of $X$ implies that Im$(X)\subset$ Ker$(X)$. Therefore, rk$(X) =$dim(Im$(X)) \leq $dim(Ker$(X))=2 N-$rk(X). This means that we can write $X$ as follows:
\be
X = \sum_{a=1}^{N} \mathsf{v}_a  \tilde{\mathsf{w}}^a\,,
\ee
where the $\mathsf{v}_a$ are $N$ linearly independent column $2 N$-vectors, and the $\tilde{\mathsf{w}}^a$ are $N$ linearly independent row $2 N$-vectors, where we have suppressed the $2 N$-dimensional indices. Squaring this we obtain:
\be
X^2  = \sum_{a, b=1}^{N} (\tilde{\mathsf{w}}^a \cdot \mathsf{v}_b) \mathsf{v}_a \tilde{\mathsf{w}}^b  = 0 \iff \tilde{\mathsf{w}}^a \cdot \mathsf{v}_b = 0\quad  \forall (a, b)\,.
\ee
This is precisely the content of the F-term equations for a would-be $U(N)$ hypermultiplet $(\mathsf{v},  \tilde{\mathsf{w}})$. Hence, 
the relations satisfied by the meson on the SQCD side will be satisfied by the field $X$ on the dual (possibly non-Lagrangian) side.

Therefore, if we recouple such a theory to the quiver simply by substituting the original meson in all couplings with this new adjoint-valued object $X$, all other equations for this branch will remain intact. In order to rediscover the ADE algebraic variety, we note that it must be described via gauge-invariant coordinates. These can be very complicated concatenations of fields connecting various nodes of the quiver. However, any gauge-invariant cycle that passes through the node in question must be built out of $X$. From this, we draw the striking conclusion that the ADE algebraic surface remains intact despite the monopole deformation.

\subsection{Monopole deformations along $U(1)$ nodes}\label{Sec:U1monop}

In the previous section, we showed how to reduce the problem of any monopole deformation with respect to  a single node to a deformation along a $U(1)$ node with two flavors, albeit at the cost of having a strongly coupled SCFT attached to it. We are now in a position to use a technique we previously developed in \cite{Collinucci:2016hpz}, dubbed `local mirror symmetry' to derive the effective theory for such a deformation. We will review this technique in what follows.

The idea is the following: Given a quiver gauge theory with a deformation by a monopole operator $\Delta W = m_i W_{i, +}$ 
corresponding to the i-th photon, we focus on this i-th node by ungauging the neighbouring nodes. In this way we end up with a $U(1)$ theory with two flavors. Mirror symmetry dual on the i-th node ``in isolation'', is tractable and powerful, as it maps the monopole superpotential deformation to a mass deformation. It is then easy to integrate out the massive modes, reapply the mirror symmetry back, and finally reinsert this resulting theory into the original quiver. The key fact is that this theory and the original one are equivalent in the IR. 

We consider a $U(1)$ node in an $\cN=4$ ADE quiver gauge theory and supplement the superpotential by the term $\Delta W = m W_{\ell,+}$, where $m$ is a parameter, and $W_{\ell, +}$ is the monopole operator charged under the 
topological $U(1)$ of the $\ell$-th node. In other words, it corresponds to the $\ell$-th dual photon.

Since the $U(1)$ node is balanced, it has $N_f=2$ flavors attached to it. In the quiver, this means that either the node is connected to other two $U(1)$ nodes or it is connected to one $U(2)$ node. The full superpotential will include the term
\begin{equation}
W \supset -\phi_\ell \sum_{i=1}^2 q_i \tilde{q}^i + \sum_{i,j=1}^2\tilde q^j{\Psi_j}^iq_i \:.
\end{equation}
Here $\phi_\ell$ is the complex scalar in the vector multiplet relative to the $U(1)$ $\ell$-th node.
When the $\ell$-th node is connceted to a $U(2)$ node, $\Psi$ is the complex scalar in the adjoint representation of $U(2)$ sitting in the vector multiplet of the nearby node. On the other hand, when the $\ell$-th node is connected to two $U(1)$ nodes, the matrix $\Psi$ is diagonal, with the diagonal entries being the complex scalars in the vector multiplets of the adjacent nodes.

Let us follow the procedure outlined above:

\begin{enumerate}
\item[1)] Ungauge the neighbouring nodes of the quiver. This results in a `local quiver theory' with a single $U(1)$ gauge node (see Figure  \ref{fig:Bloc}). 
Let us call this theory B$_{\rm loc}$.

\begin{figure}[t] 
\centering
\begin{tikzpicture}[->,thick, scale=1]
  \node[circle, draw, inner sep= 2pt](L1) at (10,0){$U(1)_\ell$};
  \node[draw, rectangle, minimum width=25pt, minimum height=25pt](L2) at (14,0){$2$};
 \path[every node/.style={font=\sffamily\small,
  		fill=white,inner sep=1pt}]
(L1) edge [bend left] node[above=2mm] {$\tilde q^1,\tilde q^2$}(L2)
(L2) edge [bend left] node[below=2mm] {$q_1,q_2$}(L1)
;

\end{tikzpicture}
\caption{Theory $B_{\rm loc}$.}
 \label{fig:Bloc}
\end{figure}

\item[2)] Apply the mirror symmetry map on this `local quiver theory' B$_{\rm loc}$, obtaining the theory A$_{\rm loc}$. 
The monopole deformation term will be mapped to an off-diagonal mass term for the matter fields in A$_{\rm loc}$, as the original node is balanced.

\item[3)] Integrate out the massive fields in A$_{\rm loc}$, leading to an effective theory ${\tilde {\rm A}_{\rm loc}}$.
Compute the mirror of ${\tilde {\rm A}_{\rm loc}}$, which we call ${\tilde {\rm B}_{\rm loc}}$.

\item[4)] Couple ${\tilde {\rm B}_{\rm loc}}$ back into the original quiver, by trading the $\ell$-th node for it.

\end{enumerate}

Let us see the details of the steps 2) and 3). The deformed superpotential  in the theory~$B_{\rm loc}$~is 
\begin{equation}\label{WBloc}
 W_{B_{\rm loc}} = -\phi_\ell \sum_{i=1,2} q_i\tilde q^i  + \mbox{tr} (\Psi \, q\, \tilde q) + m W_{\ell,+}\:,
\end{equation}
The mirror of an $\mathcal{N}=2$ $U(1)$ theory with two flavors and no superpotential is well 
known: it is again an Abelian theory with two flavors plus two neutral chiral multiplets $A_1$, $A_2$ and superpotential 
\be s_1Q\tilde{Q}+s_2P\tilde{P}.\ee 
Under the mirror map, 
the diagonal components of the meson matrix ${\mathfrak{m}_\alpha}^\beta=q_\alpha\tilde q^\beta$ are mapped to fundamental fields on the mirror side, which we call $s_1$ and $s_2$, whereas the off-diagonal components are mapped to monopole operators $w_+$ and $w_-$. The monopole operator $W_{\ell,+}$ is mapped to an off-diagonal mass term. The fields $\phi_\ell$ and $\Psi$ are gauge invariant fields which will be merely spectators in what follows.
Now, consider our gauge node as $\mathcal{N}=2$ SQED plus the neutral chirals $\phi_\ell$ and $\Psi$ 
with superpotential \eqref{WBloc} and exploit the mirror map dictionary: we find the mirror theory which is again 
SQED with two flavors (see Figure  \ref{fig:Aloc}) and superpotential 
\be\label{SQEDmirrMassDef} W_{A_{\rm loc}}=s_1Q\tilde{Q}+s_2P\tilde{P}-\phi_\ell (s_1+s_2)+\mbox{tr} (\Psi \, \mathfrak{m})+mP\tilde{Q}\:,\ee
\begin{figure}[t] 
\centering
\begin{tikzpicture}[->,thick, scale=1]
  \node[circle, draw, inner sep= 2pt](L1) at (10,0){$\widetilde{U(1)}_\ell$};
  \node[draw, rectangle, minimum width=25pt, minimum height=25pt](L2) at (14,0){$2$};
 \path[every node/.style={font=\sffamily\small,
  		fill=white,inner sep=1pt}]
(L1) edge [bend left] node[above=2mm] {$\tilde P,\tilde Q$}(L2)
(L2) edge [bend left] node[below=2mm] {$P,Q$}(L1)
;

\end{tikzpicture}
\caption{Theory $A_{\rm loc}$.}
 \label{fig:Aloc}
\end{figure}
where now $\mathfrak{m}$ is given by
\begin{equation}
\mathfrak{m}\equiv\left(\begin{array}{cc}
  s_1 & w_+ \\             
  w_- & s_2 \end{array}\right)\:.
\end{equation}
We now integrate out the massive fields $P$ and $\tilde{Q}$, keeping $\Psi$ until the end since it is 
coupled to other fields in the quiver. We are left with 
\be W_{A_{\rm loc}}^{\rm eff}=-\phi_\ell \mbox{tr}\,\mathfrak{m} + \mbox{tr} (\Psi\,\mathfrak{m}) - \frac{s_1s_2}{m}Q\tilde{P}\:.\ee 
The theory ${\tilde {\rm A}_{\rm loc}}$ in the case at hand is SQED with one flavor and the above superpotential. In order to 
complete our analysis, we now derive the mirror of 
this model and ``reconnect'' the resulting theory to the quiver. Since the mirror of SQED with one flavor (and no superpotential) is the XYZ model, we get 
a WZ model with superpotential 
\be 
W_{B_{\rm loc}}^{\rm eff}=\mathcal{X}\,\mathcal{Y}\,\mathcal{Z}-\phi_\ell \mbox{tr}\,\mathfrak{m} + \mbox{tr} (\Psi\,\mathfrak{m}) - \frac{s_1s_2}{m}\mathcal{X}.
\ee 
The fields $\mathcal{Y}$ and $\mathcal{Z}$ are dual to the monopole operators $w_\pm$ and are hence identified with the off-diagonal components  of the field $\mathfrak{m}$. Then, after integrating out the massive field $\phi_\ell$, the effective superpotential can be rewritten as 
\be \label{modpot}
W_{B_{\rm loc}}^{\rm eff}= \mbox{tr}(\Psi\, \tilde{\mathfrak{m}})-\frac{\mathcal{X}}{m}\text{det}\tilde{\mathfrak{m}}\:,
\ee
where  $\tilde{\mathfrak{m}}$ is the traceless part of $\mathfrak{m}$.
Notice that all the above terms are $U(2)$ invariant. 
If instead of the generic $U(2)$ matrix $\Psi$, we take the matrix 
\begin{equation}
\left(\begin{array}{cc}
  \phi_{\ell-1} & 0 \\             
  0 & \phi_{\ell+1} \end{array}\right)\:,
\end{equation}
we obtain the case when the $U(1)$ node is between other two $U(1)$ nodes.

Note, that in \cite{Benini:2017dud}, a technique was developed to handle a monopole deformation on a $U(N)$ theory with $N+1$ flavors, which generalizes this result.\footnote{Using our local mirror symmetry procedure we can actually argue that the determinant superpotential proposed in \cite{Benini:2017dud} is actually generated for arbitrary $N$: in fact, by giving diagonal vev to the meson $M$, the theory with gauge group $U(N)$ can be higgsed to $U(1)$ with two flavors and determinant superpotential whose coefficient is equal to that of the original superpotential times the determinant of $\langle M\rangle$. As we have just seen the $N=1$ superpotential is non-zero. Hence, also the original superpotential (for generic $N$) must be non-vanishing.}

In appendix \ref{app:DNU1extnode}, we carry out this strategy explicitly for the Abelian nodes of the $D_N$ series. The technique, however, can be applied for any Abelian node of the $E$ series \emph{mutatis mutandis}.

\section{$U(2)$ nodes}\label{Sec:U2node}

Having explained our general strategy for deforming $U(N)$ nodes, we now want to start with the special case of a monopole deformation for one $U(2)$ node inside a quiver. If the node is balanced, as it is in quiver ADE theories, after ungauging the nearby nodes, one obtains a $U(2)$ gauge theory with four flavors. Unfortunately in this case the procedure used for the Abelian node is not useful anymore. In fact, it would require understanding non-Abelian $\cN=2$ mirror symmetry, that is not only technically difficult, but also prone to instanton corrections. Here we follow a slightly different procedure. We first replace the $U(2)$ local theory with a theory containing a $U(1)$ gauge factor and then we follow the abelian procedure of the previous section to describe the deformed theory.

\subsection{$U(2)$ vs $SU(2)\times U(1)$ dual theories }\label{Sec:BBduality}

We now present the dual 3d $\mathcal{N}=4$ theories crucial for describing monopole deformations along a $U(2)$ node.
\begin{description}
\item {\it \textbf{Theory 1}}: $U(2)$ gauge theory with four flavors. It has $SU(4)$ global symmetry rotating the four flavors.
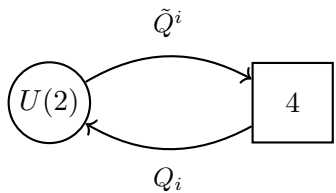
\begin{center}
\begin{tikzpicture}[->,thick, scale=0.8]
  \node[circle, draw, inner sep= 2pt](L1) at (10,0){$U(2)$};
  \node[draw, rectangle, minimum width=30pt, minimum height=30pt](L2) at (14,0){$4$};
 \path[every node/.style={font=\sffamily\small,
  		fill=white,inner sep=1pt}]
(L1) edge [bend left] node[above=2mm] {$\tilde Q^i$} (L2)
(L2) edge [bend left] node[below=2mm] {$Q_i$}(L1)
;
\end{tikzpicture}
\end{center}
The Higgs branch is parametrized by the gauge invariant meson matrix ${M_i}^j=Q_i^a\tilde Q^j_a$, that is in the bifundamental representation of the $SU(4)$ flavor group. The $\mathcal{N}=4$ superpotential of the theory is
\begin{equation}
  W = Q_i^a \breve{\Psi}_a^b \tilde Q^i_b \:
\end{equation}
(where $\breve{\Psi}$ is the adjoint scalar in the $U(2)$ vector multiplet). The corresponding 
F-term equations are
\begin{equation}
  Q_i^a \breve{\Psi}_a^b =0 \,,\qquad  \breve{\Psi}_a^b \tilde Q^i_b=0 \,,\qquad Q_i^a \tilde Q^i_b=0\:.
\end{equation}
The last equations constrain the meson matrix to satisfy
\begin{equation}
\mbox{tr} M =0 \,\,,\qquad\qquad M^2=0 \:.
\end{equation}
\item {\it \textbf{Theory 2}}: $U(1)\times SU(2)$ gauge theory with three flavors in the foundamental of $SU(2)$ and one in the bifundamental of $U(1)\times SU(2)$. This theory has an $SO(6)$ flavor symmetry rotating the three fundamental fields.
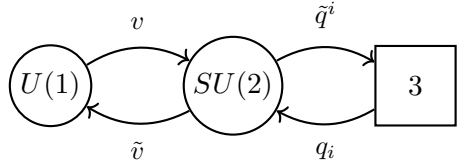
\begin{center}
\begin{tikzpicture}[->,thick, scale=0.8]
  \node[circle, draw, inner sep= 2pt](L0) at (7,0){$U(1)$};
  \node[circle, draw, inner sep= 2pt](L1) at (10,0){$SU(2)$};
  \node[draw, rectangle, minimum width=30pt, minimum height=30pt](L2) at (13,0){$3$};
 \path[every node/.style={font=\sffamily\small,
  		fill=white,inner sep=1pt}]
(L0) edge [bend left] node[above=2mm] {$ v$} (L1)
(L1) edge [bend left] node[below=2mm] {$\tilde v$}(L0)
(L1) edge [bend left] node[above=2mm] {$\tilde q^i$} (L2)
(L2) edge [bend left] node[below=2mm] {$q_i$}(L1)
;
\end{tikzpicture}
\end{center}
The superpotential is
\begin{equation}\label{W2SU2U1}
  W =  q_i^\alpha {\Phi_\alpha}^\beta \tilde q^i_\beta + v^\alpha {\Phi_\alpha}^\beta \tilde v_\beta - \phi \,v^\alpha\tilde v_\alpha\:,
\end{equation}
that gives the following F-terms equations\footnote{Notice that $\Phi$ is a traceless $2\times 2$ matrix. This is the reason why its F-term constrains only the traceless part of $q_i^\alpha \tilde q^i_\beta+v^\alpha\tilde v_\beta$.}
\begin{equation}
  q_i^\alpha \Phi_\alpha^\beta =0 \,,\qquad  \Phi_\alpha^\beta \tilde q^i_\beta=0 \,,\qquad v^\alpha (\Phi_\alpha^\beta -\phi \,\delta_\alpha^\beta)=0 \,,\qquad  (\Phi_\alpha^\beta -\phi \, \delta_\alpha^\beta) \tilde v_\beta=0 \,,
\end{equation}
\begin{equation}\label{FtermsTh2}
  q_i^\alpha \tilde q^i_\beta+v^\alpha\tilde v_\beta - \frac12 (  q_i^\gamma \tilde q^i_\gamma+v^\gamma\tilde v_\gamma)\delta^\alpha_\beta=0 \,, \qquad\qquad v^\alpha\tilde v_\alpha=0 \:.
\end{equation}
In particular we can rewrite the last one as $\epsilon^{\beta\alpha}v_\alpha\tilde{v}_\beta=0$ which means that the matrix $v_\alpha \tilde{v}_\beta$ is symmetric (we  define $v_\alpha\equiv \epsilon_{\alpha\beta}v^\beta$).
The gauge invariant coordinates are now given by the singlet 
\be
 ((v\tilde v)) \equiv v^\alpha\tilde v_\alpha 
\ee 
(where $((...))$ means that the $SU(2)$ indices are contracted) and the meson matrix in the Adjoint representation of the $SO(6)$ flavor group:
\begin{equation}\label{mathcalM}
{\cal M} = \left(\begin{array}{ccc}
q_i^\alpha \tilde q^j_\alpha && q_i^\alpha\epsilon_{\alpha\beta}q_k^\beta \\ \\  -\tilde q^\ell_\alpha\epsilon^{\alpha\beta} \tilde q^j_\beta  &&  - \tilde q^\ell_\alpha q_k^\alpha \\
\end{array}\right) \equiv
\left(\begin{array}{ccc}
{A_i}^j && b_{ik} \\ \\ - c^{\ell j}  &&  - {(A^t)^\ell}_k \\
\end{array}\right) \:,
\end{equation}
where $A$ is a $3\times 3$ complex matrix, while $b$ and $c$ are antisymmetric matrices.
\end{description}

The two theories have the same Higgs and Coulomb branches \cite{Gaiotto:2008ak}. In particular, the topological current relative to the $U(2)$ node of the first theory is mapped to the topological symmetry relative to the $U(1)$ gauge group factor in the second theory. Correspondingly the monopoles operators of the two theories (with the same R-charge) that have equal charge with respect to the topological symmetry are exchanged by the duality map.

\

Let us  work out the duality map between the two Higgs branches: we first find the correspondence between the invariant coordinates of the two theories and second we show that the relations match as well. We will crucially use the fact that the two spaces have the same flavor group.
We begin by finding the isomorphism between the groups $SU(4)$ and $SO(6)$ (or more precisely $Spin(6)$). An element $g\in SU(4)$ has a canonical action on $\C^4$. This allows to define an action on $\C^6=\C^4\wedge\C^4$ as well. Define the basis $\{e^i\wedge e^j\}_{ij=12,13,14,34,42,23}$ of $\C^6$, by using the basis $\{e^i\}_{i=1,...,4}$  of $\C^4$. Given $g\in SU(4)$, one can define a map $\rho(g)\in Spin(6)$ by
\begin{equation}
  \rho(g) \,\mathsf{v}= \mathsf{v}_{ij}\,\rho(g) \,  e^i\wedge e^j = \mathsf{v}_{ij} (g\,e^i)\wedge (g\, e^j)\:.
\end{equation}
If we now pass to the corresponding Lie algebra, taken $H\in su(4)$ and $\mathsf{v}\in\C^6$, one defines $\rho(H)\in so(6)$ by
\begin{equation}
  \rho(H) \,\mathsf{v} = \mathsf{v}_{ij}\,\rho(H) \, ( e^i\wedge e^j ) \:,
\end{equation}
where $\rho(H)$ acts on the basis elements as
\begin{eqnarray}
\rho(H) \, ( e^i\wedge e^j ) &=&  \left[ (H\,e^i)\wedge e^j +e^i\wedge (H\,e^j)\right] =\left[ {H_k}^i\,e^k\wedge e^j +{H_\ell}^j\, e^i\wedge e^\ell\right]  \nonumber \\
&=& \sum_{k\ell\in I}  e^k\wedge e^\ell \left( {H_k}^i\delta_\ell^j- {H_\ell}^i\delta_k^j +  {H_\ell}^j\delta_k^i -  {H_k}^j\delta_\ell^i \right)\:,
\end{eqnarray}
where $I=\{12,13,14,34,42,23\}$.
One can then apply this map to the traceless part of the meson matrix $M$, i.e. $M=\frac{\mbox{tr}M}{4}\,\mathbb{1}_4 + \tilde{M}$, with $\tilde{M}$ a matrix  in the adjoint of $SU(4)$. Let us write $\tilde{M}$ in a block diagonal form, as a map from $\mathbb{C}\oplus \mathbb{C}^3$ to itself:
\begin{equation}
\tilde{M} = \left(\begin{array}{ccc}
-\mbox{tr}B  && x^t  \\ y  &&  B  \\
\end{array}\right) \:.
\end{equation}
We can work out the duality map by identifying the $6\times 6$ matrix $\rho(\tilde{M})$ with the matrix \eqref{mathcalM}:
\begin{equation}
  B \leftrightarrow A - \frac{\mbox{tr}A}{2} \mathbb{1}_3 \:, \qquad   x^i \leftrightarrow u^i\equiv \frac12\epsilon^{ijk}b_{jk} \:, \qquad y_i \leftrightarrow w_i\equiv \frac12\epsilon_{ijk}c^{jk} \:.
\end{equation}
The remaining invariant coordinates, i.e. tr$M$ and $((v\tilde{v}))$ respectively are mapped to each other:
\begin{equation}
  \frac{\mbox{tr}\,M}{2} \,\,\leftrightarrow\,\, ((v\tilde{v}))\:.
\end{equation}
We can invert this map to show that in {\it Theory 2} we can collect the gauge invariant coordinates into a $4\times 4$ meson matrix 
\begin{equation}
\mathfrak{M}= X + \frac{((v\tilde{v}))}{2} \,\mathbb{1}_4= \left(\begin{array}{ccc}
\frac{\mbox{tr}A}{2}  && u^t  \\ w  &&  A-\frac{\mbox{tr}A}{2}\mathbb{1}_3    \\
\end{array}\right) + \frac{((v\tilde{v}))}{2}
\left(\begin{array}{ccc}
1  &&  \\   &&  \mathbb{1}_3    \\
\end{array}\right) 
 \:,
\end{equation}
where $X=\rho^{-1}(\mathcal{M})$ is a traceless $4\times 4$ matrix.

We now want to show that by imposing the F-term conditions \eqref{FtermsTh2} on this $4\times 4$  matrix $\mathfrak{M}$ we get tr$\mathfrak{M}=0$ and $\mathfrak{M}^2=0$. First we impose $((v\tilde v))=0$, that immediately gives tr$\mathfrak{M}=0$. Then, $\mathfrak{M}^2=X^2$, where
\begin{equation}
X^2 = \left(\begin{array}{ccc}
\left(\frac{\mbox{tr}A}{2}\right)^2+u^t\cdot w  && u^t\cdot A  \\ A\cdot w  &&  \left(A-\frac{\mbox{tr}A}{2}\mathbb{1}_3  \right)^2+w\cdot u^t  \\
\end{array}\right) \:.
\end{equation}
Imposing now \eqref{FtermsTh2} and using the fact that $v_a\tilde v_b$ is a symmetric matrix, we obtain $X^2=0$ (remember that tr$A=q_i^\gamma\tilde q^i_\gamma$). In fact:
\begin{eqnarray}
u^t\cdot w &=& \frac14 \epsilon^{ijk}q_j^\alpha\epsilon_{\alpha\beta}q_k^\beta \, \epsilon_{imn}\tilde q^m_\gamma\epsilon^{\gamma\delta}\tilde 					q^n_\delta =  \frac14 q_j^\alpha\epsilon_{\alpha\beta}q_k^\beta \tilde q^m_\gamma\epsilon^{\gamma\delta}\tilde q^n_\delta \, (\delta_m^j				\delta_n^k-\delta_n^j\delta_m^k) =  \nonumber \\
		&=& \frac12 \epsilon_{\alpha\beta}\epsilon^{\gamma\delta}\, q_j^\alpha\tilde{q}^j_\gamma q_k^\beta\tilde{q}^k_\delta = 
			\frac12 \epsilon_{\alpha\beta}\epsilon^{\gamma\delta}\, \left( v^\alpha\tilde v_\gamma - \frac{\mbox{tr}A}{2}\delta_\gamma^\alpha \right) 				\left( v^\beta\tilde v_\delta - \frac{\mbox{tr}A}{2}\delta_\delta^\beta \right)   =  \\ 
 		&=& \frac12 \epsilon_{\alpha\beta}\epsilon^{\gamma\delta}\, v^\alpha\tilde{v}_\gamma v^\beta\tilde{v}_\delta +\epsilon_{\beta\alpha}							\epsilon^{\alpha\delta}\, v^\beta\tilde v_\delta \, \frac{\mbox{tr}A}{2} -\frac12\epsilon_{\beta\alpha}\epsilon^{\alpha\beta} \left( 									\frac{\mbox{tr}A}{2} \right)^2 = - \left( \frac{\mbox{tr}A}{2} \right)^2 \:,\nonumber \\
\label{relutA} (u^t\cdot A)^m &=& \frac12\epsilon^{ijk}q_j^\alpha\epsilon_{\alpha\beta}q_k^\beta \, q_i^\gamma\tilde q^m_\gamma =  \frac12 						(\epsilon^{ijk}q_j^\alpha q_k^\beta  q_i^\gamma)\epsilon_{\alpha\beta}\tilde q^m_\gamma =  0 \:,\\
\label{relAw} (A\cdot w)_m &=& \frac12 q_m^\gamma\tilde q^i_\gamma \epsilon_{ijk}\tilde q^j_\alpha\epsilon^{\alpha\beta}\tilde q^k_\beta  = 						\frac12  q_m^\gamma \epsilon^{\alpha\beta}(\epsilon_{ijk}\tilde q^i_\gamma \tilde q^j_\alpha \tilde q^k_\beta) =  0 \:, 
\end{eqnarray}
\begin{eqnarray}
\label{relwut} {(w\cdot u^t)_i}^\ell &=& \frac14 \epsilon_{ijk}\tilde q_\gamma^j \epsilon^{\gamma\delta} \tilde q_\delta^k \, \epsilon^{\ell m n}q_m^				\alpha\epsilon_{\alpha\beta} q_n^\beta = \nonumber\\ 
		&=&  \frac14 \tilde q_\gamma^j \epsilon^{\gamma\delta} \tilde q_\delta^k \, q_m^\alpha\epsilon_{\alpha\beta} q_n^\beta \,\delta_i^									\ell(\delta_j^m\delta_k^n-\delta_j^n\delta_k^m)
			+	\frac12 \tilde q_\gamma^j \epsilon^{\gamma\delta} \tilde q_\delta^k \, q_m^\alpha\epsilon_{\alpha\beta} q_n^\beta \, \delta_i^n(\delta_j^				\ell\delta_k^m-\delta_j^m\delta_k^\ell) = \nonumber\\
 		&=& ( u^t\cdot w )\,\delta_i^\ell - \epsilon_{\alpha\beta}\epsilon^{\gamma\delta} q_i^\beta\tilde{q}_\gamma^\ell\tilde{v}_\delta v^\alpha
			+\epsilon_{\alpha\beta}\epsilon^{\gamma\delta} q_i^\beta\tilde{q}_\gamma^\ell\delta_\delta^\alpha 	\,\left(\frac{\mbox{tr}A}{2}\right) =	 						 \nonumber\\
  		&=&		( u^t\cdot w )\,\delta_i^\ell+ \epsilon^{\gamma\delta} q_i^\beta\tilde{q}_\gamma^\ell\tilde{v}_\delta v_\beta 
  			+ \frac{\mbox{tr}A}{2}\,q_i^\alpha\tilde{q}_\alpha^\ell 		  =  \\
		&=&  ( u^t\cdot w )\,\delta_i^\ell+ 	 \epsilon^{\gamma\delta} q_i^\beta\tilde{q}_\gamma^\ell\tilde{v}_\beta v_\delta  	+ \frac{\mbox{tr}A}{2}					\,q_i^\alpha\tilde{q}_\alpha^\ell  = \nonumber \\
        &=&	  		- \left( \frac{\mbox{tr}A}{2} \right)^2\,\delta_i^\ell + (q_i^\beta\tilde v_\beta)\, (v^\gamma\tilde q^\ell_\gamma) + 
        	\frac{\mbox{tr}A}{2}\,q_i^\alpha\tilde{q}_\alpha^\ell  \:, \nonumber\\
{\left[\left(A-\frac{\mbox{tr}A}{2}\mathbb{1}_3  \right)^2\right]_i}^\ell &=&\left( q_i^\alpha\tilde q^j_\alpha - \frac{\mbox{tr}A}{2}\delta_i^j 							\right) 				\left( q_j^\beta\tilde q^\ell_\beta- \frac{\mbox{tr}A}{2}\delta_j^\ell \right) =
			\nonumber\\
		&=&  q_i^\alpha\left( \tilde q_\alpha^j q^\beta_j - \frac{\mbox{tr}A}{2}\delta_\alpha^\beta \right) \tilde q_\beta^\ell - \frac{\mbox{tr}A}{2} \left( 			q_i^\alpha\tilde q^\ell_\alpha- \frac{\mbox{tr}A}{2}\delta_i^\ell \right) =  \\
		&=& -   (q_i^\alpha\tilde v_\alpha)\, (v^\beta\tilde q^\ell_\beta) - \frac{\mbox{tr}A}{2} q_i^\alpha\tilde q^\ell_\alpha + \left(\frac{\mbox{tr}A}{2}				\right)^2\delta_i^\ell \:.\nonumber
\end{eqnarray}
In the relation \eqref{relutA} and \eqref{relAw} we used the fact that $\alpha,\beta,\gamma$ run only from 1 to 2. Hence $\gamma$ must be equal to either $\alpha$ or $\beta$, say for example $\alpha=\gamma=1$, and then we have a contraction of an even with an odd combination (in the example $\epsilon^{ijk}q_j^1q_i^1=0$). In the relation \eqref{relwut} we expressed the product of epsilon tensors in terms of delta functions
$$ \epsilon_{ijk}\epsilon^{\ell mn} =\delta_i^\ell(\delta_j^m\delta_k^n-\delta_j^n\delta_k^m) - \delta_i^m(\delta_j^\ell\delta_k^n-\delta_j^n\delta_k^\ell)+  \delta_i^n(\delta_j^\ell\delta_k^m-\delta_j^m\delta_k^\ell) \:, $$
and noticed that the first term produces a term proportional to $u^t\cdot w$, while the other two terms are equal to each other.

\

As regards the Coulomb branches of the two theories: In each case there is one topological $U(1)_J$ symmetry with R-charge one monopole operators charged under it, enhancing the symmetry to $SU(2)$. These monopole operators are mapped to each other.

\subsection{The external $U(2)$ node of the $E_7$ quiver}

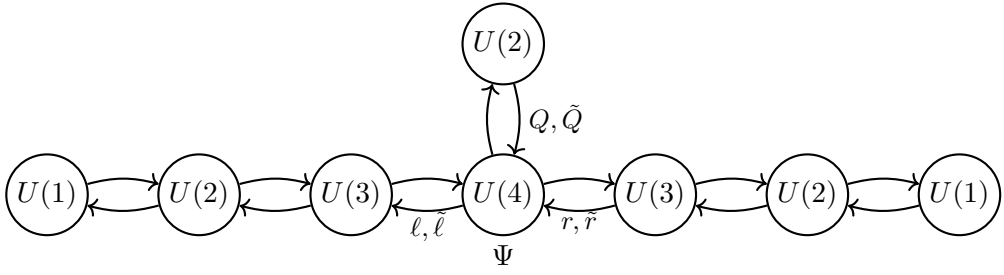
\begin{figure}[ht!]
\begin{center}
\begin{tikzpicture}[->, thick, scale=0.8]
  \node[circle, draw, inner sep= 2pt](V2) at (-5.5,0){$U(1)$};
  \node[circle, draw, inner sep= 2pt](V3) at (-3,0){$U(2)$};
  \node[circle, draw, inner sep= 2pt](V4) at (-.5,0){$U(3)$};
  \node[circle, draw, inner sep= 2pt](V5) at (2,0){$U(4)$};
  \node[circle, draw, inner sep= 2pt](V6) at (4.5,0){$U(3)$};
    \node[circle, draw, inner sep= 2pt](V7) at (7,0){$U(2)$};
  \node[circle, draw, inner sep= 2pt](V1) at (9.5,0){$U(1)$};
    \node[circle, draw, inner sep= 2pt](V8) at (2,2.5){$U(2)$};
  \node at (2,-1){$\Psi$};
  
 \path[every node/.style={font=\sffamily\small,
  		fill=white,inner sep=1pt}]
(V2) edge [bend left=15]   (V3)
(V3) edge [bend left=15]   (V2)
(V3) edge [bend left=15]   (V4)
(V4) edge [bend left=15]   (V3)
(V4) edge [bend left=15]  node[below=4mm] {$\ell,\tilde \ell$} (V5)
(V5) edge [bend left=15]   (V4)
(V5) edge [bend left=15]  node[below=4mm] {$r,\tilde r$} (V6)
(V6) edge [bend left=15]   (V5)
(V6) edge [bend left=15]   (V7)
(V7) edge [bend left=15]   (V6)
(V5) edge [bend left=15]  node[right=5mm] {$Q,\tilde Q$} (V8)
(V8) edge [bend left=15]   (V5)
(V1) edge [bend left=15]   (V7)
(V7) edge [bend left=15]   (V1)

;

\end{tikzpicture}
\caption{$E_7$ quiver.}\label{Fig:E7quiv}
\end{center}
\end{figure}

One external node in the $E_7$ quiver is a $U(2)$ node attached to the $U(4)$ node (see Figure \ref{Fig:E7quiv}).

By ungauging the $U(4)$ node we obtain the {\it Theory 1} above. We can then apply the duality map and obtain the {\it Theory 2} with $U(1)\times SU(2)$ gauge group, one bifundamental hypermultiplet $(v,\tilde v)$ and three hypermultiplets in the fundamental representation of $SU(2)$. 

\begin{figure}[ht!]
\begin{center}
\begin{tikzpicture}[->,thick, scale=0.8]
  \node[circle, draw, inner sep= 2pt](V2) at (-5.5,0){$U(1)$};
  \node[circle, draw, inner sep= 2pt](V3) at (-3,0){$U(2)$};
  \node[circle, draw, inner sep= 2pt](V4) at (-.5,0){$U(3)$};
  \node[circle, draw, inner sep= 2pt](V5) at (2,0){$U(4)$};
  \node[circle, draw, inner sep= 2pt](V6) at (4.5,0){$U(3)$};
    \node[circle, draw, inner sep= 2pt](V7) at (7,0){$U(2)$};
  \node[circle, draw, inner sep= 2pt](V1) at (9.5,0){$U(1)$};
    \node[circle, draw, inner sep= 2pt](V8) at (2,2.5){{\small $SU(2)$}};
    \node[circle, draw, inner sep= 2pt](V9) at (2,5){$U(1)$};
  \node at (2,-1){$\tilde \Psi+\frac{\psi}{2} \mathbb{1}_4$};
  \node at (3,5){$\phi'$};

 \path[every node/.style={font=\sffamily\small,
  		fill=white,inner sep=1pt}]
(V2) edge [bend left=15]   (V3)
(V3) edge [bend left=15]   (V2)
(V3) edge [bend left=15]   (V4)
(V4) edge [bend left=15]   (V3)
(V4) edge [bend left=15] node[below=4mm] {$\ell,\tilde \ell$}  (V5)
(V5) edge [bend left=15]   (V4)
(V5) edge [bend left=15] node[below=4mm] {$r,\tilde r$}  (V6)
(V6) edge [bend left=15]   (V5)
(V6) edge [bend left=15]   (V7)
(V7) edge [bend left=15]   (V6)
(V5) edge [bend left=15]  node[right=5mm] {$q,\tilde q$} (V8)
(V9) edge [bend left=15]  (V8)
(V8) edge [bend left=15]  node[right=5mm] {$v,\tilde v$} (V9)
(V8) edge [bend left=15]   (V5)
(V1) edge [bend left=15]   (V7)
(V7) edge [bend left=15]   (V1)

;

\end{tikzpicture}
\caption{Modified $E_7$ quiver (with $\phi'=\phi+\psi$ the only $U(1)$ scalar coupled to $v,\tilde{v}$).} 
\label{fig:modE7quiv}
\end{center}
\end{figure}
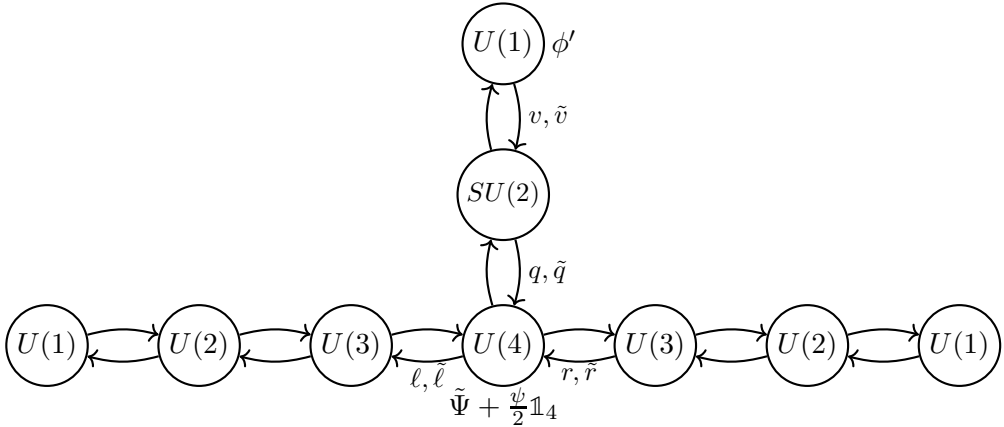

By gauging the $SU(4)$ flavor symmetry one can attach this theory back to the quiver. 
 The fields $q_i,\tilde q^i$ are uncharged with respect to the diagonal $U(1)_d$ of $U(4)$. The field $v,\tilde{v}$ may be charged with respect to this $U(1)_d$ (in fact we will see that this indeed is the case), but 
 this charge is irrelevant. In fact, one can redefine its generator by adding the generator of the isolated $U(1)$ and making $v,\tilde{v}$ neutral with respect to the new $U(1)_d$.\footnote{As we will see shortly, the fields $v,\tilde v$ are coupled to $U(1)_d$. This is encoded in the superpotential coupling $\psi ((v\tilde{v}))$, where $\psi$ is the complex scalar in the $U(1)_d$ vector multiplet. On the other hand, the local $U(1)\times SU(2)$ theory has already an analogous coupling, i.e. $-\phi ((v\tilde{v}))$. One can then define $\phi'=\phi-\psi$ as the only fields coupled to $((v\tilde v))$. This is equivalent to the redefinition of the $U(1)_d$ generator.}
We can then write the new quiver as in Figure  \ref{fig:modE7quiv}, 
where now the fields $q,\tilde q$ couple only to the $SU(4)$ subgroup of $U(4)$, while $\ell,\tilde{\ell}$ and $r,\tilde{r}$ still couple also to the $U(1)_d$ factor of $U(4)$.

Now we are able to prove that the new theory reproduces the $E_7$ singularity in its Higgs branch. The Higgs branch is described by all the gauge invariants modulo relations. The gauge invariants depends on $Q,\tilde Q$ only via the meson combinations ${M_i}^j=Q_i^a\tilde Q_a^j$.
These combinations satisfy the conditions $M^2=0$ and tr$M$=0 and are coupled to the adjoint field $\Psi$ relative to the $U(4)$ node through the superpotential
\begin{equation}\label{U4U2W}
W \supset \tilde Q^j_a{\Psi_j}^i Q^a_i = \mbox{tr} \,\Psi M    \:.
\end{equation}
All the relations important for defining the $E_7$ singularities and involving ${M_i}^j$ come from differentiating the superpotential with respect to $\Psi$.
One can also write the superpotential by separating the traceless part of the $4\times 4$ matrices, i.e. taking $M=\tilde{M} + \frac{\mbox{tr}M}{4}\mathbb{1}_4$ and $\Psi=\frac{\psi}{2}\mathbb{1}_4+\tilde \Psi$:
\begin{equation}\label{U4U2Wpieces}
W \supset \tilde Q^j_a{\Psi_j}^i Q^a_i = \mbox{tr} \,\tilde \Psi \tilde M  +\tfrac12 \psi \mbox{tr}M  \:.
\end{equation}

We now consider the `modified $E_7$ quiver'. We apply the explicit map between the variables ${M_i}^j$ and $\mathcal{M}$,$((v\tilde{v}))$
to the superpotential \eqref{U4U2W}. Hence in the `modified $E_7$ theory' we will have the superpotential
\begin{equation}\label{U4U2Wth2}
W \supset \mbox{tr} \,\Psi \mathfrak{M} = \mbox{tr} \,\tilde \Psi X + \psi ((v\tilde{v}))\:.
\end{equation}
From this we immediately see that $(v,\tilde v)$ couples to the diagonal $U(1)$ factor of $U(4)$. 
This superpotential manifestly generates the same relation for $\mathfrak{M}$ as the $M$ had in the dual $E_7$ theory.

We now show that \eqref{U4U2Wth2} is the superpotential that one would canonically write for the 3d $\cN=4$ dual theory. We need to concentrate on the first term, as the second one is already in the canonical form. In particular we will see that tr$\tilde{\Psi}X$ in \eqref{U4U2Wth2} can be written in terms of the fundamental fields $q,\tilde{q}$. Let us consider $\Psi=\frac{\psi}{2}\mathbb{1}_4+\tilde \Psi$: the traceless part is sent by the $\rho$ map to a $6\times 6$ antisymmetric matrix, while $\psi$ is a singlet of the $SO(6)$ flavor group:
\begin{equation}
\tilde{\Psi} = \left(\begin{array}{ccc}
\frac{\mbox{tr}\varphi}{2}  && u_\varphi^t  \\ w_\varphi  &&  \varphi-\mathbb{1}_3\frac{\mbox{tr}\varphi}{2}    \\ 
\end{array}\right) 
\qquad\rightarrow\qquad
\Upsilon\equiv \rho(\tilde{\Psi}) = \left(\begin{array}{ccc}
\varphi  && r  \\ \\  \ell  &&  -\varphi^t   \\ 
\end{array}\right)\:,
\end{equation}
with $v_\varphi^i=\frac12\epsilon^{ijk}r_{jk}$ and $w_{\varphi i}=\frac12\epsilon_{ijk}\ell^{jk}$. One can show that
\begin{equation}
\mbox{tr} \,\tilde \Psi X = v_\varphi^t\cdot w+v^t\cdot w_\varphi+\mbox{tr}\varphi A = \frac12( \mbox{tr}rc + \mbox{tr} \ell b ) +\mbox{tr}\varphi A = \frac12 \mbox{tr} \left[\Upsilon\mathcal{M} \right] \:. 
\end{equation}
The fields $q,\,\tilde q$ can be arranged into a ${\bf 6}$ representation of $SU(4)\cong SO(6)$
\begin{equation}\label{qpsiso6}
 q_i^\alpha\,, \tilde q^i_\beta \,\,\,\,\, i=1,...,3  \qquad \rightarrow \qquad \psi_m^\alpha  \,\,\,\,\, m=1,...,6 \qquad \mbox{with}\qquad \psi_i^\alpha=q_i^\alpha \,, \psi_{i+3}^{\alpha}= \epsilon^{\alpha\beta}\tilde q^i_\beta\:.
\end{equation}
In terms of $\psi_m$, the meson matrix is written as $\mathcal{M}_{mn}=\psi_m^\alpha\epsilon_{\alpha\beta}\psi_n^\beta$. Hence the superpotential is now
\begin{equation}\label{WE7mod}
W\supset \frac12 \mbox{tr} \left[\Upsilon\mathcal{M} \right]  = \frac12 \Upsilon^{mn}\psi_m^\alpha\epsilon_{\alpha\beta}\psi_n^\beta \:.
\end{equation}
This is consistent with the standard coupling of the $SO(6)$ gauge field with a field in the ${\bf 6}$ representation.

As already anticipated, the Higgs branch of the dual theory is the same as the $E_7$ quiver gauge theory: The superpotential for the latter is 
\begin{eqnarray}
W_{\rm E_7} = 
	-Q_i^a \breve{\Psi}_a^b \tilde Q^i_b +\ \mbox{tr} \, \Psi \,M + ...
\end{eqnarray}
where the ``$...$'' means terms that do not involve fields belonging to the $U(2)$ and the $U(4)$ nodes.
On the other hand, we have just shown that the total superpotential of the `modified $E_7$' quiver gauge theory can be written as
\begin{eqnarray}
W_{\rm modif \, E_7} =
	q_i^\alpha \Phi_\alpha^\beta \tilde q^i_\beta + v^\alpha \Phi_\alpha^\beta \tilde v^\beta - \phi v^\alpha\tilde v_\alpha +\ \mbox{tr} \,\Psi \, \mathfrak{M} + ...
\end{eqnarray}
with ``$...$'' the same terms as before. 
As explained in Section \ref{Sec:GoalStrategy1}, the fact that $M$ and $\mathfrak{M}$ (both traceless and squaring to zero) couple in the same way to the rest of the quiver implies that the two Higgs branches are the same.

\subsection{An internal $U(2)$ node of the $D_N$ quiver}\label{sec:InternalU2DN}

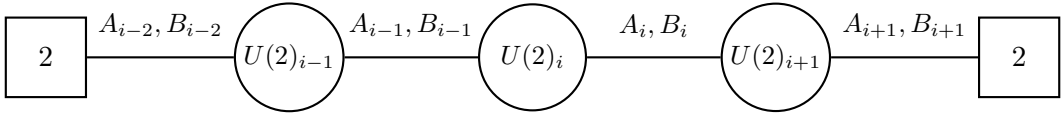
\begin{figure}[ht!]
\begin{center}
\begin{tikzpicture}[thick, scale=0.8]
  \node[draw, rectangle, minimum width=30pt, minimum height=30pt](L0L) at (2,0){$2$};
  \node[draw, rectangle, minimum width=30pt, minimum height=30pt](L0R) at (18,0){$2$};
  \node[circle, draw, inner sep= 2pt](L1) at (6,0){{\small $U(2)_{i-1}$}};
  \node[circle, draw, inner sep= 2pt,, minimum height=40pt](L2) at (10,0){{\small $U(2)_i$}};
  \node[circle, draw, inner sep= 2pt](L3) at (14,0){{\small $U(2)_{i+1}$}};
 \path[every node/.style={font=\sffamily\small,
  		fill=white,inner sep=1pt}]
(L1) edge node[above=2mm] {$A_{i-1},B_{i-1}$}  (L2)
(L2) edge node[above=2mm] {$A_i,B_i$} (L3)
(L1) edge node[above=2mm]{$A_{i-2},B_{i-2}$} (L0L)
(L3) edge node[above=2mm]{$A_{i+1},B_{i+1}$} (L0R)
;
\end{tikzpicture}
\end{center}
\caption{$U(2)$ nodes in the $D_N$ quiver.}\label{Fig:intU2}
\end{figure}

We now consider a segment of the internal chain of $U(2)$ nodes in the $D_N$ quiver (see Figure~\ref{Fig:intU2}).
We want to replace the $U(2)_i$ node by the $SU(2)\times U(1)$ dual theory. We then ungauge the nearby $U(2)$ node, replace the $U(2)$ gauge theory with four flavors with the dual theory and then gauge the $SU(2)\times SU(2)\times U(1)_r\cong SO(4)\times SO(2)$ subgroup of $SO(6)$, where the $U(1)_r$ factor is the relative $U(1)$ gauge symmetry of the two nearby nodes. We need also to gauge the diagonal $U(1)_d$ of the two nodes. 

We obtain the same result if we gauge the full $U(4)$ group, analogously to the $E_7$ case, and then Higgs to the $U(2)\times U(2)$ subgroup. We would then obtain the superpotential \eqref{U4U2Wth2} with now some components of $\Psi$ set to zero, i.e.
\be
W \supset \mbox{tr} \, \Psi\, \mathfrak{M} \:, 
\ee
with 
\begin{eqnarray}
\Psi &=& \frac{\psi}{2} \, \mathbb{1}_4+
\left(\begin{array}{cccc}
\frac{\mbox{tr}\varphi}{2}  & u_\varphi^1  &&   \\  w_{\varphi 1}  & {\varphi_{1}}^1 -\frac{\mbox{tr}\varphi}{2} &&    \\
&&   {\varphi_{2}}^2 -\frac{\mbox{tr}\varphi}{2} &  {\varphi_{2}}^3 \\
&&   {\varphi_{3}}^2  &  {\varphi_{3}}^3 -\frac{\mbox{tr}\varphi}{2} \\
\end{array}\right) \\
&=& \frac{\psi}{2} \left(\begin{array}{cc}
 \mathbb{1}_2 & \\ & \mathbb{1}_2 \\
\end{array}\right) +\frac{{\varphi_1}^1}{2} \left(\begin{array}{cc}
 \mathbb{1}_2 & \\ & -\mathbb{1}_2 \\
\end{array}\right) + \left(\begin{array}{cc}
 \tilde{\Psi}^{i-1} & \\ & \tilde{\Psi}^{i+1} \\
\end{array}\right)
\end{eqnarray}
where we identify $\psi=\varphi^{i-1}+\varphi^{i+1}$ and ${\varphi_1}^1=\varphi^{i-1}-\varphi^{i+1}$ with the diagonal and relative combination of the $U(1)_{i-1}$ and $U(1)_{i+1}$ vector multiplet scalars, and 
$$
\tilde{\Psi}^{i-1} = \left(\begin{array}{cc}
 \frac{{\varphi_2}^2+{\varphi_3}^3}{2} & u_\varphi^1 \\ w_{\varphi 1} & -\frac{{\varphi_2}^2+{\varphi_3}^3}{2} \\
\end{array}\right) \qquad\mbox{and}\qquad 
\tilde{\Psi}^{i+1} = \left(\begin{array}{cc}
 \frac{{\varphi_2}^2-{\varphi_3}^3}{2} & {\varphi_2}^3 \\ {\varphi_3}^2 & \frac{{\varphi_3}^3-{\varphi_2}^2}{2} \\
\end{array}\right)
$$ 
with trasless matrix scalars in the vector multiplet of $SU(2)_{i-1}$ and $SU(2)_{i+1}$ respectively.
We apply the map $\rho$ on the traceless part of $\Psi$ and we obtain the field $\Upsilon\equiv \rho(\tilde \Psi)$ in the form
\be\label{IupSO4SO2}
\Upsilon = \left(\begin{array}{cccccc}
{\varphi_{1}}^1 & & & & & \\
& {\varphi_{2}}^2 & {\varphi_{2}}^3 & & 0 & r_{23} \\  
& {\varphi_{3}}^2 & {\varphi_{3}}^3 & & -r_{23} & 0 \\
& & & -  {\varphi_{1}}^1 & & \\
& 0 & \ell^{23} & & -{\varphi_{2}}^2 & -{\varphi_{3}}^2 \\
& -\ell^{23} & 0 & & -{\varphi_{2}}^3 & -{\varphi_{3}}^3 \\
\end{array}\right) \:,
\ee
where it is manifest the $SO(4)\times SO(2)$ structure (we call the two pieces $\Upsilon_{\rm SO(4)}$ and $\Upsilon_{\rm SO(2)}$).
The three flavors $q_i,\tilde{q}^i$ ($i=1,2,3$) splits into two flavors $q_i,\tilde{q}^i$ ($i=2,3$) and one flavor $(q_1,\tilde{q}^1)$.
As we did in \eqref{qpsiso6}, one can substitute the flavor fields $q_i,\tilde{q}^i$ ($i=2,3$) with
\begin{equation}
\psi_m^\alpha  \,\,\,\,\, m=1,...,4 \:,\qquad \mbox{where}\qquad \psi_{i-1}^\alpha=q_{i}^\alpha\,,\,\,\,\,\,\,\,\, \psi^{\alpha}_{i+1}= \epsilon^{\alpha\beta}\tilde q^{i}_\beta  \,\,\,\,\, i=2,3   \:.
\end{equation}

Inserting \eqref{IupSO4SO2} into \eqref{WE7mod} the superpotential becomes
\begin{eqnarray}
 \tfrac12 \mbox{tr} \Upsilon \mathcal{M} &=& \tfrac12 \mbox{tr}
{\footnotesize \left(\begin{array}{cccccc}
{\varphi_{1}}^1 & & & & & \\
& {\varphi_{2}}^2 & {\varphi_{2}}^3 & & 0 & r_{23} \\  
& {\varphi_{3}}^2 & {\varphi_{3}}^3 & & -r_{23} & 0 \\
& & & -  {\varphi_{1}}^1 & & \\
& 0 & \ell^{23} & & -{\varphi_{2}}^2 & -{\varphi_{3}}^2 \\
& -\ell^{23} & 0 & & -{\varphi_{2}}^3 & -{\varphi_{3}}^3 \\
\end{array}\right)
\left(\begin{array}{cccccc}
{A_{1}}^1 & & & & & \\
& {A_{2}}^2 & {A_{2}}^3 & & 0 & b_{23} \\  
& {A_{3}}^2 & {A_{3}}^3 & & -b_{23} & 0 \\
& & & -  {A_{1}}^1 & & \\
& 0 & c^{23} & & -{A_{2}}^2 & -{A_{3}}^2 \\
& -c^{23} & 0 & & -{A_{2}}^3 & -{A_{3}}^3 \\
\end{array}\right) }\nonumber\\  \\
&=& {\varphi_1}^1 q_1^\alpha\tilde q_\alpha +\tfrac12 \Upsilon_{SO(4)}^{mn}\psi_m^\alpha\epsilon_{\alpha\beta}\psi_n^\beta \:. \nonumber
\end{eqnarray}
We immediately read that $q_1,\tilde{q}^1$ couples to the relative $U(1)_r\cong SO(2)$ of the two nearby nodes, while the $\psi_m$ couple to the $SU(2)\times SU(2)\cong SO(4)$ gauge fields. Moreover, from the $E_7$ case we know that none of them couples to the diagonal $U(1)_d$ group. Only the flavors $(v,\tilde{v})$ couples to $U(1)_d$ analogously to the $E_7$ case, but its generator can be redefined to cancel this coupling (equivalently we defined $\phi=\phi'-{\varphi_1}^1$). 

We can also write the vector representation of $SO(4)$ as a bifundamental representation of $SU(2)\times SU(2)$: 
\begin{equation}
 \psi_m^\alpha  \,\,\,\,\, m=1,...,4 \qquad \rightarrow \qquad Q^{\alpha\ell r}  \,\,\,\,\, a,\ell,r=1,2 \qquad \mbox{with}\qquad Q^{\alpha \ell r}=\psi_m^\alpha(\sigma^m)^{\ell r}\:,
\end{equation}
where $\sigma^m=(\sigma^i,\mathbb{1}_2)$ (the three Pauli matrices and the identity matrix). The triple of indices $(abc)$ is relative to the three $SU(2)$ group that the fields couple. Then the total superpotential including the nearby nodes can be written as 
\begin{eqnarray}
W &=&   \phi \,v^\alpha\tilde{v}_\alpha + \varphi^{i-1} ( v^\alpha\tilde{v}_\alpha + q_1^\alpha\tilde q_\alpha - A_{i-2}B_{i-2} ) + \varphi^{i+1} ( v^\alpha\tilde{v}_\alpha + A_{i+1}B_{i+1} - q_1^\alpha\tilde q_\alpha ) \nonumber\\
    &&   + Q^{\alpha\ell r} {\Phi_\alpha}^{\alpha'} Q_{\alpha'\ell r} + Q^{\alpha\ell r} {\tilde{\Psi}_{\ell}}^{i-1\, \ell'} Q_{\alpha\ell' r}+ Q^{\alpha\ell r} {\tilde{\Psi}_{r}}^{i+1\, r'} Q_{\alpha\ell r'}
\end{eqnarray}
where we used that ${\varphi_1}^1=\varphi^{i-1}-\varphi^{i+1}$ and $\psi=\varphi^{i-1}+\varphi^{i+1}$
This is the $\mathcal{N}=4$ superpotential one would write once the matter fields are given. As we have just seen, this superpotential can be arranged into the form
\begin{equation}
 W \supset \mbox{tr} \Psi \mathfrak{M} \:.
\end{equation}
with $\Psi$ in the block diagonal form. This coupling appears also in the $D_N$ superpotential as
\begin{equation}
 W \supset \mbox{tr} \Psi M \:.
\end{equation}
again with $\Psi$ in the block diagonal form. Again this implies 
that the Higgs branch is not modified by  replacing the $U(2)$ node with the $SU(2)\times U(1)$ quiver.

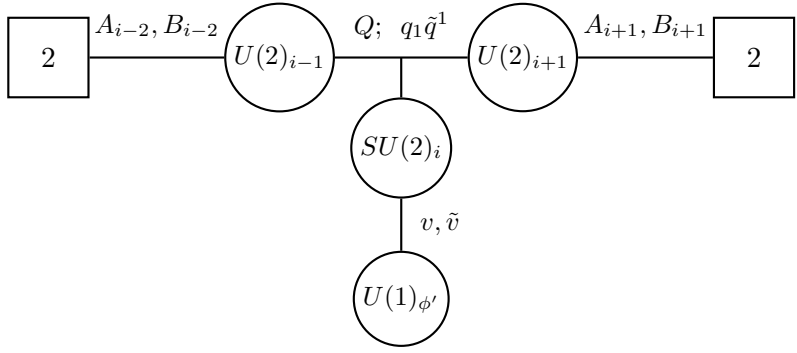
\begin{figure}[ht]
\begin{center}
\begin{tikzpicture}[thick, scale=0.8]
  \node[draw, rectangle, minimum width=30pt, minimum height=30pt](L0L) at (2.2,2){$2$};
  \node[draw, rectangle, minimum width=30pt, minimum height=30pt](L0R) at (13.8,2){$2$};
  \node[circle, draw, inner sep= 2pt](L1) at (6,2){{\small $U(2)_{i-1}$}};
  \node[circle, draw, inner sep= 2pt](L2) at (8,0.5){{\small $SU(2)_i$}};
  \node[circle, draw, inner sep= 2pt](L3) at (10,2){{\small $U(2)_{i+1}$}};
  \node[circle, draw, inner sep= 2pt](L5) at (8,-2){{\small $U(1)_{\phi'}$}};
 \path[every node/.style={font=\sffamily\small,
  		fill=white,inner sep=1pt}]
(L1) edge node[above=2mm] {$Q;\,\,\, q_1\tilde q^1$}  (L3)
(L2) edge node[right=2mm]{$v,\tilde v$} (L5)
(L2) edge (8,2)
(L1) edge node[above=2mm]{$A_{i-2},B_{i-2}$} (L0L)
(L3) edge node[above=2mm]{$A_{i+1},B_{i+1}$} (L0R)
;
\end{tikzpicture}
\end{center}
\caption{Modified $D_N$ quiver (with $\phi'=\phi+\varphi^{i-1}+\varphi^{i+1}$ the scalar coupled to $v,\tilde{v}$). In the diagram it is meant that $Q$ is not charged under $U(1)_{i\pm 1}$, while $q_1,\tilde q^1$ are singlets under $SU(2)_{i\pm 1}$.
}\label{Fig:ModDN}
\end{figure}

We can also redefine the scalar field $\phi$ in the vector multiplet of the $U(1)$ as $\phi=\phi'- \varphi^{i-1}- \varphi^{i+1}$. This corresponds to a redefinition of the $U(1)_{i-1}$ and $U(1)_{i+1}$ generators such that $v,\tilde v$ are not charged under the new generators. After this change, we can represent the `modified $D_N$ chain' as in Figure  \ref{Fig:ModDN}, where it is meant that the fields $q_1,\tilde{q}^1$ are charged only under the relative $U(1)$ symmetry of the two nodes (and in the fundamental representation of the $SU(2)_i$), while the fields $Q$ couple only to $SU(2)_{i\pm 1}$.\footnote{Equivalently, the diagonal $U(1)$ generators $t_{i-1}$ and $t_{i+1}$ have ben replaced by $t_{i-1}'=t_{i-1}-t_\phi$ and $t_{i+1}'=t_{i+1}-t_\phi$, where $t_\phi$ is the generator relative to the $U(1)$ node.}

\subsection{The $U(2)$ node of $D_4$}\label{sec:U2D4}

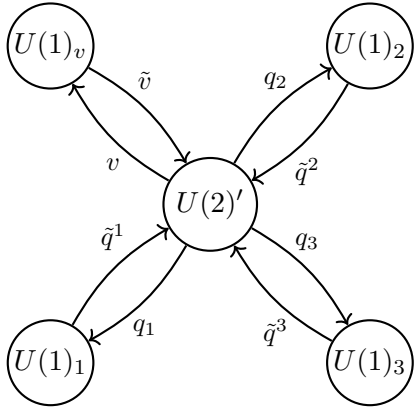
\begin{figure}
\begin{center}
\begin{tikzpicture}[->, every node/.style={circle,draw},thick, scale=0.7]
  \node[inner sep=1](L1) at (-6,3){$U(1)_v$};
  \node[inner sep=1](L2) at (-6,-3){$U(1)_1$};
  \node[inner sep=1, minimum height=35pt](V1) at (-3,0){$U(2)'$};
  \node[inner sep=1](L3) at (0,-3) {$U(1)_3$};
    \node[inner sep=1](L4) at (0,3) {$U(1)_2$};

 \path[every node/.style={font=\sffamily\small,
  		fill=white,inner sep=1pt}]
(L1) edge [bend left=15] node[above=2mm] {$\tilde v$} (V1)
(V1) edge [bend left=15] node[below=2mm] {$v$}(L1)
(L2) edge [bend left=15] node[above=2mm] {$\tilde q^1$} (V1)
(V1) edge [bend left=15] node[below=2mm] {$q_1$}(L2)
(L3) edge [bend left=15] node[below=2mm] {$\tilde q^3$} (V1)
(V1) edge [bend left=15] node[above=2mm] {$q_3$}(L3)
(L4) edge [bend left=15] node[below=2mm] {$\tilde q^2$} (V1)
(V1) edge [bend left=15] node[above=2mm] {$q_2$}(L4)
;
\end{tikzpicture}
\end{center}
\caption{Dual D$_4$ quiver. } \label{fig:d4quiverdual}
\end{figure}

It is instructive to consider one further simple case, i.e. the $D_4$ quiver gauge theory.
\begin{figure}[ht]
\begin{center}
\begin{tikzpicture}[->, every node/.style={circle,draw},thick, scale=0.7]
  \node[inner sep=1](L1) at (-6,3){$U(1)_u$};
  \node[inner sep=1](L2) at (-6,-3){$U(1)_p$};
  \node[inner sep=1, minimum height=35pt](V1) at (-3,0){$U(2)$};
  \node[inner sep=1](L3) at (0,-3) {$U(1)_t$};
    \node[inner sep=1](L4) at (0,3) {$U(1)_s$};

 \path[every node/.style={font=\sffamily\small,
  		fill=white,inner sep=1pt}]
(L1) edge [bend left=15] node[above=2mm] {$\tilde u$} (V1)
(V1) edge [bend left=15] node[below=2mm] {$u$}(L1)
(L2) edge [bend left=15] node[above=2mm] {$\tilde p$} (V1)
(V1) edge [bend left=15] node[below=2mm] {$p$}(L2)
(L3) edge [bend left=15] node[below=2mm] {$\tilde t$} (V1)
(V1) edge [bend left=15] node[above=2mm] {$t$}(L3)
(L4) edge [bend left=15] node[below=2mm] {$\tilde s$} (V1)
(V1) edge [bend left=15] node[above=2mm] {$s$}(L4)
;
\end{tikzpicture}
\end{center}
\caption{D$_4$ quiver. } \label{fig:d4quiver}
\end{figure}
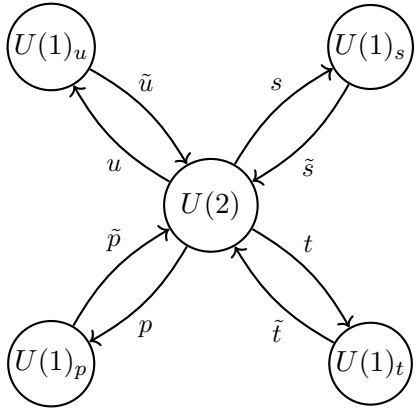
We apply the duality map to the $U(2)$ central node. To connect the dual theory to the four $U(1)$ nodes, we write the superpotential term
\be \label{WD4modif}
W \supset \mbox{tr}\, \Psi \mathfrak{M} \:,
\ee
where now 
\begin{eqnarray}
\Psi &=&  \left(\begin{array}{cccc}
 \phi_u &&& \\ & \phi_p && \\ && \phi_s & \\ &&& \phi_t \\
\end{array}\right) \:,
\end{eqnarray}
with $\phi_u,\phi_p,\phi_s,\phi_t$ the four complex scalars in the vector multiplets of the external $U(1)$ nodes. Plugging the expression of $\mathfrak{M}$ into \eqref{WD4modif}, and adding also the term $-\phi ((v\tilde v))$ to complete the superpotential, we obtain
\begin{eqnarray}
W &=& ((v\tilde v))\left( -\phi +\frac{\phi_u+\phi_p+\phi_s+\phi_t}{2}\right) + q_1\tilde q^1 \left(\frac{\phi_u+\phi_p-\phi_s-\phi_t}{2}\right) \nonumber \\
   &&  + q_2\tilde q^2 \left(\frac{\phi_u-\phi_p+\phi_s-\phi_t}{2}\right)   + q_3\tilde q^3 \left(\frac{\phi_u-\phi_p-\phi_s+\phi_t}{2}\right) \:.
\end{eqnarray}
With a proper redefinition of the scalar fields (that corresponds to a redefinition of the $U(1)$ generators), we can write the superpotential as
\begin{equation}
W = ((v\tilde v))\left( \phi_{U(2)'}-\phi \right) + q_1\tilde q^1 \left(\phi_{U(2)'}-\phi_1 \right) + q_2\tilde q^2 \left(\phi_{U(2)'}-\phi_2 \right)   + q_3\tilde q^3 \left(\phi_{U(2)'}-\phi_3 \right) \:,
\end{equation}
hence recovering the $D_N$ quiver (see Figure  \ref{fig:d4quiverdual}). 
Hence the $D_N$ theory is self-dual under the replacement of the central nodes. Notice however that the diagonal $U(1)$ of the original $U(2)$ has been mapped to the $U(1)$ relative to the external $U(1)_v$ node.
In this example the fact that the Higgs branch remains the same is manifest.

\subsection{Monopole deformation along the $U(2)$ node}\label{U2monopDef}

We now want to deform the superpotential of an ADE quiver gauge theory by switching on a monopole deformation 
$$\Delta W=m V_-\:,$$
where $V_-$ corresponds to the simple root of a $U(2)$ node; in other words, $V_-$ is the R-charge one monopole operator charged under the topological $U(1)_J$ corresponding to the $U(2)$ node. As said before, it is not easy to study this deformation in the original theory. Our strategy is first to replace the $U(2)$ node by the dual $SU(2)\times U(1)$ dual theory. 
We then map the R-charge one monopole operators of one theory to the one in the other theory. 
In particular, the monopole operator $V_-$ we are interested in is mapped to the R-charge one monopole operator relative to the $U(1)$ node, say $W_+$ (see Section \ref{Sec:montsu2} for a proof of this).
Hence the deformed superpotential is
\begin{equation}
W = q_i^\alpha \Phi_\alpha^\beta \tilde q^i_\beta + v^\alpha \Phi_\alpha^\beta \tilde v_\beta - \phi v^\alpha\tilde v_\alpha  +  m W_+\:.
\end{equation}
The effect of such a deformation has been studied in \cite{Collinucci:2016hpz} and it is summarized in Section \ref{Sec:U1monop}:  it removes the $U(1)$ node, keeps the $SU(2)$ meson ${\mathfrak{m}_\alpha}^\beta\leftrightarrow \tilde{v}_\alpha v^\beta$ as a fundamental field. The effective superpotential is
\begin{equation}
W^{\rm eff} = 
 \mbox{tr} \, \left[\left(\tilde{\mathfrak{m}}-\mu_{\rm su(2)}\right) \Phi\right]  -\frac{\mathcal{X}}{m}\mbox{det}\,\tilde{\mathfrak{m}}\,,
\end{equation}
where $\tilde{\mathfrak{m}}$ is traceless part of $\mathfrak{m}$ and ${(\mu_{\rm su(2)})_\beta}^\alpha\equiv-\tilde{q}^i_\beta q_i^\alpha$ denotes the $SU(2)$ moment map.
Notice that we did not touch the fields charged under the $SU(4)$ symmetry in the `modified ADE quiver'. As an example, we can apply this procedure to the external $U(2)$ node of the $E_7$ quiver. 
We can write the relevant terms for the $E_7$ effective superpotential as\footnote{Notice that the tracelessness of $\tilde{\mathfrak{m}}$ set to zero the coupling with the field $\psi\sim \tr\Psi$.} 
\begin{equation}\label{supEffE7quiv}
W^{\rm eff}_{\rm E_7 \, quiv}  \supset  \mbox{tr} \, \left[\left(\tilde{\mathfrak{m}}-\mu_{\rm su(2)}\right) \Phi\right]  -\frac{\mathcal{X}}{m}\mbox{det}\,\tilde{\mathfrak{m}} + \tr X \tilde{\Psi} \:.
\end{equation}
This superpotential gives the same Higgs branch as the undeformed theory. In fact, the gauge invariant generators and the relations among them remain unaltered.
In particular, the first term in \eqref{supEffE7quiv} still implies $X^2=0$ (remember that $X$ is written in terms of $q_i^\alpha,\tilde q^j_\beta$ hidden in $\mu_{\rm su(2)}$), while the last term is responsible for the relations leading to the $E_7$ singularity.

\section{$U(3)$ nodes and $E_6$ Minahan-Nemeschansky theory} \label{Sec:U3node}

Having studied monopole deformations on $U(1)$ nodes in our previous work, and on $U(2)$ nodes in the previous section, we now move on to the case of $U(3)$. These show up in the exceptional Dynkin quivers, and we will see that their dual descriptions involve non-Lagrangian theories. 

The basic move is to ungauge the nearby nodes, which yields a $U(3)$ gauge theory with six flavors. Through an Argyres Seiberg duality, we will replace this theory with a dual theory with a $U(1)$ factor, where the abelian procedure can be applied to describe the deformed theory.

\subsection{Argyres Seiberg duality with gauged $U(1)$}

We now find the dual theory that should replace the $U(3)$ node. We start by describing the parent Argyres Seiberg (AS) duality \cite{Argyres:2007cn} and then modify this by simply gauging one $U(1)$ factor of the global symmetry. Let us briefly review the AS duality.

\begin{paragraph}{Lagrangian side:}  On the Lagrangian side we have an $\mathcal{N}=2, d=4$ $SU(3)$ gauge theory with $N_f=6$. The vector multiplet contains an adjoint scalar ${\tilde{\Psi}_a}^b$. The hypers are $(Q_i^{a}, \tilde Q^i_{a})$, with $a=1, \ldots, 3$ the color indices, and $i=1, \ldots, 6$ the flavor indices. The flavor symmetry is $G_f = U(6)$. We can construct a meson, a baryon and an anti-baryon:
\be
{M_i}^j = Q_i^{a} \tilde Q^j_{a}\,, \quad \tilde B^{i j k} = \epsilon^{a b c}\, \tilde Q^i_{a} \tilde Q^j_{b} \tilde Q^k_{c}\,, \quad  B_{i j k} = \epsilon_{a b c}\,  Q_i^{a}  Q_j^{b}  Q_k^{c}
\ee
The $\mathcal{N}=2$ superpotential reads
\be
W_{SU(3)} =   Q_i^{a} \,{\tilde{\Psi}_a}^b \, \tilde Q^i_{b} 
\ee
It will be useful to decompose the meson in trace plus traceless part, so let us define
\be
\tilde M \equiv M-\tfrac{1}{6} ({\rm tr}M) \mathbb{1}_6\,. 
\ee
If we promote the gauge group to $U(3)$, which is our case of interest, then the scalar in the vector multiplet, which we call $\breve{\Psi}$, gains a traceful part.
Now the superpotential becomes
\be
W_{U(3)} =Q_i^{a} \,{\breve{\Psi}_a}^b \, \tilde Q^i_{b}  \:,
\ee
and the F-term for tr$\breve{\Psi}$ tells us that ${\rm tr}(M)=0$.
\end{paragraph}
 
\begin{paragraph}{Non-Lagrangian side:} Argyres and Seiberg found the S-dual to the theory described above. It is an $SU(2)$ gauge theory coupled weakly to the $E_6$ Minahan-Nemeschansky (MN) theory \cite{Minahan:1996fg}, plus a hyper in the fundamental $(v, \tilde v)$ of $SU(2)$. 

The Higgs branch of the $E_6$ theory is generated by moment map operators transforming in the ${\bf 78}$ of the $E_6$ flavor symmetry. In order to construct the S-dual to the previous theory, we observe that $E_6 \supset SU(6) \times SU(2)$, and couple the $SU(2)$ global current to the $SU(2)$ gauge symmetry. This leaves as a flavor group $SU(6)$, plus a $U(1)$ that acts on the doublet as 
$(v, \tilde v) \mapsto (e^{-3 i \theta} v, e^{3 i \theta} \tilde v)$, thereby matching the expected $U(6)$ global symmetry.

Let us write out explicitly the matter content of this theory. Aside from the $(v, \tilde v)$, we have an adjoint $SU(2)$ scalar $\Phi$, to which it couples via
\be
W =  v^\alpha\, {\Phi_\alpha}^\beta\, \tilde v_\beta\,.
\ee
The ${\bf 78}$ decomposes under $SU(6) \times SU(2)$ as 
\be
78 \rightarrow (35, 1)  \oplus (20,2) \oplus (1, 3)\,.
\ee
Let us use the conventions of Gaiotto, Tachikawa and Neitzke \cite{Gaiotto:2008nz}, and write these fields as:
\be
{X^i}_j\,, \quad Y_\alpha^{[i j k]}\,, \quad Z_{\alpha \beta}
\ee
where $i=1, \dots, 6$, and $\alpha=1,2$. $Z$ satisfies the following relation
\be \label{zelim}
Z_{\alpha \beta} + v_{(\alpha} \tilde v_{\beta)}=0\,,
\ee
which means it can always be eliminated. As we will show in a moment, the following are all gauge invariant operators that can be generated:
\be
((v \tilde v))\,, \quad {X^i}_j\,, \quad((Y^{i j k} \tilde v))\,, \quad((Y_{i j k}  v))\,,
\ee
where $((...))$ means that the $SU(2)$ indices are contracted and the flavor indices are raised and lowered with an epsilon tensor. 
\end{paragraph}

The identification of gauge invariant operators across the duality is as follows:
\be
{\rm tr}M \leftrightarrow \frac{((v \tilde v))}{3}\,, \quad {\tilde M^i}_j \leftrightarrow {X^i}_j\,, \quad \tilde B^{i j k} \leftrightarrow ((Y^{i j k} \tilde v))\,, \quad  B_{i j k} \leftrightarrow ((Y_{i j k}  v))
\ee
where we previously defined $\tilde M$ as the traceless part of $M$. Here we see that $v$ and $\tilde v$ carry the baryonic $U(1)$ charge. The relations for the Higgs branch derived in \cite{Gaiotto:2008nz} are the following:
\begin{eqnarray}
0&=& {X^i}_j Z_{\alpha \beta}+\tfrac{1}{4} Y^{i k l}_{(\alpha} Y_{|j k l| \beta)} \label{joe1}\\
0&=& {X^l}_{\{i} Y_{[j k]\} l \alpha} \label{joe2}\\
0&=& {X^{\{i}}_{l} Y^{[j k]\} l}_\alpha \label{joe3} \\
0&=& Y^{i j k}_\alpha Z_{\beta \gamma} \epsilon^{\alpha \beta}+{X^{[i}}_l Y^{j k] l}_\gamma \label{joe4}\\
0&=& (Y^{i j m}_\alpha Y_{k l m \beta} \epsilon^{\alpha \beta}-4 {X^{[i}}_{[k} {X^{j]}}_{l]})|_{(0,1,0,1,0)}\label{joe5}\\
0&=& {X^{i}}_k {X^k}_j -\tfrac{1}{6} {\delta^i}_j {X^k}_l {X^l}_k \label{joe6}\\
0&=& Y^{i j k}_\alpha Y_{i j k \beta} \epsilon^{\alpha \beta}+24 Z_{\alpha \beta} Z_{\gamma \delta} \epsilon^{\alpha \gamma} \epsilon^{\beta \gamma}\label{joe7}\\
0&=& {X^{i}}_j {X^j}_i+3 Z_{\alpha \beta} Z_{\gamma \delta} \epsilon^{\alpha \gamma} \epsilon^{\beta \delta} \label{joe8}
\end{eqnarray}
Where the $(0,1,0,1,0)$ subscript in \eqref{joe4} indicates the weights of the highest weight under $SU(6)$.
The third relation \eqref{joe3} represents the following projection:
\be      \Yvcentermath1
\ytableausetup
 {mathmode, boxsize=1.2em}
{X^{i}}_{l} Y^{[j k] l} = \by
j \\ k
\ey
\otimes
\begin{ytableau}
i
\end{ytableau}
\vert_{(1,1,0,0,0)} = 
\begin{ytableau} j& i\\k \end{ytableau}
= {X^{i}}_{l} Y^{[j k] l} - {X^{[i}}_{l} Y^{j k] l} \:.
\end{equation}
Here, we are using conventions whereby we symmetrize all indices in the same row of a Young tableau \emph{first, and then} anti-symmetrize along columns.
The second relation \eqref{joe2} represents a dual projection:
\begin{equation}
{X^l}_{\{i} Y_{[j k]\}}=      \Yvcentermath1
 \left.   \yng(1,1,1,1,1) \otimes \yng(1,1,1,1) \right\vert_{(0,1,0,1,0)} = \yng(2,2,2,2,1)  = {X^l}_{i} Y_{[j k]}-{X^l}_{[i} Y_{j k]} \:.
 \end{equation}
Finally, for the fifth relation \eqref{joe5}, we first raise all $SU(6)$ indices with the epsilon tensor, and then impose the following projector:
\begin{equation}
     \Yvcentermath1
  \left.   \yng(1,1,1,1) \otimes \yng(1,1) \right\vert_{(0,1,0,1,0)} = \yng(2,2,1,1) \:\:\:.
 \end{equation}
The list of possible $SU(2)$ basic gauge invariant operators is the following:
\be
((v \tilde v))\,, \quad {X^i}_j\,, \quad((Y^{i j k} \tilde v))\,, \quad((Y_{i j k}  v))\,, \quad ((Y^{i j k} Y^{l m n}))\,.
\ee
The last class of invariants can be eliminated in favor of the others, as was shown in \cite{Gaiotto:2008nz}. We will prove this here again. But first, let us decompose the general $Y_\alpha Y_\beta$ tensor in irreducible representations of $SU(6)$ in order to figure out its symmetry properties with respect to  $\alpha$ and $\beta$:
\be
\begin{aligned}
     \Yvcentermath1
Y^{i j k}_{\alpha} Y^{l m n}_{\beta} &=    \by i \\ j \\ k \\ l \\ m \\ n \ey \oplus  \by i & l \\ j & m \\ k & n \ey \oplus \by i & l \\ j & m\\ k \\ n \ey \oplus \by i & l \\ j \\ k \\ m \\ n \ey\\
&= (0,0,0,0,0) \oplus (0,0,2,0,0) \oplus (0,1,0,1,0) \oplus (1,0,0,0,1) \:.
\end{aligned}
\ee
\begin{enumerate}
\item The first term (the singlet) is the $(1^6)$ partition, i.e. the completely antisymmetric 6-tensor. 
It is clearly anti-symmetric with respect to  the exchange $\alpha \leftrightarrow \beta$, since this amounts to exchanging three anti-symmetrized $SU(6)$ indices.

\item The second term is symmetric with respect to  to the simultaneous exchange $(i, j, k) \leftrightarrow (l, m, n)$, which implies it is symmetric with respect to  $(\alpha, \beta)$. 

\item The third term has the form
\be
\by i & l \\ j & m\\ k \\ n \ey \propto Y^{[i j k}_{[\alpha} Y^{n] l m}_{\beta]}+Y^{m [k i}_{[\alpha} Y^{j n] l}_{\beta]} \:.
\ee

\item The fourth term is explicitly the following:
\be
\by i & l \\ j \\ k \\ m \\ n \ey \propto  Y^{[i j k}_\alpha Y^{m n] l}_\beta+Y^{l [j k}_\alpha Y^{i m n]}_\beta  =   2  Y^{[i j k}_{(\alpha} Y^{m n] l}_{\beta)} \:.
\ee
\end{enumerate}

To summarize, we have the following decomposition:
\be \label{su2sym}
Y^{i j k}_{\alpha} Y^{l m n}_{\beta} = (0,0,0,0,0)_{[\alpha \beta]} \oplus (0,0,2,0,0)_{(\alpha \beta)} \oplus (0,1,0,1,0)_{[\alpha \beta]} \oplus (1,0,0,0,1)_{(\alpha \beta)} \:.
\ee
This decomposition will be useful in several calculations to come. 

Let us first begin by proving that $((Y^{i j k} Y^{l m n}))$ can be eliminated in favor of the fields $X, v$ and $\tilde v$. The first thing to notice from \eqref{su2sym} is that only the $SU(6)$ singlet and $(0,1,0,1,0)$ representations contribute due to the contraction with $\epsilon^{\alpha \beta}$.
The singlet part can be eliminated in favor of $Z$, and hence $v, \tilde v$, via \eqref{joe7}, and the second part in terms of $X$ via \eqref{joe5}

In conclusion, the gauge invariant $(Y Y)$ can be eliminated, leaving us with the following list of $SU(3)$ gauge invariants to parametrize the Higgs branch:
\be \label{gaugeinv1}
((v \tilde v))\,, \quad {X^i}_j\,, \quad((Y^{i j k} \tilde v))\,, \quad((Y_{i j k}  v))\,.
\ee

\subsubsection*{Gauging a baryonic $U(1)$}

Now we would like to promote the $SU(3)$ gauge group to $U(3)$ on the Lagrangian side. This will reduce the $U(6)$ flavor symmetry to $SU(6)$. On the non-Lagrangian side, this requires us to gauge the $U(1)$, whose moment map is given by $- ((v \tilde v))$. This entails introducing a $U(1)$ vector multiplet with scalar $\phi$ and coupling it to this moment map via the term
\be \label{gaugingterm}
W = \phi ((v \tilde v))\,.
\ee
Now our list of gauge-invariants \eqref{gaugeinv1} needs to be modified. On the Lagrangian side, our baryon and anti-baryon $B$ and $\tilde B$ transform non-trivially under the $U(1)$ of $U(3)$. Hence, the only gauge-invariant we can make out of them is $\tilde B^{i j k}  B_{l m n}$. However, this can be written in terms of the meson as
\be
\tilde B^{i j k}  B_{l m n} = 6 M^{[i}_l M^{j}_m M^{k]}_{n} \:.
\ee
On the non-Lagrangian side, we need to form invariants under the newly gauged $U(1)$. Out of the list \eqref{gaugeinv1}, the only trouble makers are $((Y v))$ and $((Y \tilde v))$, which have non-zero $U(1)$ charge. The only gauge invariant we can make is $((Y^{i j k} v)) ((Y^{l m n} \tilde v))$. We will now check that this can be eliminated in terms of $(X, v, \tilde v)$.

First, we notice that the extra term \eqref{gaugingterm} in the superpotential implies $((v \tilde v))=0$. Hence, $v_{[\alpha} \tilde v_{\beta]} \sim ((v \tilde v)) \epsilon_{\alpha \beta} = 0$, which implies, $v_\alpha \tilde v_\beta = v_{(\alpha} \tilde v_{\beta)}= - Z_{\alpha \beta}$ (see \eqref{zelim}).

Let us use this last observation to rewrite $((Y^{i j k}\tilde  v)) ((Y^{l m n}  v))$ as follows:
\begin{align}
Y^{i j k}_\alpha \tilde v_\beta \epsilon^{\alpha \beta} \, Y^{l m n}_{\delta}  v_{\gamma} \epsilon^{\delta \gamma} &= -\big(Y^{i j k}_\alpha Z_{\beta \gamma} \epsilon^{\alpha \beta}\big)\,Y^{l m n}_{\delta} \epsilon^{\delta \gamma} = X^{[i}_l Y^{j k] l}_\gamma Y^{l m n}_{\delta} \epsilon^{\delta \gamma}\\
 &= -X^{[i}_l ((Y^{j k] l} Y^{l m n}))\,,
\end{align}
whereby we used \eqref{joe4} in the second step to replace the $YZ$ term in parenthesis.  As seen in the previous section, $((YY))$ can be eliminated in terms of $X, v, \tilde v$. After gauging the flavor $U(1)$, we can eliminate all dependence on $((v\tilde v))$.
Hence, we are done. The conclusion is that the Higgs branch can be parametrized purely in terms of $X^i_j$, which matches the expectation from the Lagrangian side, where the Higgs branch is parametrized by the meson matrix $\tilde M^i_j$.

As we have seen in Section \ref{Sec:GoalStrategy}, another crucial ingredient we need to prove is the nilpotency of the $X$-operator: 
If we  show that $X$ is traceless and  nilpotent, we can rewrite it as a would-be meson matrix, and all algebraic relations for the Higgs branch will follow automatically.

First of all, $X$ is traceless by definition, since $X$ is in the adjoint of $SU(6)$. Second, it is nilpotent as we now prove. Notice the following:
\be \label{elimvv}
Z_{\alpha \beta} Z_{\gamma \delta} \epsilon^{\alpha \gamma}  \epsilon^{\beta \delta} = v_\alpha \tilde v_\beta v_\gamma \tilde v_\delta \epsilon^{\alpha \gamma}  \epsilon^{\beta \delta} = 0
\ee
By \eqref{joe8} this implies that tr$X^2=0$. In turn, this implies by \eqref{joe6} that $X^i_k X^k_j=0$ as expected. Hence, we deduce that all possible relations satisfied by the meson on the Lagrangian side will be satisfied by the field $X$ on the non-Lagrangian side.

\subsection{Coupling S-dual theories into the $E$ quiver}

In this section, we briefly  look at two examples of $E$-type quivers where we extract an $U(3)$ node, S-dualize it, and couple it back into the quiver.

\begin{paragraph}{$E_8$ quiver}
In the setting of D2-branes probing ADE singularities, an external $U(3)$ gauge node shows up in the $E_8$ quiver.

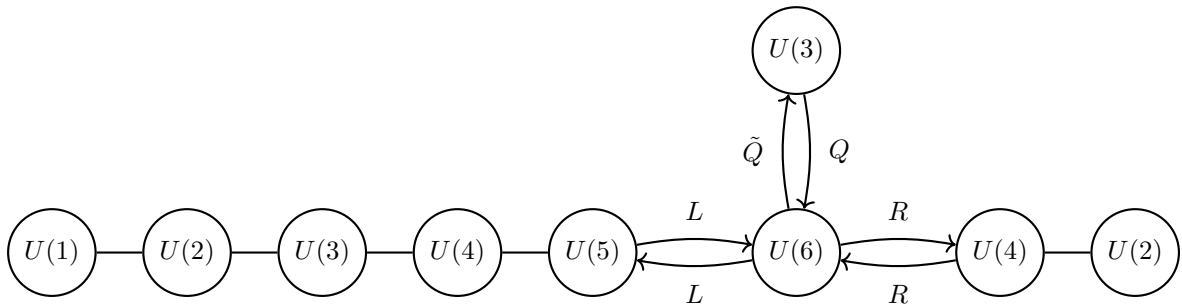
\begin{figure}[h!]
\begin{center}
\begin{tikzpicture}[->, thick]
\node[shape=circle, draw] (1) at (0,0) {{\small $U(1)$}};
\node[shape=circle, draw] (2) [right= .6cm of 1] {{\small $U(2)$}};
\node[shape=circle, draw] (3) [right= .6cm of 2] {{\small $U(3)$}};
\node[shape=circle, draw] (4) [right= .6cm of 3] {{\small $U(4)$}};
\node[shape=circle, draw] (5) [right= .6cm of 4] {{\small $U(5)$}};
\node[shape=circle, draw] (6) [right= 1.5cm of 5] {{\small $U(6)$}};
\node[shape=circle, draw] (7) [right= 1.5cm of 6] {{\small $U(4)$}};
\node[shape=circle, draw] (8) [right= .6cm of 7] {{\small $U(2)$}};
\node[shape=circle, draw] (9) [above=1.5cm of 6] {{\small $U(3)$}};
\draw[-] (1) -- (2);
\draw[-] (2) -- (3);
\draw[-] (3) -- (4);
\draw[-] (4) -- (5);
\draw[-] (7) -- (8);
 \path[every node/.style={font=\sffamily\small,
  		fill=white,inner sep=1pt}]

(5) edge [bend left=10] node[above=2mm] {$L$}(6)
(6) edge [bend left=10] node[below=2mm] {$L$} (5)
(6) edge [bend left=10] node[above=2mm] {$R$}(7)
(7) edge [bend left=10] node[below=2mm] {$R$} (6)
(6) edge [bend left=10] node[left=2mm] {$\tilde Q$}(9)
(9) edge [bend left=10] node[right=2mm] {$Q$} (6)
;
\end{tikzpicture}
\end{center}
\caption{$E_8$ quiver. We show the double arrows between the nodes only when relevant for our discussion.}\label{Fig:E8quiv}
\end{figure}
Focusing on the $U(3)$ and $U(6)$ nodes, the following superpotential is present:
\be
W = \tilde Q_i^{a} \,{\breve{\Psi}^b}_a \, Q^i_{b} + {\Psi^j}_i ( {L}^i_\lambda {\tilde{L}}^\lambda_j + {R}_\rho^i {\tilde{R}}_j^\rho)+{{\Psi}^j}_i {M^i}_j\,,
\ee
where $\breve{\Psi}$ is the scalar relative to the $U(3)$ node, while $\Psi$ to the $U(6)$ node.

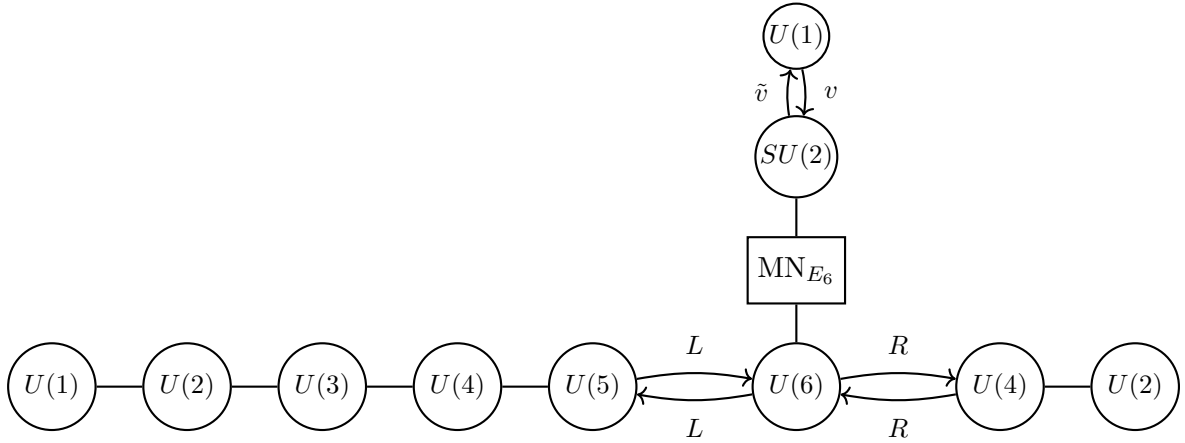
\begin{figure}[h!]
\begin{center}
\begin{tikzpicture}[->, thick]
\node[shape=circle, draw] (1) at (0,0) {{\small $U(1)$}};
\node[shape=circle, draw] (2) [right= .6cm of 1] {{\small $U(2)$}};
\node[shape=circle, draw] (3) [right= .6cm of 2] {{\small $U(3)$}};
\node[shape=circle, draw] (4) [right= .6cm of 3] {{\small $U(4)$}};
\node[shape=circle, draw] (5) [right= .6cm of 4] {{\small $U(5)$}};
\node[shape=circle, draw] (6) [right= 1.5cm of 5] {{\small $U(6)$}};
\node[shape=circle, draw] (7) [right= 1.5cm of 6] {{\small $U(4)$}};
\node[shape=circle, draw] (8) [right= .6cm of 7] {{\small $U(2)$}};
\node[shape=rectangle, draw, minimum height=25pt] (9) [above=.5cm of 6] {MN$_{E_6}$};
\node[shape=circle, draw, inner sep=0.6pt] (10) [above=.5cm of 9] {{\small $SU(2)$}};
\node[shape=circle, draw, inner sep=0.6pt] (11) [above=.6 of 10] {{\small $U(1)$}};
\draw[-] (1) -- (2);
\draw[-] (2) -- (3);
\draw[-] (3) -- (4);
\draw[-] (4) -- (5);
\draw[-] (7) -- (8);
\draw[-] (9) -- (10);
 \path[every node/.style={font=\sffamily\small,
  		fill=white,inner sep=1pt}]

(5) edge [bend left=10] node[above=2mm] {$L$}(6)
(6) edge [bend left=10] node[below=2mm] {$L$} (5)
(6) edge [bend left=10] node[above=2mm] {$R$}(7)
(7) edge [bend left=10] node[below=2mm] {$R$} (6)
(9) edge[-]  (6) 
(10) edge [bend left=10] node[left=2mm] {$\tilde v$} (11) 
(11) edge [bend left=10] node[right=2mm] {$v$} (10) 
;
\end{tikzpicture}
\end{center}
\caption{Modified $E_8$ quiver.}\label{Fig:E8quiv2}
\end{figure}

We now decouple the $U(3)$ node, perform a gauged AS duality on it, and recouple it to the quiver in the analogous way as we did for the $U(2)$ node of $E_7$ and $D_N$. The new theory 
has the following superpotential:
\be
W =  -\phi \, ((v\tilde v)) + v^\alpha{\Phi_\alpha}^\beta \tilde v_\beta +  {\Psi^j}_i ( {L}^i_\lambda {\tilde{L}}^\lambda_j + {R}_\rho^i {\tilde{R}}_j^\rho)  +  {\tilde{\Psi}^j}_i {X^i}_j + \psi\, ((v\tilde v))  \:,
\ee
where $\Psi=\tilde{\Psi}+\frac{\psi}{2}\mathbb{1}_6$ with $\tilde{\Psi}$ traceless. We can reabsorb the last term by redefining $\phi=\phi'+\psi$. The `modified quiver' is depicted in Figure  \ref{Fig:E8quiv2}.
Note, that we could try to couple further moment maps with ${{\Psi}^j}_i$. However, these would necessarily have the form $Y^{i k l}_\alpha Y_{j k l \beta} \sim {X^i}_j v_\alpha \tilde v_\beta$, which would give zero upon contracting the $SU(2)$ indices with anything.

As for the $U(2)$ cases, it is now easy to see that the Higgs branch of the `modified quiver' gauge theory is the same as the standard one: All gauge invariants are now constructed out of $X$ instead of $M$. Since $X$ is traceless and nilpotent and couples to the rest of the quiver as $M$ did, the Higgs branch generators and relations are the same in the dual theories.
\end{paragraph}

\begin{paragraph}{$E_6$ quiver}

A different example is that of the central node of the $E_6$ quiver (see Figure  \ref{Fig:E6quiv}). In this case, the $U(3)$ node couples to three $U(2)$ nodes. However, the general procedure outlined before is the same. The only difference is that now, the $SU(6)$ flavor is broken to $S(U(2) \times U(2) \times U(2))$. This breaking is inflicted by the couplings to the neighboring vector multiplets, which are not organized into a $U(6)$ vector multiplet.

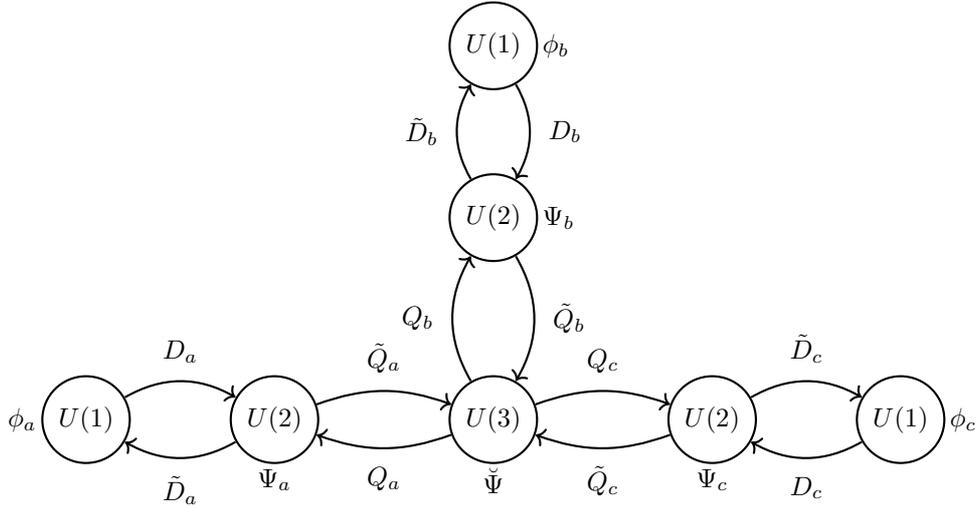
\begin{figure}[ht!]
\begin{center}
\begin{tikzpicture}[->, thick]
\node[shape=circle, draw] (1) at (0,0) {{\small $U(1)$}};
\node[shape=circle, draw] (2) [right= 1.3cm of 1] {{\small $U(2)$}};
\node[shape=circle, draw] (3) [right= 1.7cm of 2] {{\small $U(3)$}};
\node[shape=circle, draw] (4) [right= 1.7cm of 3] {{\small $U(2)$}};
\node[shape=circle, draw] (5) [right= 1.3cm of 4] {{\small $U(1)$}};
\node[shape=circle, draw] (6) [above=1.5cm of 3] {{\small $U(2)$}};
\node[shape=circle, draw] (7) [above= 1.1cm of 6] {{\small $U(1)$}};
 \path[every node/.style={font=\sffamily\small,
  		fill=white,inner sep=1pt}]
(1) edge [bend left] node[above=2mm] {$D_a$} (2)
(2) edge [bend left] node[below=2mm] {$\tilde D_a$}(1)
(2) edge [bend left=20] node[above=2mm] {$\tilde Q_a$} (3)
(3) edge [bend left=20] node[below=2mm] {$Q_a$}(2)
(3) edge [bend left=20] node[above=2mm] {$Q_c$} (4)
(4) edge [bend left=20] node[below=2mm] {$\tilde Q_c$}(3)
(4) edge [bend left] node[above=2mm] {$\tilde D_c$} (5)
(5) edge [bend left] node[below=2mm] {$D_c$}(4)
(3) edge [bend left] node[left=2mm] {$Q_b$} (6)
(6) edge [bend left] node[right=2mm] {$\tilde Q_b$}(3)
(6) edge [bend left] node[left=2mm] {$\tilde D_b$} (7)
(7) edge [bend left] node[right=2mm] {$D_b$}(6)
(1) node[left=6mm] {$\phi_a$}
(2) node[below=6mm] {$\Psi_a$}
(3) node[below=6mm] {$\breve{\Psi}$}
(4) node[below=6mm] {$\Psi_c$}
(6) node[right=6mm] {$\Psi_b$}
(7) node[right=6mm] {$\phi_b$}
(5) node[right=6mm] {$\phi_c$}
;
\end{tikzpicture}
\end{center}
\caption{$E_6$ quiver.}\label{Fig:E6quiv}
\end{figure}

The superpotential is
\be
W = \sum_{k=a,b,c} \left[ \phi_k D_k \tilde D_k+\mbox{tr} \Psi_k (\tilde D_k D_k+\tilde Q_k Q_k) + \mbox{tr} \breve{\Psi} Q_k \tilde Q_k\right] \:.
\ee 
Here, $D \tilde D = D^\alpha \tilde D_\alpha$, $\tilde D D = \tilde D_\alpha  D^\beta$ and so on.

The first step we need to take is to ungauge the three $U(2)$ nodes. We then substitute the $U(3)$ gauge theory with six flavors with the dual non-Lagrangian theory. Finally we recouple it to the three $U(2)$ vector multiplets.
Practically this is done (like in the $D_N$ case) by considering again the coupling 
$$
W\supset {\Psi^j}_i \left({X^i}_j + \frac{((v\tilde{v}))}{3}\delta^i_j\right).
$$ 
However, now the $6\times 6$ matrix $\Psi$ is taken to be block diagonal, with three $2\times 2$ block corresponding to the three scalars $\Psi_k$ relative to the adjacent $U(2)$ nodes.
In particular, we need to decompose $X^i_j$ into $S(U(2) \times U(2) \times U(2))$ representations.
As before, the fact that the traceless matrix field $X$ square to zero ($X^2=0$) implies that the F-term relations coming from differentiating with respect to $\Psi_k$ do not change the Higgs branch relations that lead to the $E_6$ singularity.

\end{paragraph}

\subsection{Deforming by a monopole operator}
We are now in a position to ask what happens when we deform the $G=U(3), N_F = 6$ theory by a monopole operator. On the non-Lagrangian side, this will amount to deforming the $U(1)$ node by a monopole operator. This node is connected to an $SU(2)$ node that is connected to the non-Lagrangian block.
Hence the monopole deformation works analogous to the $U(2)$ case described in Section \ref{U2monopDef}.

More precisely, on the Lagrangian side, we are interested in deforming the superpotential by a monopole operator:
\be
\Delta_{\rm Lag} W = m V_-
\ee
with $V_-$ the R-charge one monopole operator charged under the topological $U(1)_J$ symmetry associated with the $U(3)$ node. Since global symmetries must match across the S-duality, the only candidate for this $U(1)_J$ on the non-Lagrangian side is the shift symmetry of the dual photon of the $U(1)$ under which the doublet $(v^\alpha, \tilde v_\beta)$ is charged.

On the non-Lagrangian side, we must deform our superpotential by a monopole operator on the $U(1)$ node.
\be
\Delta_{\rm non-Lag}= m W_{+} \:.
\ee

As in Section \ref{U2monopDef}, the $U(1)$ node is connected to an $SU(2)$ node and we can proceed by the same steps we did there. As a result, the
$U(1)$ node disappears from the deformed theory and  the $SU(2)$ meson  ${\mathfrak{m}_\alpha}^\beta\leftrightarrow \tilde{v}_\alpha v^\beta$ becomes a fundamental field. The local effective theory is coupled to the rest of of the quiver by the superpotential
\be 
W^{\rm eff}= \mbox{tr} \, [(\tilde{\mathfrak{m}}-\mu_{\rm su(2)}) \Phi]  -\frac{\mathcal{X}}{m}\mbox{det}\,\tilde{\mathfrak{m}}+ \mbox{tr}\tilde\Psi \,X +...\:,
\ee
where we have integrated the field $\phi'=\phi+\psi$ out,  $\tilde{\mathfrak{m}}$ is the traceless part of $\mathfrak{m}$, $(\mu_{\rm su(2)})_{\alpha\beta}=Z_{\alpha\beta}$ is the $SU(2)$ moment map and ``$...$'' means the rest of the `modified $E_N$ quiver'.

Again the question is, how is the Higgs branch modified after this whole operation? The new matrix $\tilde{\mathfrak{m}}$ has no non-zero gauge invariants, so it does not contribute as a coordinate for the Higgs branch. The only way in which it affects things, is via its coupling to the $SU(2)$ gauge theory, which is in turn coupled to the Minahan Nemeschansky theory. Concretely, it means that we have to replace all instances of $\tilde v_\alpha  v^\beta$ by ${\tilde{\mathfrak{m}}_\alpha}^ \beta$ in the relations \eqref{zelim} and \eqref{joe1} through \eqref{joe8}.
Since we've already showed that the Higgs branch of the undeformed theory was entirely parametrized by $X^i_j$ (i.e. had no $(v, \tilde v)$ dependence), the conclusion is that it does not change after this deformation.

\section{$U(N)$ nodes and class S trinions} 
\label{Sec:trinions}

In this section we generalize the discussion of the previous sections to $U(N)$ SQCD with $N>3$. The dual frame we find involves 
the class $\mathcal{S}$ three-punctured sphere $R_{0,N}$ so we need to introduce some machinery to understand the chiral ring of these models. 
We will start by reviewing the chiral ring of $T_N$ theory and then discuss the operation usually called in the literature 
``closure of puncture'', which allows us to flow from $T_N$ theory to $R_{0,N}$.

\subsection{Chiral ring relations for $T_N$ theory and its descendants} 

We are interested in studying the Higgs branch of (the dimensional reduction of) $R_{0,N}$ theory. Since the Higgs branch does not 
change under dimensional reduction, we can work directly in four dimensions; the conclusions will apply to the three-dimensional 
theory as well. Our starting point is the set of chiral ring relations for $T_N$ theory discussed in \cite{Maruyoshi:2013hja},\cite{Tachikawa:2015bga}. 
$R_{0,N}$ is then obtained from $T_N$ by partially closing one of the punctures, i.e. giving nilpotent vev to the corresponding $SU(N)$ 
moment map, so we need to understand how the vev for the moment map affects the chiral ring relations.

$T_N$ theory has a global symmetry $SU(N)_A\times SU(N)_B\times SU(N)_C$. Its chiral ring includes operators of dimension two $\mu_{A,B,C}$ in the adjoint representation of the three $SU(N)$ factors and operators $Q^{(k)}$ transforming in the $\Lambda^k$ (rank-k antisymmetric) representation of each $SU(N)$ symmetry, with dimension $k(N-k)$ for $k=1,\dots,N-1$. The most important ones for us will be the operators labelled by $k=1$ and $k=N-1$, which correspond to 
chirals $Q_{ijk}$, $\widetilde{Q}^{ijk}$ in the trifundamental and antitrifundamental of $SU(N)^3$ respectively. They satisfy the following chiral ring relations: 
\be \label{eq1}\tr\mu_A^k=\tr\mu_B^k=\tr\mu_C^k,\ee
\be \label{eq2}(\mu_A)_i^{i'}Q_{i'jk}=(\mu_B)_j^{j'}Q_{ij'k}=(\mu_C)_k^{k'}Q_{ijk'},\ee
\be \label{eq3}Q_{ijk}\widetilde{Q}^{lmk}=\sum_s \nu_s\sum_{n=0}^{N-s-1}(\mu_A^{N-s-1-n})_i^l(\mu_B^n)_j^m.\ee
In the last equation $\nu_s$ denote the coefficients of the characteristic polynomial of $\mu$: $\text{det}(x-\mu)=\sum_s \nu_sx^{N-s}$. 
Because of (\ref{eq1}), the polynomial does not depend on which $\mu$ operator we use. Notice that $\nu_0=1$ and $\nu_1=0$.

\subsection{Nilpotent vev and $R_{0,N}$ theory}

To derive the theory $R_{0,N}$ from $T_N$ we should give a nilpotent vev to one $\mu$ operator (say $\mu_C$) of the form 
\be\langle\mu_C\rangle=\left(\begin{array}{c|c} 
J_{N-2} & \\
\hline 
 & \\
\end{array}\right)\ee 
where $J_{N-2}$ is a Jordan block of size $N-2$. The corresponding nilpotent orbit can be labelled by an embedding $\rho:SU(2)\rightarrow SU(N)$. This vev breaks the $SU(N)_C$ global symmetry to $SU(2)\times U(1)$ and as a result most 
of the components of $\mu_C$ become the lowest components of Goldstone multiplets and decouple. Indeed the vev also breaks the original $SU(2)$ R-symmetry of the theory, which mixes with the $SU(2)$ subgroup defined by $\rho$ to give the new R-symmetry group in the infrared. For example, the Cartan generator $I_3$ is redefined as follows: 
\be\label{nuovaR}I_3\rightarrow I_3-\rho(\sigma_3),\ee 
where $\rho(\sigma_3)$ can be taken of the form 
$$\rho(\sigma_3)=\text{Diag}\left(\frac{N-3}{2},\frac{N-5}{2},\dots,\frac{3-N}{2},0,0\right)\:.$$
In order to analyze the resulting 
theory in the IR, we can expand around the vev keeping only the components which remain coupled to the theory. The resulting $\mu_C$ 
can be written in the form (see \cite{Gadde:2013fma} for details)
\be\label{newc}\mu_C=\left(\begin{array}{ccccc|c}
  \alpha & 1 & 0 & \dots & 0 & 0\\
  M_{1} & \alpha & 1 & \dots & 0 & \\
  M_{2} & M_{1} & \ddots & \ddots & 0 & \vdots\\
  \vdots & \ddots & \ddots & \alpha & 1 & 0\\
  M_{N-3} & \dots & M_{2} & M_{1} & \alpha & \beta\quad\gamma\\ 
  \hline 
  \begin{array}{c} \delta \\ \epsilon\end{array} & 0 & & \dots &  0 & \mu_{\rm su(2)}+\tilde{\alpha}\mathbb{1}_2   
\end{array}\right)\:.\ee
The traceless condition implies $\tilde{\alpha}=-(N-2)\alpha/2$ and $\mu_{\rm su(2)}$ denotes the $SU(2)$ moment map. The trifundamental 
$Q_{ijk}$ decomposes under $SU(2)$ as one doublet ($k=N-1,N$) and $N-2$ singlets ($k=1,\dots,N-2$). 
We will now argue that the $N-2$ $SU(2)$ singlets are not independent in the chiral ring.
The argument is essentially the same as in \cite{Maruyoshi:2013hja}: the chiral ring relations valid for $T_N$ hold for $R_{0,N}$ as well, provided we replace all 
occurrences of $\mu_C$ with (\ref{newc}). From (\ref{eq2}) we find 
$$(\mu_{B}^{n})_{j}^{j'}Q_{ij'1}=(\mu_{C}^{n})_{1}^{k}Q_{ijk}\quad \forall n.$$ 
Using (\ref{newc}), we see that the above equation (setting $n=1$) implies that $Q_{ij2}$ can be written in terms of $Q_{ij1}$, 
$\alpha$ and $\mu_B$ (or $\mu_A$), so is not a generator of the chiral ring. Analogously, for $n<N-2$ we deduce that 
$Q_{ijn+1}$ can be written in terms of $Q_{ijk}$ with $k\leq n$ and the other fields, leading to the conclusion that the only 
generator in the chiral ring is $Q_{ij1}$. An analogous argument allows to express $\widetilde{Q}^{ijk}$ with $k<N-2$ in terms 
of $\widetilde{Q}^{i,j,N-2}$. 

A key property of $R_{0,N}$ theory is that the manifest $SU(2)\times U(1)\times SU(N)^2$ global symmetry actually enhances to 
$SU(2)\times SU(2N)$ and our next task is to construct the moment map of $SU(2N)$ explicitly. We will state the result first and 
then provide evidence for our claim. Our proposal is as follows: 
\be\label{mmap}
X=\left(\begin{array}{c|c}\mu_A-\tilde{\alpha}\mathbb{1}_N & Q_{ij1} \\ 
\hline -\widetilde{Q}^{i,j,N-2} & \tilde{\alpha}\mathbb{1}_N-\mu_B^t\end{array}\right)\ee
where we called the moment map of $SU(2N)$ $X$ instead of $\mu_{SU(2N)}$ (as it will play the same role of the field $X$ in Sections \ref{Sec:U2node} and \ref{Sec:U3node}).
First of all, notice that all the fields appearing in the matrix $X$ have charge one under the redefined Cartan of $SU(2)_R$ (\ref{nuovaR}). A more stringent check can be obtain as follows:
indeed, the duality with SQCD discussed previously implies that the $SU(2N)$ moment map of $R_{0,N}$ should be identified with 
the traceless part of the meson in the gauge theory, whose square is proportional to the identity as dictated by F-term 
equations. If our claim is correct, the same should be true for $X$ 
as defined above and we will now see that this is precisely implied by equations (\ref{eq1}-\ref{eq3}). 

Let's start by squaring (\ref{mmap}); it is easy to see that the off-diagonal blocks are identically zero thanks to (\ref{eq2}), 
so we are left with 
\be\label{mmap2} X^2 
=\left(\begin{array}{c|c}\mu_A^2-2\tilde{\alpha}\mu_A+\tilde{\alpha}^2\mathbb{1}_N-Q_{il1}\widetilde{Q}^{j,l,N-2} & 0 \\ 
\hline 0 & -\widetilde{Q}^{l,i,N-2}Q_{lj1}+\tilde{\alpha}^2\mathbb{1}_N-2\tilde{\alpha}\mu_B^t+(\mu_B^t)^2\end{array}\right).\ee
Let us now discuss the term $Q_{il1}\widetilde{Q}^{j,l,N-2}$. Using (\ref{eq3}) we can rewrite it as 
\be\label{sum1}Q_{il1}\widetilde{Q}^{j,l,N-2}=\sum_s\nu_s\sum_{n=0}^{N-s-1}(\mu_C^{N-s-1-n})_1^{N-2}(\mu_A^n)_i^j.\ee 
Now, given the form (\ref{newc}) of $\mu_C$, we can observe that $(\mu_C^k)_1^{N-2}$ is nonzero only for $k\geq N-3$, consequently 
in the above formula we find nontrivial contributions only for $s=0,2$ ($\nu_1$ is zero). We can then rewrite the r.h.s. of (\ref{sum1}) as 
\be Q_{il1}\widetilde{Q}^{j,l,N-2}=(\mu_C^{N-3})_1^{N-2}(\nu_2\delta_i^j+(\mu_A^2)_i^j)+(\mu_C^{N-2})_1^{N-2}(\mu_A)_i^j+(\mu_C^{N-1})_1^{N-2}\delta_i^j.\ee
A detailed calculation using (\ref{newc}) leads to the following identities: 
$$\nu_2=-\frac{\tr\mu_C^2}{2}=-\frac{1}{2}\left(\tr\mu_{\rm su(2)}^2+(2N-6)M_1+\frac{N^2-2N}{2}\alpha^2\right)\,,$$
$$(\mu_C^{N-1})_1^{N-2}=(N-3)M_1+\frac{(N-2)(N-1)}{2}\alpha^2\,,$$ 
$$(\mu_C^{N-2})_1^{N-2}=(N-2)\alpha\,\quad\mbox{and}\qquad (\mu_C^{N-3})_1^{N-2}=1\,.$$
Plugging these equations into (\ref{mmap2}), one can easily see that the upper left block reduces to $\frac{\tr\mu_{\rm su(2)}^2}{2}$ times the identity. 
This argument applies also to the second diagonal block in (\ref{mmap2}) (one simply needs to replace $\mu_A$ with $\mu_B$ 
and take the transpose) with identical conclusion. As a result, we find the equation 
\be\label{nmap} X^2 
=\frac{\tr\mu_{\rm su(2)}^2}{2}\mathbb{1}_{2N},\ee 
which matches precisely the gauge theory expectation. 

If we now gauge the baryon number getting $U(N)$ SQCD, the chiral ring relation on the meson $M$ becomes $M^2=0$ and this should 
indeed occur in the dual frame involving $R_{0,N}$ as well. Since from the discussion of the previous section we know that gauging 
the baryon number amounts to gauging the $U(1)$ symmetry carried by the $SU(2)$ doublet, we have to check that $X^2$ 
is set to zero when we perform the $U(1)$ gauging. After the gauging the superpotential becomes 
$$W=v^\alpha {\Phi_\alpha}^\beta\tilde{v}_\beta - \tr(\Phi\mu_{\rm su(2)})-\phi\,v^\alpha\tilde{v}_\alpha,$$ 
where $v,\tilde{v}$ and $\Phi$ are the $SU(2)$ doublets and adjoint chiral respectively and $\phi$ is the chiral in the abelian 
4d $\mathcal{N}=2$ vector multiplet. Of course the last term is present only if the $U(1)$ symmetry is gauged. The F-term for $\Phi$ 
tells us that the traceless part of $\tilde{v}_\alpha v^\beta$ is equal to $(\mu_{\rm su(2)})_\alpha^\beta$, and the F-term for $\phi$ says instead that $\tilde{v}_\alpha v^\beta$ 
squares to zero. Overall, this implies that $\mu_{\rm su(2)}$ squares to zero and using (\ref{nmap}) we reach the desired conclusion 
\be\label{meson1} X^2 
=0.\ee
As shown in Section \ref{Sec:GoalStrategy}, the fact that the $SU(2N)$ moment map squares to zero  implies that we do not modify the Higgs branch 
of the E-type quiver by replacing a $U(N)$ gauge node with our dual description involving $R_{0,N}$.

\subsection{Modified E type quivers} 
\label{generalized-ade} 



In order to discuss exceptional quivers, it is convenient to consider separately two cases: in the first case we replace only gauge nodes 
in the tails; in the second case we modify the trivalent node. 

\subsubsection*{Changing the tails}

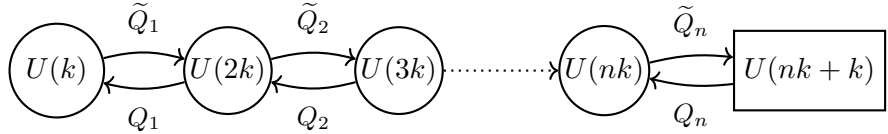
\begin{figure}[ht!]
\begin{center}
\begin{tikzpicture}[->, every node/.style={circle,draw},thick, scale=0.9]
  \node(V1) at (-10,0){$U(k)$};
  \node[inner sep=.8](V2) at (-7.5,0){$U(2k)$}; 
  \node[inner sep=.8](V3) at (-5,0){$U(3k)$};
  \node[inner sep=.8](VN+1) at (-2,0){$U(nk)$};
  \node[rectangle, draw, minimum height=30pt, thick](V4) at (1,0){$U(nk+k)$};

 \path[every node/.style={font=\sffamily\small,
  		fill=white,inner sep=1pt}]
(V1) edge [bend left=15] node[above=2mm] {$\widetilde{Q}_1$} (V2)
(V2) edge [bend left=15] node[below=2mm] {$Q_1$}(V1)
(V2) edge [bend left=15] node[above=2mm] {$\widetilde{Q}_2$} (V3)
(V3) edge [bend left=15] node[below=2mm] {$Q_2$} (V2)
(V3) edge [dotted] (VN+1)
(VN+1) edge [bend left=10] node[above=2mm] {$\widetilde{Q}_n$} (V4)
(V4) edge [bend left=10] node[below=2mm] {$Q_n$}(VN+1)
;

\end{tikzpicture}
\caption{Linear tail of length $n+1$. The rectangle denotes a global symmetry which is gauged in the E-type quiver (central node).}
\label{figtail}
\end{center}
\end{figure}
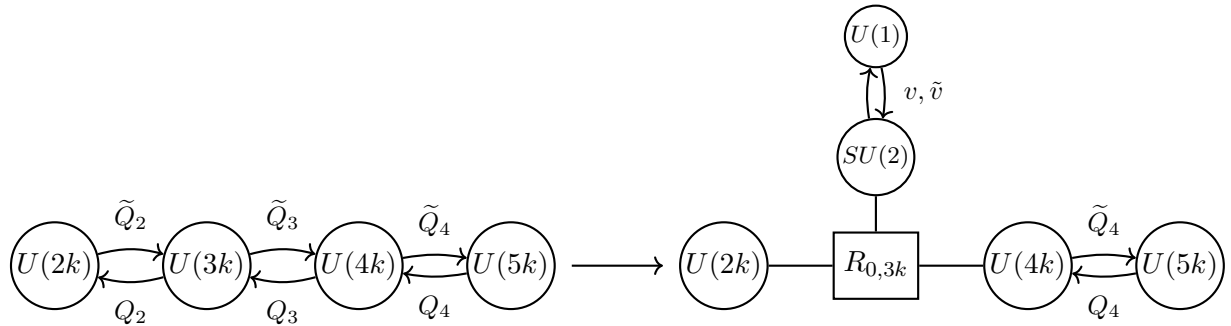
\begin{figure}[ht!]
\centering
\begin{tikzpicture}[->, every node/.style={circle,draw},thick, scale=0.8]

  \node[inner sep=.8](V2) at (-18,0){$U(2k)$}; 
  \node[inner sep=.8](V3) at (-15.5,0){$U(3k)$};
  \node[inner sep=.8](VN+1) at (-13,0){$U(4k)$};
  \node[inner sep=.8](V4) at (-10.5,0){$U(5k)$};

  \node[inner sep=.8](L2) at (-7,0){$U(2k)$}; 
  \node[rectangle, draw, minimum height=25pt, thick](L3) at (-4.5,0){$R_{0,3k}$}; 
\node[inner sep=.8](L8) at (-4.5,1.8){{\footnotesize $SU(2)$}};
\node[inner sep=.8](L9) at (-4.5,3.8){{\footnotesize $U(1)$}};
  \node[inner sep=.8](L4) at (-2,0){$U(4k)$};
  \node[inner sep=.8](L5) at (.5,0){$U(5k)$};

\draw[=>] (-9.5,0)--  (-8,0);

\draw[-] (L2) -- (L3);
\draw[-] (L3) -- (L4);
\draw[-] (L3) -- (L8);

 \path[every node/.style={font=\sffamily\small,
  		fill=white,inner sep=1pt}]
(V2) edge [bend left=15] node[above=2mm] {$\widetilde{Q}_2$} (V3)
(V3) edge [bend left=15] node[below=2mm] {$Q_2$} (V2)
(V3) edge [bend left=15] node[above=2mm] {$\widetilde{Q}_3$} (VN+1)
(VN+1) edge [bend left=15] node[below=2mm] {$Q_3$} (V3)
(VN+1) edge [bend left=10] node[above=2mm] {$\widetilde{Q}_4$} (V4)
(V4) edge [bend left=10] node[below=2mm] {$Q_4$}(VN+1)
(L8) edge [bend left=10] (L9)
(L9) edge [bend left=10] node[right=2mm] {$v,\tilde v$} (L8)

(L4) edge [bend left=10] node[above=2mm] {$\widetilde{Q}_4$} (L5)
(L5) edge [bend left=10] node[below=2mm] {$Q_4$}(L4)
;

\end{tikzpicture} 
\caption{We replace the $U(3k)$ gauge node with non-Lagrangian dual theory in the quiver tail.}
\label{figlin}
\end{figure}

Let us concentrate on a single linear tail of the form represented in Figure  \ref{figtail}.
All tails in the E-type quiver have this form. Let us now replace say the node $U(3k)$ as in Figure  \ref{figlin} with the 
non-Lagrangiam theory. 

The meson of $SU(6k)$ can be written as follows 
\be\label{mmap3}X_{SU(6k)}=\left(\begin{array}{c|c}M_{2k} & A \\ 
\hline B & M_{4k}\end{array}\right)\:,\ee
where $M_{2k}$ and $M_{4k}$ transform in the adjoint of the $SU(2k)$ and $SU(4k)$ subgroups which are gauged in the linear tail theory, 
whereas $A$ and $B$ are bifundamentals. 
The 3d $\mathcal{N}=4$ theory couples to the rest of the quiver by the superpotential term 
\be\label{mes3} 
W\supset \tr \Psi X_{SU(6k)} =\tr\Psi_{2k}M_{2k}+\tr\Psi_{4k}M_{4k}\:,
\ee
where $\Psi$ takes a block diagonal form, with the two adjoint fields $\Psi_{2k}$ and $\Psi_{4k}$ of the adjacent nodes in the diagonal.
Again, this coupling implies, together with $X_{SU(6k)}^2=0$, that the Higgs branch is unchanged after replacing the $U(3k)$ node.

\subsubsection*{Changing the trivalent node} 

The trivalent node is a $U(N)$ gauge theory with $SU(2N)$ global symmetry, of which a 
$SU(n_1)\times SU(n_2)\times SU(n_3)$ subgroup ($n_1+n_2+n_3=2N$) is gauged. 
\begin{figure}[ht!]
\centering
\begin{tikzpicture}[->, every node/.style={circle,draw},thick, scale=0.8]

  \node[inner sep=.8](V1) at (-17,2){$U(n_1)$}; 
  \node[inner sep=.8](VN) at (-14,2){$U(N)$};
  \node[inner sep=.8](V2) at (-14,-0.5){$U(n_2)$};
  \node[inner sep=.8](V3) at (-11,2){$U(n_3)$};

  \node[inner sep=.8](L2) at (-7,1){$U(n_1)$}; 
  \node[rectangle, draw, minimum height=25pt, thick](L3) at (-4.5,1){$R_{0,N}$}; 
  \node[inner sep=.8](L8) at (-4.5,2.8){{\footnotesize $SU(2)$}};
\node[inner sep=.8](L9) at (-4.5,4.8){{\footnotesize $U(1)$}};
  \node[inner sep=.8](L4) at (-2,1){$U(n_3)$};
  \node[inner sep=.8](L5) at (-4.5,-1.5){$U(n_2)$};

\draw[=>] (-10,1)--  (-8.5,1);

\draw[-] (L2) -- (L3);
\draw[-] (L4) -- (L3);
\draw[-] (L5) -- (L3);
\draw[-] (L3) -- (L8);

 \path[every node/.style={font=\sffamily\small,
  		fill=white,inner sep=1pt}]
(V1) edge [bend left=15] node[above=1mm] {$\widetilde{Q}_1$} (VN)
(VN) edge [bend left=15] node[below=1mm] {$Q_1$}(V1)
(V2) edge [bend left=15] node[left=1mm] {$\widetilde{Q}_2$} (VN)
(VN) edge [bend left=15] node[right=1mm] {$Q_2$} (V2)
(V3) edge [bend left=15] node[below=1mm] {$\widetilde{Q}_3$} (VN)
(VN) edge [bend left=15] node[above=1mm] {$Q_3$} (V3)
(L8) edge [bend left=10] (L9)
(L9) edge [bend left=10] node[right=2mm] {$v,\tilde v$} (L8)

;

\end{tikzpicture} 
\caption{We replace the $U(N)$ central node with the non-Lagrangian dual theory in the quiver.}
\label{figcentral}
\end{figure}
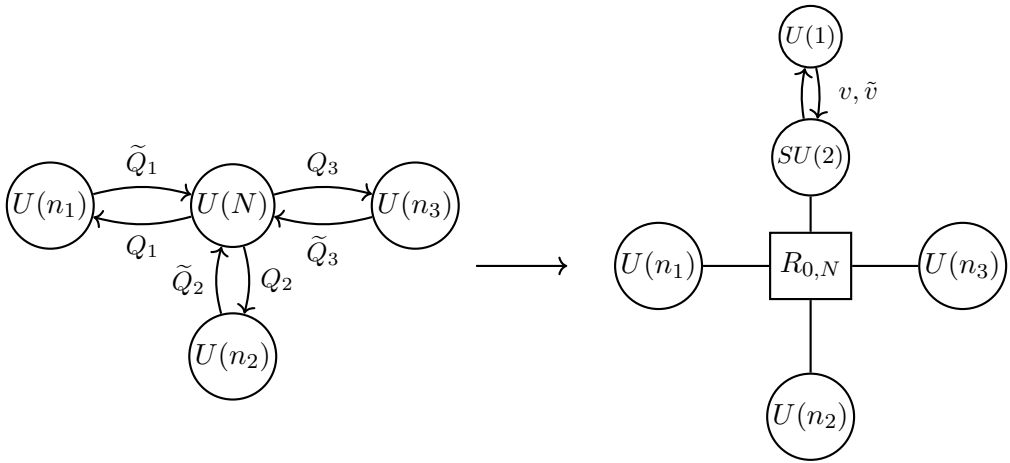

\noindent In this case it is convenient to write the $SU(2N)$ meson of $R_{0,N}$ as a block matrix of the form 
\be X_{SU(2N)}=\left(\begin{array}{c|c|c}M_{11} & M_{12} & M_{13} \\ 
\hline M_{21} & M_{22} & M_{23} \\ 
\hline M_{31} & M_{32} & M_{33}\end{array}\right).\ee 
The blocks $M_{ii}$ transform in the adjoint of $SU(n_i)$ and the off-diagonal blocks are bifundamentals. 

Again we couple the dual node to the quiver by the superpotential term
\be
W\supset \tr \Psi X_{SU(N)}\:,  
\ee
where now $\Psi$ has three non-zero block along the diagonal, i.e. the adjoint scalars $\Psi_{n_1}$, $\Psi_{n_2}$ and $\Psi_{n_3}$ of the adjacent nodes.
This replacement does not change the Higgs branch.

\subsection{Deforming by a monopole operator}

So far we have only discussed the $\mathcal{N}=4$ theory, without introducing the monopole deformation. As already seen, this can be easily handled using our duality. Since the analysis is basically identical to that of the previous sections, we will be brief. Once we replace a $U(N)$ gauge node with the $R_{0,N}$ theory coupled to the $U(1)\times SU(2)$ quiver, the monopole deformation should be turned on at the $U(1)$ node so, using the results discussed in Section \ref{Sec:U1monop}, we conclude that the net effect of the extra superpotential term is to replace the abelian tail with a chiral multiplet $\mathfrak{m}$ in the adjoint of the $SU(2)$ gauge group. Its interactions with the other fields are described by the superpotential 
\be\mathcal{W}=\phi\,\tr \mathfrak{m} + \mathcal{X}\,\text{det}\mathfrak{m} + \tr((\mathfrak{m}-\mu_{\rm su(2)})\Phi).\ee 
Here $\phi$  and $\Phi$ are the chirals in the $\mathcal{N}=4$ $U(1)$ and $SU(2)$ vector multiplets respectively. The above interactions imply that the matrix $\mathfrak{m}$ is traceless and singular, so it squares to zero. The F-term for $\Phi$ in turn imply that the $SU(2)$ moment map of $R_{0,N}$ satisfies the same constraint and consequently the same is still true for the $SU(2N)$ moment map.

\subsection{Baryonic operators in $R_{0,N}$ theory} 

Even though in this paper we are mostly interested in $U(N)$ SQCD, where baryons are not gauge invariant, let us see how we can get (the dual of) baryonic operators in $R_{0,N}$ theory. 

We start by noticing that from $Q^{ijk}$ we get an operator transforming in the $({\bf 2},{\bf N},{\bf N})$ of $SU(2)\times SU(N)^2$ with charge $(N-1)/2$ under $I_3$ (defined as in (\ref{nuovaR})). More in general, from every operator $Q^{(k)}$ we get an operator with charge $(N-1)/2$ under $I_3$ which transforms as a doublet of $SU(2)$ and in the $(\Lambda^k,\Lambda^k)$ of $SU(N)^2$. In order to see this, let us write the indices of $Q^{(k)}$ explicitly: 
$$Q^{(k)}=Q^{[a_1,\dots,a_k],[b_1,\dots,b_k],[c_1,\dots,c_k]}.$$ 
If we now set $[c_1,\dots,c_k]=[1,2,\dots,k-1,i]$ $(i=N-1,N)$ we see that the resulting operator transforms as stated above under the global symmetry of the theory and its charge under~(\ref{nuovaR})~is 
$$\frac{1}{2}\left(k(N-k)-\sum_{i=1}^{k-1}(N-1-2i)\right)=\frac{N-1}{2}.$$
Furthermore, from (\ref{newc}) we see that the components $\beta, \gamma, \delta, \epsilon$ of $\mu_C$ fit into two doublets of $SU(2)$, always with $I_3$ charge $(N-1)/2$. 

The above analysis suggests that all these operators fit into a single irrep of $SU(2N)$ of dimension $\binom{2N}{N}$, which is the dimension of the rank-N antisymmetric representation of $SU(2N)$. Notice that under $SU(N)^2$ it decomposes as 
$$\Lambda^N=\bigoplus_i (\Lambda^i,\Lambda^{N-i}),$$ 
exactly reproducing the set of $N+1$ operators found above. We conclude that $R_{0,N}$ includes a chiral operator of dimension $N-1$ transforming in the $({\bf 2},\Lambda^N)$ of $SU(2)\times SU(2N)$ which we call $Y^{\alpha}_{[a_1,\dots,a_N]}$. This operator is known to exist in the case $N=3$ and was discussed in the previous section. In order to get $SU(2)$ invariants, we have to contract the $SU(2)$ index with the doublet $v,\tilde{v}$. In this way we get two gauge invariant operators ($((Yv))$ and $((Y\tilde{v}))$) of $I_3$ charge $N/2$. These have the same R-charge and transform in the same representation under $SU(2N)$ as the baryons $B$ and $\widetilde{B}$ in $SU(N)$ SQCD. 

Since in SQCD mesons and baryons generate the whole Higgs branch, we do not expect other Higgs branch operators in $R_{0,N}$. In order to prove this, perhaps along the lines of the argument for $X_{SU(2N)}$ given in the previous section, we would need chiral ring relations involving $Q^{(k)}$ operators, which are currently not known. A chiral ring relation generalizing (\ref{eq2}), consistent with those given in \cite{Tachikawa:2015bga}, which would be helpful in the proof is 
$$\mu_A^{k-1}Q^{(k)}=\mu_B^{k-1}Q^{(k)}=\mu_C^{k-1}Q^{(k)},$$ 
where the indices of $\mu_i$ are contracted with those of the $Q^{(k)}$ operator. We will not attempt to prove them in this paper.

\section{New three-dimensional dualities} \label{Sec:dualnew}

All the models discussed in this paper are quiver theories which display (before the superpotential deformation) an ADE global symmetry which is not manifest from the Lagrangian representation of the theory, rather it arises quantum mechanically due to the presence of monopole operators of R-charge one. When we turn on a monopole deformation at two different nodes in the quiver, the resulting theories naively seem different. However from the perspective of the mirror theory, in which the ADE symmetry is manifest, this just corresponds to turning on a mass deformation along two different simple roots of the Lie algebra and these are indeed related by a Weyl transformation. 

From this discussion we learn that the two resulting theories are dual to each other, although the Lagrangian presentation of theory obscures this fact. The purpose of this section is to prove this duality field theoretically: we will see that the theory we get by deforming the $\mathcal{N}=4$ theory with a monopole superpotential does not depend on the particular gauge node we choose. 

Actually, we expect the duality to be even more general: a well known fact is that for all simply laced Lie algebras we can obtain any root starting from a given one by acting with the Weyl group. The conclusion is that the duality still holds if we turn on the deformation along a generic root, not necessarily a simple one. This corresponds to turning on a superpotential term proportional to a monopole charged under the topological $U(1)$ symmetry of multiple gauge nodes. At present we do not know how to handle such superpotentials. However, using the machinery developed in this paper, we will show this duality for the subsector of monopole operators magnetically charged under a single quiver node, which in turn correspond to the simple roots of the ADE algebras.

\subsection{D-type quivers}

Consider the $D_N$ quiver gauge theory (see Figure  \ref{DNquiverapp}).
As said above, from mirror symmetry considerations one expects the following: deforming the $D_N$ gauge theory superpotential by switching on a monopole operator charged under one node  
should produce the same effective theory as deforming the $D_N$ gauge theory superpotential by a monopole operator charged under a different node. 

This statement is far from being obvious in the underlying quiver field theory.
In this section we will prove it, by showing the equivalence of one $U(1)$ node with the adjacent $U(2)$ node and the equivalence of two adjacent $U(2)$ nodes.

\subsubsection*{The external $U(1)$ node is equivalent to adjacent $U(2)$ node}

Let us consider {\it Theory 1} and {\it Theory 2} of Section \ref{Sec:BBduality}. 
The two theories have the same Higgs and Coulomb branches. In particular, the topological symmetry relative to the diagonal $U(2)$ node of {\it Theory 1} is mapped to the topological symmetry relative to the $U(1)$ gauge group factor in {\it Theory 2}. Correspondingly the monopole operators of the two theories that have equal charge with respect to the topological symmetry (and the same R-charge) are exchanged.

We now  modify both theories by gauging a $U(1)$ subgroup of the $SU(4)\cong SO(6)$ global symmetry, such that the surviving global symmetry is $SU(3)$. Correspondingly we need to add the proper term to the superpotential necessary to preserve 3d $\mathcal{N}=4$ supersymmetry.
\begin{description}
\item {\it Theory 1}': Consider the subgroup $ SU(3)\times U(1)_{\varphi}\subset SU(4)$, where the $U(1)_{\varphi}$ is rotating one of the four flavors. Gauging the $U(1)_\varphi$ factor corresponds then  to pick up one of the four flavors, say $(\tilde Q^1,Q_1)$, and add the superpotential term
\begin{equation}
  W \supset \varphi Q_1 \tilde{Q}^1\:,
\end{equation}
where the gauging introduces the scalar $\varphi$ that completes the $3d$ $\mathcal{N}=4$ vector multiplet.
The corresponding theory is represented in the following quiver:
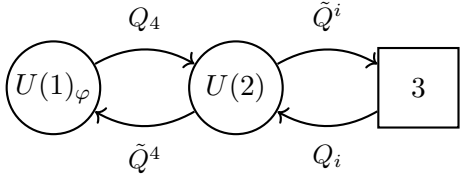
\begin{center}
\begin{tikzpicture}[->,thick, scale=0.8]
  \node[circle, draw, inner sep= 1pt, minimum width=35pt](L0) at (7,0){$U(1)_\varphi$};
  \node[circle, draw, inner sep= 1pt, minimum width=35pt](L1) at (10,0){$U(2)$};
  \node[draw, rectangle, minimum width=30pt, minimum height=30pt](L2) at (13,0){$3$};
 \path[every node/.style={font=\sffamily\small,
  		fill=white,inner sep=1pt}]
(L0) edge [bend left] node[above=2mm] {$Q_4$} (L1)
(L1) edge [bend left] node[below=2mm] {$\tilde Q^4$}(L0)
(L1) edge [bend left] node[above=2mm] {$\tilde Q^i$} (L2)
(L2) edge [bend left] node[below=2mm] {$Q_i$}(L1)
;
\end{tikzpicture}
\end{center}
where now $i=2,...,4$.
\item {\it Theory 2}': Applying the duality map between {\it Theory 1} and {\it Theory 2}, we see that now $U(1)_\varphi$
rotates all the fundamental fields $q_i$ with the same phase (and $\tilde{q}^i$ with the opposite phase). Gauging the $U(1)_\varphi$ factor corresponds then  to add the superpotential term
\begin{equation}
  W \supset \varphi \sum_{i=1}^3 q_i \tilde{q}^i\:.
\end{equation}
with $\varphi$ again the scalar in the $\mathcal{N}=4$ vector multiplet.
The corresponding quiver is:
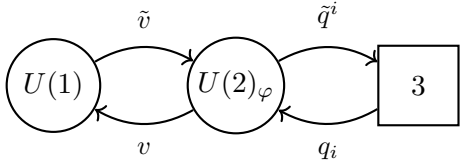
\begin{center}
\begin{tikzpicture}[->,thick, scale=0.8]
  \node[circle, draw, inner sep= 2pt, minimum width=35pt](L0) at (7,0){$U(1)$};
  \node[circle, draw, inner sep= 2pt, minimum width=35pt](L1) at (10,0){$U(2)_\varphi$};
  \node[draw, rectangle, minimum width=30pt, minimum height=30pt](L2) at (13,0){$3$};
 \path[every node/.style={font=\sffamily\small,
  		fill=white,inner sep=1pt}]
(L0) edge [bend left] node[above=2mm] {$\tilde{v}$} (L1)
(L1) edge [bend left] node[below=2mm] {$v$}(L0)
(L1) edge [bend left] node[above=2mm] {$\tilde q^i$} (L2)
(L2) edge [bend left] node[below=2mm] {$q_i$}(L1)
;
\end{tikzpicture}
\end{center}
where now the gauged $U(1)_\varphi$ is the diagonal abelian factor of $U(2)$, that for this reason we call $U(2)_\varphi$.
\end{description}
We see that now the two theories are identical.  In particular, the global symmetry of the Coulomb branch is now $SU(3)$ (as the nodes are balanced) and the (previous) duality exchanges the two topological $U(1)$ generators corresponding to the two nodes. As a consequence, the monopole operators of R-charge $r$ corresponding to the $U(2)$ factor in one theory are mapped to the monopole operators of R-charge $r$ corresponding to the $U(1)$ factor in the other theory. 

This result allows us the show the equivalence of one $U(1)$ node of the $D_N$ quiver with the adjacent $U(2)$ node: Deforming the $D_N$ superpotential by an R-charge one monopole operator relative to the external $U(1)$ node of the $D_N$ quiver gauge theory is equivalent to deforming it by an R-charge one monopole operator charged under the topological $U(1)$ of the adjacent $U(2)$ node. 

\subsubsection*{Internal adjacent $U(2)$ nodes are equivalent}\label{Sec:InternalU2nodes}
To prove the equivalences of the other nodes, we need to introduce another duality. Consider two $SU(2)$ gauge groups with bifundamentals. A hypermultiplet $(q,\tilde q)$ in the bifundamental representations of two $SU(2)$ gauge groups enjoys an $SU(2)$ flavor symmetry that rotates the two half-hypermultiplets. We will represent this as

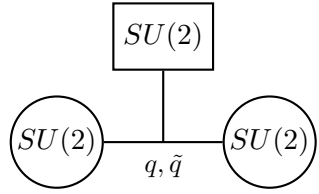
\begin{center}
\begin{tikzpicture}[thick, scale=0.7]
  \node[circle, draw, inner sep= 1pt](L0) at (6,0){$SU(2)$};
  \node[circle, draw, inner sep= 1pt](L1) at (10,0){$SU(2)$};
  \node[draw, rectangle, minimum width=20pt, minimum height=25pt](L3) at (8,2){$SU(2)$};
 \path[every node/.style={font=\sffamily\small,
  		fill=white,inner sep=1pt}]
(L0) edge  (L1)
(L3) edge node[below=6mm] {$q,\tilde q$} (8,0)
;
\end{tikzpicture}
\end{center}

Let us now consider the following theory with $SU(2)$ gauge group and four fundamental fields with the structure:
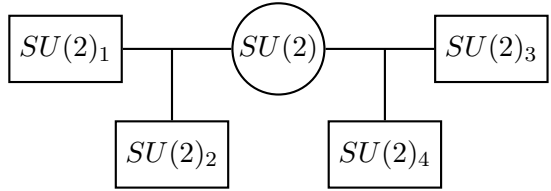
\begin{center}
\begin{tikzpicture}[thick, scale=0.7]
  \node[draw, rectangle, minimum width=30pt, minimum height=25pt](L1) at (6,0){$SU(2)_1$};
  \node[circle, draw, inner sep= 1pt](L2) at (10,0){$SU(2)$};
  \node[draw, rectangle, minimum width=30pt, minimum height=25pt](L3) at (14,0){$SU(2)_3$};
  \node[draw, rectangle, minimum width=30pt, minimum height=25pt](L5) at (8,-2){$SU(2)_{2}$};
  \node[draw, rectangle, minimum width=30pt, minimum height=25pt](L6) at (12,-2){$SU(2)_{4}$};
 \path[every node/.style={font=\sffamily\small,
  		fill=white,inner sep=1pt}]
(L1) edge  (L2)
(L2) edge (L3)
(L5) edge (8,0)
(L6) edge (12,0)

;
\end{tikzpicture}
\end{center}

This is dual to the theory with the same gauge group but represented by a different `quiver' (see \cite{Gaiotto:2009we})
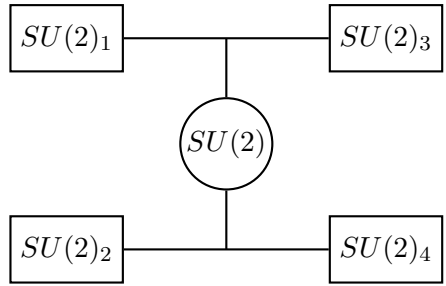
\vspace*{0.5cm}
\begin{center}
\begin{tikzpicture}[thick, scale=0.7]
  \node[draw, rectangle, minimum width=30pt, minimum height=25pt](L1) at (6,2){$SU(2)_1$};
  \node[circle, draw, inner sep= 1pt](L2) at (9,0){$SU(2)$};
  \node[draw, rectangle, minimum width=30pt, minimum height=25pt](L3) at (12,2){$SU(2)_3$};
  \node[draw, rectangle, minimum width=30pt, minimum height=25pt](L5) at (6,-2){$SU(2)_{2}$};
  \node[draw, rectangle, minimum width=30pt, minimum height=25pt](L6) at (12,-2){$SU(2)_{4}$};
 \path[every node/.style={font=\sffamily\small,
  		fill=white,inner sep=1pt}]
(L1) edge  (L3)
(L5) edge (L6)
(L2) edge (9,2)
(L2) edge (9,-2)

;
\end{tikzpicture}
\end{center}

\

We now apply the duality just described to construct three dual theories:
\begin{description}
\item {\it Theory 1}: We start with a three dimensional $\mathcal{N}=4$ theory with gauge group $SU(2)^2$, with bifundamental and fundamental fields as represented in the following quiver 
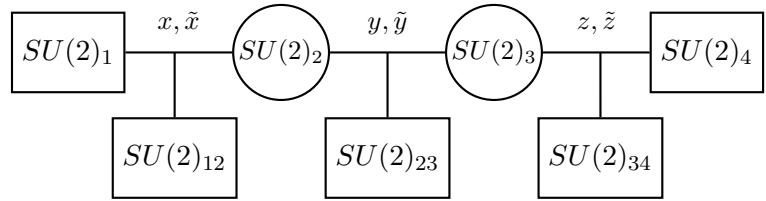
\vspace*{0.7cm}
\begin{center}
\begin{tikzpicture}[thick, scale=0.7]
  \node[draw, rectangle, minimum width=30pt, minimum height=30pt](L1) at (6,0){$SU(2)_1$};
  \node[circle, draw, inner sep= 1pt](L2) at (10,0){{\small $SU(2)_2$}};
  \node[circle, draw, inner sep= 1pt](L3) at (14,0){{\small $SU(2)_3$}};
  \node[draw, rectangle, minimum width=30pt, minimum height=30pt](L4) at (18,0){$SU(2)_4$};
  \node[draw, rectangle, minimum width=30pt, minimum height=30pt](L5) at (8,-2){$SU(2)_{12}$};
  \node[draw, rectangle, minimum width=30pt, minimum height=30pt](L6) at (12,-2){$SU(2)_{23}$};
  \node[draw, rectangle, minimum width=30pt, minimum height=30pt](L7) at (16,-2){$SU(2)_{34}$};
 \path[every node/.style={font=\sffamily\small,
  		fill=white,inner sep=1pt}]
(L1) edge  node[above=2mm] {$x,\tilde x$} (L2)
(L2) edge node[above=2mm] {$y,\tilde y$} (L3)
(L3) edge node[above=2mm] {$z,\tilde z$} (L4)
(L5) edge (8,0)
(L6) edge (12,0)
(L7) edge (16,0)
;
\end{tikzpicture}
\end{center}
\vspace*{0.7cm}
Applying the duality one obtains:
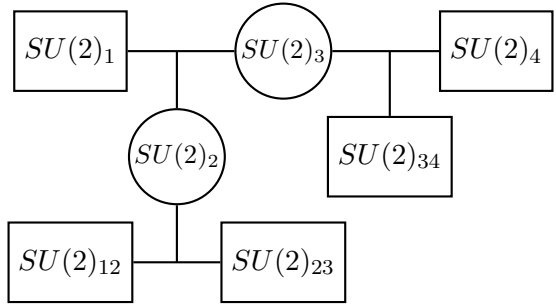
\vspace*{0.7cm}
\begin{center}
\begin{tikzpicture}[thick, scale=0.7]
 \node[draw, rectangle, minimum width=30pt, minimum height=30pt](L1) at (14,-4){$SU(2)_{23}$};
  \node[draw, rectangle, minimum width=30pt, minimum height=30pt](L2) at (10,0){$SU(2)_1$};
  \node[circle, draw, inner sep= 1pt](L3) at (14,0){{\small $SU(2)_3$}};
  \node[draw, rectangle, minimum width=30pt, minimum height=30pt](L4) at (18,0){$SU(2)_4$};
  \node[draw, rectangle, minimum width=30pt, minimum height=30pt](L5) at (10,-4){$SU(2)_{12}$};
  \node[circle, draw, inner sep= 1pt](L6) at (12,-2){{\small $SU(2)_2$}};
  \node[draw, rectangle, minimum width=30pt, minimum height=30pt](L7) at (16,-2){$SU(2)_{34}$};
 \path[every node/.style={font=\sffamily\small,
  		fill=white,inner sep=1pt}]
(L1) edge  (L5)
(L2) edge (L3)
(L3) edge (L4)
(12,-4) edge (L6)
(L6) edge (12,0)
(L7) edge (16,0)
;
\end{tikzpicture}
\end{center}
\vspace*{0.7cm}

\item {\it Theory 2}:  We now apply  the duality further, obtaining:
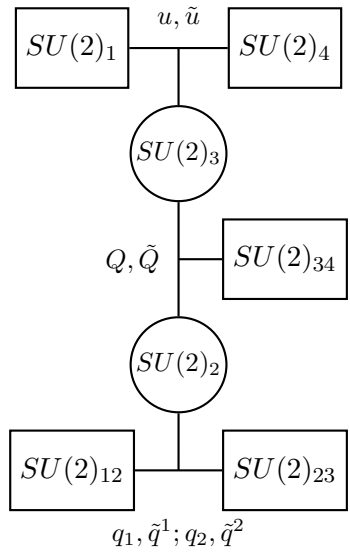
\vspace*{0.7cm}
\begin{center}
\begin{tikzpicture}[thick, scale=0.7]
 \node[draw, rectangle, minimum width=30pt, minimum height=30pt](L1) at (14,-8){$SU(2)_{23}$};
  \node[draw, rectangle, minimum width=30pt, minimum height=30pt](L2) at (10,0){$SU(2)_1$};
  \node[draw, rectangle, minimum width=30pt, minimum height=30pt](L3) at (14,0){$SU(2)_4$};
 \node[circle, draw, inner sep= 1pt](L4) at (12,-2){{\small $SU(2)_3$}};
  \node[draw, rectangle, minimum width=30pt, minimum height=30pt](L5) at (10,-8){$SU(2)_{12}$};
  \node[circle, draw, inner sep= 1pt](L6) at (12,-6){{\small $SU(2)_2$}};
  \node[draw, rectangle, minimum width=30pt, minimum height=30pt](L7) at (14,-4){$SU(2)_{34}$};
 \path[every node/.style={font=\sffamily\small,
  		fill=white,inner sep=1pt}]
(L1) edge node[below=6mm] {$q_1,\tilde q^1; q_2,\tilde q^2$}  (L5)
(L2) edge node[above=2mm] {$u,\tilde u$}  (L3)
(L6) edge node[left=2mm] {$Q,\tilde Q$} (L4)
(12,-8) edge (L6)
(L4) edge (12,0)
(L7) edge (12,-4)
;
\end{tikzpicture}
\end{center}
\vspace*{0.7cm}

\item {\it Theory 3}: Finally applying further the duality, we obtain the same theory but with the $SU(2)$ flavor groups arranged in a different way:
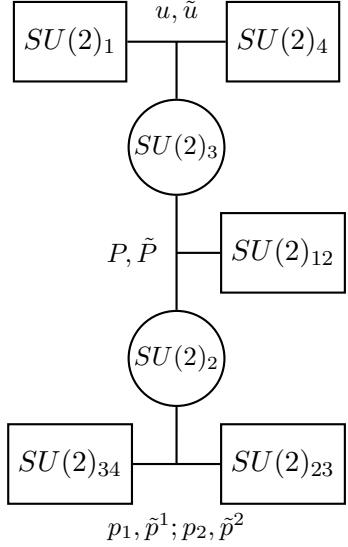
\vspace*{0.7cm}
\begin{center}
\begin{tikzpicture}[thick, scale=0.7]
 \node[draw, rectangle, minimum width=30pt, minimum height=30pt](L1) at (14,-8){$SU(2)_{23}$};
  \node[draw, rectangle, minimum width=30pt, minimum height=30pt](L2) at (10,0){$SU(2)_1$};
  \node[draw, rectangle, minimum width=30pt, minimum height=30pt](L3) at (14,0){$SU(2)_4$};
 \node[circle, draw, inner sep= 1pt](L4) at (12,-2){{\small $SU(2)_3$}};
  \node[draw, rectangle, minimum width=30pt, minimum height=30pt](L5) at (10,-8){$SU(2)_{34}$};
  \node[circle, draw, inner sep= 1pt](L6) at (12,-6){{\small $SU(2)_2$}};
  \node[draw, rectangle, minimum width=30pt, minimum height=30pt](L7) at (14,-4){$SU(2)_{12}$};
 \path[every node/.style={font=\sffamily\small,
  		fill=white,inner sep=1pt}]
(L1) edge node[below=6mm] {$p_1,\tilde p^1; p_2,\tilde p^2$}  (L5)
(L2) edge node[above=2mm] {$u,\tilde u$}  (L3)
(L6) edge node[left=2mm] {$P,\tilde P$} (L4)
(12,-8) edge (L6)
(L4) edge (12,0)
(L7) edge (12,-4)
;
\end{tikzpicture}
\end{center}

\end{description}
\vspace*{0.7cm}

We are now ready to approach the $D_N$ series. Take {\it Theory 1} and gauge $U(1)$ symmetries such that the gauge groups $SU(2)_2$ and $SU(2)_3$ are promoted to $U(2)$ groups. This will break a number of flavor symmetries. Of course we want to preserve $SU(2)_1$ and $SU(2)_4$ that will be gauged once we attach the resulting theory to the $D_N$ quiver.

We start with $SU(2)_2$. Promoting this to $U(2)_2$ breaks the flavor symmetries $SU(2)_{12}$ and $SU(2)_{23}$, while preserving $SU(2)_1$. Analogously, gauging $SU(2)_3$ to $U(2)_3$ breaks the flavor symmetries $SU(2)_{23}$ and $SU(2)_{34}$, while preserving $SU(2)_4$.
This reflects into the new superpotential couplings that should be introduce to preserve $\mathcal{N}=4$ supersymmetry:
\begin{equation}\label{Th1GInt}
  W \supset \varphi_2 (x \tilde{x} - y \tilde y) + \varphi_3 (y \tilde{y} - z \tilde z)\:.
\end{equation}
One can check that they break the flavor symmetries as required.
The final theory can be represented as
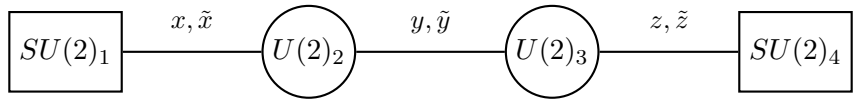
\vspace*{0.5cm}
\begin{center}
\begin{tikzpicture}[thick, scale=0.8]
  \node[draw, rectangle, minimum width=30pt, minimum height=30pt](L1) at (6,0){$SU(2)_1$};
  \node[circle, draw, inner sep= 2pt](L2) at (10,0){$U(2)_2$};
  \node[circle, draw, inner sep= 2pt](L3) at (14,0){$U(2)_3$};
  \node[draw, rectangle, minimum width=30pt, minimum height=30pt](L4) at (18,0){$SU(2)_4$};
 \path[every node/.style={font=\sffamily\small,
  		fill=white,inner sep=1pt}]
(L1) edge  node[above=2mm] {$x,\tilde x$} (L2)
(L2) edge node[above=2mm] {$y,\tilde y$} (L3)
(L3) edge node[above=2mm] {$z,\tilde z$} (L4)

;
\end{tikzpicture}
\end{center}
\vspace*{0.5cm}
that is a segment of the $U(2)$ chain in the $D_N$ quiver.

Following the duality described above, we see what this gauging  produces in 
{\it Theory 2}. Here we must gauge the $U(1)$s that reproduce the same effect on the flavor symmetry (i.e. the flavor symmetry of the resulting theories must be the same). The superpotential that encodes this is
\begin{equation}
  W \supset \varphi_2 q_1\tilde q^1  + \varphi_3 (q_1 \tilde{q}^1 + q_2 \tilde q^2 + Q\tilde Q)\:.
\end{equation}
The first coupling breaks the symmetries $SU(2)_{12}$ and $SU(2)_{23}$ keeping $SU(2)_{34}$, while the second coupling  breaks the symmetries $SU(2)_{23}$ and $SU(2)_{34}$ keeping $SU(2)_{12}$. In particular, from the second coupling we can read that the $SU(2)_2$ gauge group is enhanced to $U(2)$. This theory is represented in the following figure
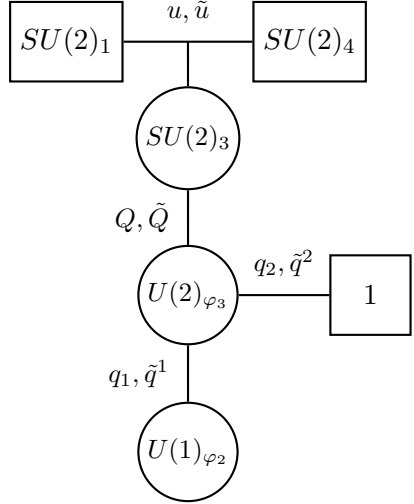
\vspace*{0.7cm}
\begin{center}
\begin{tikzpicture}[thick, scale=0.8]
 \node[draw, rectangle, minimum width=30pt, minimum height=30pt](L1) at (15,-4.2){$1$};
  \node[draw, rectangle, minimum width=30pt, minimum height=30pt](L2) at (10,0){$SU(2)_1$};
  \node[draw, rectangle, minimum width=30pt, minimum height=30pt](L3) at (14,0){$SU(2)_4$};
 \node[circle, draw, inner sep= 2pt](L4) at (12,-1.6){{\small $SU(2)_3$}};
  \node[circle, draw, inner sep= 2pt](L5) at (12,-6.8){{\small $U(1)_{\varphi_2}$}};
  \node[circle, draw, inner sep= 2pt](L6) at (12,-4.2){{\small $U(2)_{\varphi_3}$}};
 \path[every node/.style={font=\sffamily\small,
  		fill=white,inner sep=1pt}]
(L1) edge node[above=2mm] {$q_2,\tilde q^2$}  (L6)
(L5) edge node[left=2mm] {$q_1,\tilde q^1$}  (L6)
(L2) edge node[above=2mm] {$u,\tilde u$}  (L3)
(L6) edge node[left=2mm] {$Q,\tilde Q$} (L4)
(L4) edge (12,0)
;
\end{tikzpicture}
\end{center}
\vspace*{0.7cm}

In  {\it Theory 3} we should repeat the same procedure, but now we should take into account that the flavor symmetries are in different positions. The superpotential that reproduces this is
\begin{equation}
  W \supset \varphi_2 (p_1 \tilde{p}^1 + p_2 \tilde p^2 + P\tilde P)  + \varphi_3 p_1 \tilde{p}^1\:.
\end{equation}
Again, the first coupling breaks the symmetries $SU(2)_{12}$ and $SU(2)_{23}$ keeping $SU(2)_{34}$, while the second coupling  breaks the symmetries $SU(2)_{23}$ and $SU(2)_{34}$ keeping $SU(2)_{12}$. In particular, now it is the first coupling that shows the enhancement from $SU(2)_2$ to $U(2)_2$. This theory is represented in the following figure
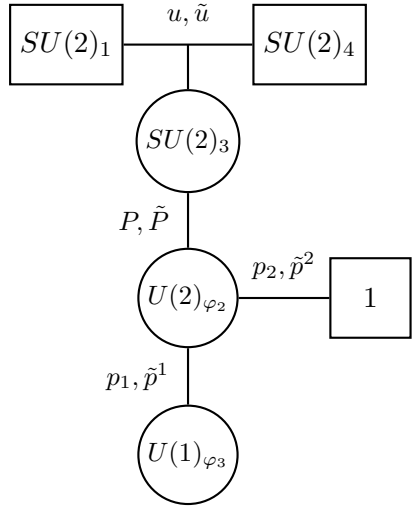
\begin{center}
\begin{tikzpicture}[thick, scale=0.8]
 \node[draw, rectangle, minimum width=30pt, minimum height=30pt](L1) at (15,-4.2){$1$};
  \node[draw, rectangle, minimum width=30pt, minimum height=30pt](L2) at (10,0){$SU(2)_1$};
  \node[draw, rectangle, minimum width=30pt, minimum height=30pt](L3) at (14,0){$SU(2)_4$};
 \node[circle, draw, inner sep= 2pt](L4) at (12,-1.6){{\small $SU(2)_3$}};
  \node[circle, draw, inner sep= 2pt](L5) at (12,-6.8){{\small $U(1)_{\varphi_3}$}};
  \node[circle, draw, inner sep= 2pt](L6) at (12,-4.2){{\small $U(2)_{\varphi_2}$}};
 \path[every node/.style={font=\sffamily\small,
  		fill=white,inner sep=1pt}]
(L1) edge node[above=2mm] {$p_2,\tilde p^2$}  (L6)
(L5) edge node[left=2mm] {$p_1,\tilde p^1$}  (L6)
(L2) edge node[above=2mm] {$u,\tilde u$}  (L3)
(L6) edge node[left=2mm] {$P,\tilde P$} (L4)
(L4) edge (12,0)
;
\end{tikzpicture}
\end{center}

Analogously with the previous section, {\it Theory 2} and {\it Theory 3} are the same theory. In {\it Theory 2} the topological symmetry corresponding to the $U(1)_{\varphi_2}$ node is mapped to the topological symmetry of the $U(2)_2$ node of {\it Theory 1}, while in {\it Theory 3} the topological symmetry corresponding to the $U(1)_{\varphi_3}$ node is mapped to the topological symmetry of the $U(2)_3$ node of {\it Theory 1}.

Hence deforming the {\it Theory 2$=$3} by switching on the monopole operator relative to the $U(1)$ node is equivalent to switching on the monopole relative to either $U(2)_2$ or $U(2)_3$ in {\it Theory 1}: hence the two are equivalent.

\subsection{E-type quivers}

As in the $D_N$ case, the Coulomb branch global symmetry of the E-type quiver implies that by turning on the monopole deformation 
at two different gauge nodes we get equivalent theories, although from the quiver description of the theory this fact is not 
obvious\footnote{The statement refers to the minimal nilpotent orbit only. Equivalently, we turn on the monopole superpotential 
deformation at a single gauge node.}. The purpose of the present section is to shed light on this duality. 

\begin{figure}[ht!]
\begin{center}
\begin{tikzpicture}[->, thick,state/.style={circle, draw, minimum size=29pt}]
\node[state] (1) at (0,0) {$1$};
\node[state] (2) [right= .6cm of 1] {$2$};
\node[state] (3) [right= .6cm of 2] {$J$};
\node[state] (4) [right= .6cm of 3] {\tiny{$J+2$}};
\node[state] (5) [right= .6cm of 4] {\tiny{$N-2$}};
\node[state] (6) [right= .6cm of 5] {$N$}; 
\node[shape=rectangle, draw, minimum size=25pt] (7) [above= .6cm of 6] {$3$}; 
\node[state] (8) [right= .6cm of 6] {\tiny{$N-1$}}; 
\node[state] (9) [right= .6cm of 8] {$1$};

\draw[-] (1) -- (2);
\draw[-][dotted] (2) -- (3);
\draw[-] (3) -- (4);
\draw[-][dotted] (4) -- (5);
\draw[-] (5) -- (6); 
\draw[-] (6) -- (7); 
\draw[-] (6) -- (8); 
\draw[-][dotted] (8) -- (9);
;

\end{tikzpicture}
\caption{The mirror dual to $\boxed{N-2k}-U(N-k)-U(N)-\boxed{N+k}$.}
\label{mirrorHW}
\end{center} 
\end{figure}
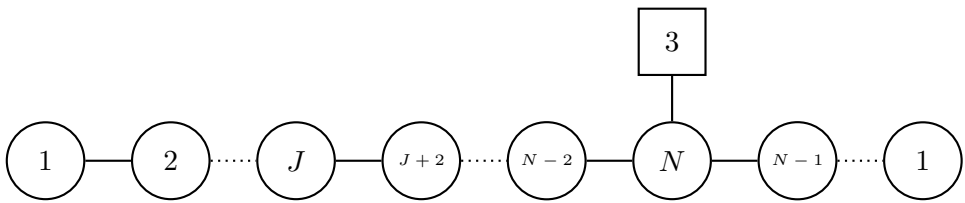

Our strategy is the following: we restrict to a linear tail with only balanced (in the sense of \cite{Gaiotto:2008ak}) unitary 
gauge groups and argue that turning on the monopole deformation at neighboring gauge groups leads to equivalent theories. By repeatedly applying this duality, we get to the desired conclusion. Without loss of generality, we can focus on a subquiver with two gauge nodes. This is always of the form 
$$\boxed{N-2k}-U(N-k)-U(N)-\boxed{N+k}$$
The mirror theory can be easily read out from the Hanany-Witten brane setup \cite{Hanany:1996ie} and is the linear quiver in Figure \ref{mirrorHW},

\noindent where $J=N-2k$. The topological abelian symmetry enhances to $SU(N+k)\times SU(N-2k)\times U(1)$ (the $U(1)$ factor is the topological symmetry of the $U(J)$ node, which is not balanced). The three fundamentals of the central node give an extra $SU(3)$ global symmetry, matching the Coulomb branch symmetry of the original theory. 

\begin{figure}[ht!]
\begin{center}
\begin{tikzpicture}[->, thick, state/.style={circle, draw, minimum size=29pt}]
\node[state] (1) at (0,0) {$1$};
\node[state] (2) [right= .6cm of 1] {$2$};
\node[state] (3) [right= .6cm of 2] {$J$};
\node[state] (4) [right= .6cm of 3] {\tiny{$J+2$}};
\node[state] (5) [right= .6cm of 4] {\tiny{$N-2$}};
\node[state] (6) [right= .6cm of 5] {$N$}; 
\node[shape=rectangle, draw, minimum size=24pt] (7) [above= .7cm of 6] {$1$}; 
\node[state] (8) [right= .6cm of 6] {\tiny{$N-1$}}; 
\node[state] (9) [right= .6cm of 8] {$1$};
\node[state] (10) [above= .4cm of 5] {$1$};
\node[state] (11) [above= .4cm of 8] {$1$};

\draw[-] (1) -- (2);
\draw[-][dotted] (2) -- (3);
\draw[-] (3) -- (4);
\draw[-][dotted] (4) -- (5);
\draw[-] (5) -- (6); 
\draw[-] (6) -- (7); 
\draw[-] (6) -- (8); 
\draw[-][dotted] (8) -- (9); 
\draw[-] (6) -- (10);
\draw[-] (6) -- (11);
;

\end{tikzpicture}
\caption{The mirror dual of $\boxed{N-2k}-SU(N-k)-SU(N)-\boxed{N+k}$.}
\label{starshaped}
\end{center}
\end{figure}
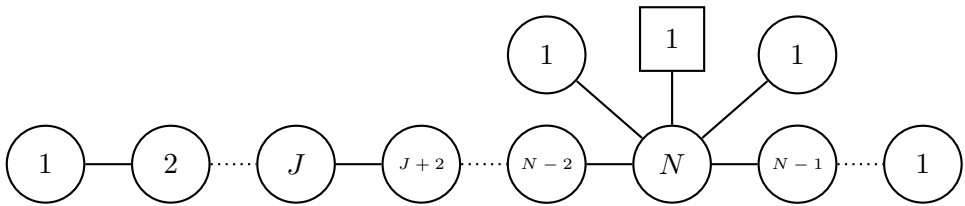

If we instead consider the model $$\boxed{N-2k}-SU(N-k)-SU(N)-\boxed{N+k}$$ the gauge nodes are not balanced anymore and there is 
no enhancement of the global symmetry due to monopole operators. The mirror is known to be the star-shaped quiver \cite{Benini:2010uu} in Figure \ref{starshaped}.

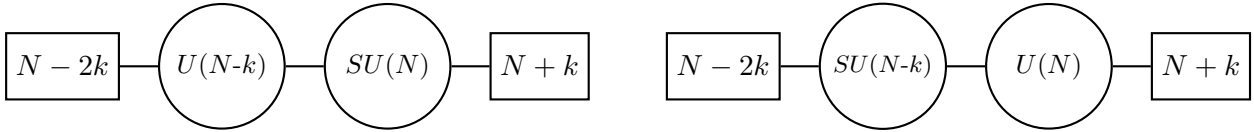
\begin{figure}[ht!]
\begin{center}
\begin{tikzpicture}[->, thick, state/.style={circle, draw, minimum size=47pt}]
\node[shape=rectangle, draw, minimum height=25pt] (1) at (0,0) {$N-2k$};
\node[state, shape=circle, draw] (2) [right= .5cm of 1] {\small{$U(N$-$k)$}};
\node[state, shape=circle, draw] (3) [right= .5cm of 2] {\small{$SU(N)$}};
\node[shape=rectangle, draw, minimum height=25pt] (4) [right= .5cm of 3] {$N+k$};
\node[shape=rectangle, draw, minimum height=25pt] (5) [right= 1cm of 4] {$N-2k$}; 
\node[state, shape=circle, draw] (6) [right= .5cm of 5] {\footnotesize{$SU(N$-$k)$}}; 
\node[state, shape=circle, draw] (7) [right= .5cm of 6] {{\small $U(N)$}}; 
\node[shape=rectangle, draw, minimum height=25pt] (8) [right= .5cm of 7] {$N+k$};

\draw[-] (1) -- (2);
\draw[-] (2) -- (3);
\draw[-] (3) -- (4);
\draw[-] (5) -- (6); 
\draw[-] (6) -- (7); 
\draw[-] (7) -- (8); 
;

\end{tikzpicture} 
\caption{The two linear quivers with a single balanced gauge node.}
\label{duality1}
\end{center} 
\end{figure}

The difference with respect to  the linear quiver discussed above is that the Cartan subgroup of $SU(3)$ is now gauged. We can ungauge one of the abelian factors simply by gauging the corresponding topological $U(1)$ symmetry, which amounts on the mirror side to gauging one of the two independent ``baryonic'' $U(1)$ symmetries acting on the matter fields. Depending on which $U(1)$ subgroup we gauge, we end up with one of the two models in Figure \ref{duality1}.

\begin{figure}[ht!]
\begin{center}
\begin{tikzpicture}[->, thick, state/.style={circle, draw, minimum size=29pt}]
\node[state, shape=circle, draw] (1) at (0,0) {$1$};
\node[state, shape=circle, draw] (2) [right= .6cm of 1] {$2$};
\node[state, shape=circle, draw] (3) [right= .6cm of 2] {$J$};
\node[state, shape=circle, draw] (4) [right= .6cm of 3] {\tiny{$J+2$}};
\node[state, shape=circle, draw] (5) [right= .6cm of 4] {\tiny{$N-2$}};
\node[state, shape=circle, draw] (6) [right= .6cm of 5] {$N$}; 
\node[shape=rectangle, draw, minimum size=24pt] (7) [above= .5cm of 6] {$2$}; 
\node[state, shape=circle, draw] (8) [right= .6cm of 6] {\tiny{$N-1$}}; 
\node[state, shape=circle, draw] (9) [right= .6cm of 8] {$1$};
\node[state, shape=circle, draw] (10) [below= .5cm of 6] {$1$};

\draw[-] (1) -- (2);
\draw[-][dotted] (2) -- (3);
\draw[-] (3) -- (4);
\draw[-][dotted] (4) -- (5);
\draw[-] (5) -- (6); 
\draw[-] (6) -- (7); 
\draw[-] (6) -- (8); 
\draw[-][dotted] (8) -- (9); 
\draw[-] (6) -- (10);
;

\end{tikzpicture}
\caption{The common mirror of the quivers in Figure \ref{duality1}.}
\label{quivernn}
\end{center} 
\end{figure}
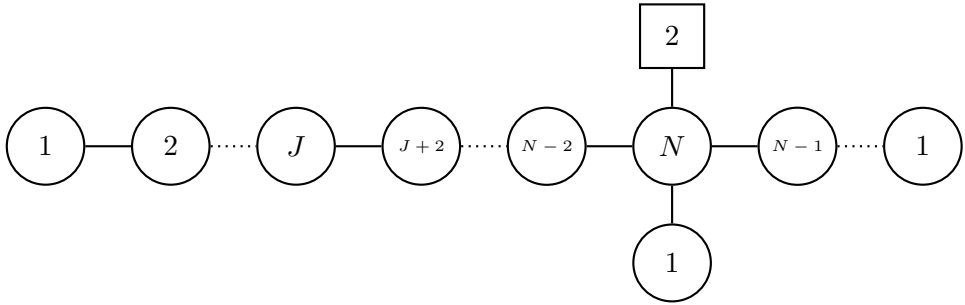

These two linear quivers have a single monopole operator of R-charge one and are actually equivalent theories (so in particular the R-charge one monopole operators are mapped to each other under this ``duality''). The easiest way to see the equivalence is perhaps to notice that they are both mirror dual to the theory in Figure~\ref{quivernn}.

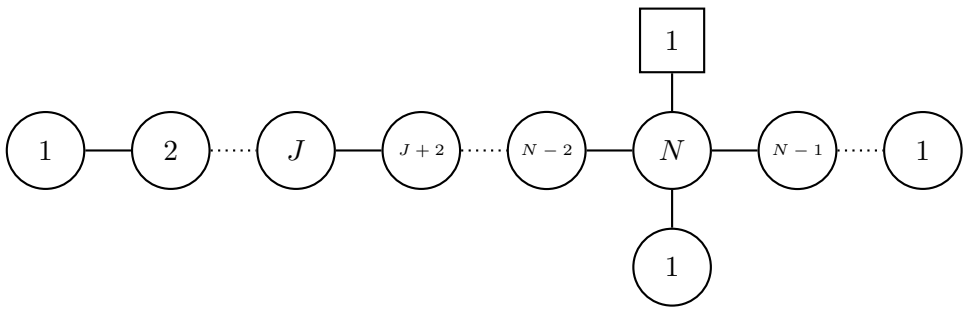
\begin{figure}[ht!]
\begin{center}
\begin{tikzpicture}[->, thick, state/.style={circle, draw, minimum size=29pt}]
\node[state, shape=circle, draw] (1) at (0,0) {$1$};
\node[state, shape=circle, draw] (2) [right= .6cm of 1] {$2$};
\node[state, shape=circle, draw] (3) [right= .6cm of 2] {$J$};
\node[state, shape=circle, draw] (4) [right= .6cm of 3] {\tiny{$J+2$}};
\node[state, shape=circle, draw] (5) [right= .6cm of 4] {\tiny{$N-2$}};
\node[state, shape=circle, draw] (6) [right= .6cm of 5] {$N$}; 
\node[shape=rectangle, draw, minimum size=24pt] (7) [above= .5cm of 6] {$1$}; 
\node[state, shape=circle, draw] (8) [right= .6cm of 6] {\tiny{$N-1$}}; 
\node[state, shape=circle, draw] (9) [right= .6cm of 8] {$1$};
\node[state, shape=circle, draw] (10) [below= .5cm of 6] {$1$};

\draw[-] (1) -- (2);
\draw[-][dotted] (2) -- (3);
\draw[-] (3) -- (4);
\draw[-][dotted] (4) -- (5);
\draw[-] (5) -- (6); 
\draw[-] (6) -- (7); 
\draw[-] (6) -- (8); 
\draw[-][dotted] (8) -- (9); 
\draw[-] (6) -- (10);
;

\end{tikzpicture}
\caption{The mirror of $SU(N-k)$ $N_f=2N-2k$ SQCD plus a decoupled $SU(N)\times SU(k)$ bifundamental.}
\label{mirrornew}
\end{center} 
\end{figure}

\noindent This fact can be argued as follows: let us consider the star-shaped quiver in Figure~\ref{mirrornew}.
\noindent According to \cite{Benini:2010uu}, this is the mirror dual of the (dimensional reduction of) $A_{N-1}$ class $\mathcal{S}$ theory labelled by a sphere with one full puncture, two minimal and one with partition $(2^k,1^{N-2k})$. This theory is known to be $SU(N-k)$ SQCD with $2N-2k$ flavors plus a decoupled free sector describing a hypermultiplet in the bifundamental of $SU(N)\times SU(k)$. This decoupled sector can be recovered directly in the star-shaped quiver by applying the analysis of \cite{Gaiotto:2008ak}: the central node is unbalanced ($U(N)$ with $2N-1$ flavors), it has  a sequence of $k-1$ balanced nodes on its left and a sequence of $N-1$ balanced nodes on its right. In such a situation we get $Nk$ monopole operators with R-charge 1/2. 

\begin{figure}[ht!]
\begin{center}
\begin{tikzpicture}[->, thick, state/.style={circle, draw, minimum size=29pt}]
\node[state, shape=circle, draw] (1) at (0,0) {$1$};
\node[state, shape=circle, draw] (2) [right= .6cm of 1] {$2$};
\node[state, shape=circle, draw] (3) [right= .6cm of 2] {$J$};
\node[state, shape=circle, draw] (4) [right= .6cm of 3] {\tiny{$J+2$}};
\node[state, shape=circle, draw] (5) [right= .6cm of 4] {\tiny{$N-2$}};
\node[state, shape=circle, draw] (6) [right= .6cm of 5] {$N$}; 
\node[shape=rectangle, draw, minimum size=24pt] (7) [above= .5cm of 6] {$2$}; 
\node[state, shape=circle, draw] (8) [right= .6cm of 6] {\tiny{$N-1$}}; 
\node[state, shape=circle, draw] (9) [right= .6cm of 8] {$1$};

\draw[-] (1) -- (2);
\draw[-][dotted] (2) -- (3);
\draw[-] (3) -- (4);
\draw[-][dotted] (4) -- (5);
\draw[-] (5) -- (6); 
\draw[-] (6) -- (7); 
\draw[-] (6) -- (8); 
\draw[-][dotted] (8) -- (9); 
;

\end{tikzpicture}
\caption{The mirror of $U(N-k)$ $N_f=2N-2k$ SQCD plus a decoupled $SU(N)\times SU(k)$ bifundamental.}
\label{mirror2}
\end{center} 
\end{figure}
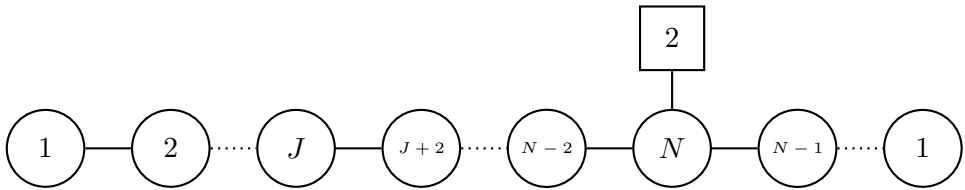

If we wish to gauge the baryon number and study $U(N-k)$ instead of $SU(N-k)$ SQCD, on the mirror side we simply have to ungauge the $U(1)$ node connected to the central node, ending up with the linear quiver in Figure~\ref{mirror2}.

\noindent In this way we get two flavors in the fundamental of the central $U(N)$ node, which is consistent with the enhancement of the topological symmetry of $U(N-k)$ SQCD to $SU(2)$.

We now recover the theory 
$$\boxed{N-2k}-SU(N-k)-SU(N)-\boxed{N+k}$$
from $SU(N-k)$ $N_f=2N-2k$ SQCD plus a decoupled $SU(N)\times SU(k)$ bifundamental by gauging the $SU(N)$ global symmetry and adding $N$ flavors of $SU(N)$. At the level of the mirror star-shaped quiver (in 
Figure \ref{mirrornew}), this operation is easy to describe since we just need to introduce a $U(1)$ node connected to the central node. This in fact leads to the quiver In Figure~\ref{starshaped}. Again, if we want instead to perform a $U(N)$ gauging as opposed to $SU(N)$, we should add in the star-shaped quiver \ref{mirrornew} a flavor in the fundamental of the central $U(N)$ node. As a consistency check for this claim, notice in fact that, starting from the mirror dual of $U(N-k)$ SQCD (see Figure \ref{mirror2}) and adding a flavor to the central node, we get precisely the mirror dual of the $U(N-k)\times U(N)$ gauge theory (Figure \ref{mirrorHW}). 

By applying the above rules, we can check that the two theories in Figure \ref{duality1} have the same mirror: this follows simply by performing an $SU(N)$ gauging of the $U(N-k)$ theory (Figure \ref{mirror2}) or a $U(N)$ gauging of the $SU(N-k)$ theory (Figure \ref{mirrornew}). For $N=2$ and $k=1$ this duality reduces to that we exploited in the study of monopole deformations of $U(2)$ SQCD with four flavors.

Once we have established this fact, the desired result follows easily: an off-diagonal mass term for the two $U(N)$ 
fundamentals in the star-shaped quiver \ref{quivernn} is mapped on the mirror side (both models in Figure \ref{duality1}) to a superpotential deformation involving the monopole operator of R-charge one. If we now gauge the leftover ``baryonic'' $U(1)$ symmetry group in the theory in  Figure~\ref{duality1}, we obtain precisely the duality we were looking for.

\section{$S^3_b$ partition function and monopole operators} \label{Sec:squasheds3}
In the previous sections we exploited the 3d duality ``induced'' by S-duality in four dimensions to understand the effect of a monopole superpotential at a non-abelian node in the quiver.  
Here we want to check the duality at the level of the partition function on the three-sphere.

\subsection{Basic $\mathcal{N}=4$ mirror symmetry and the pentagon identity} \label{Sec:sqed1}
In this subsection, we review the building block of $\mathcal{N}=4$ mirror symmetry in the context of the partition function on the squashed sphere. At the level of the latter, it will reduce to what is known as the \emph{pentagon identity}. This has been discussed in various works in the literature, but it is useful to repeat it here for convenience, and in order to setup our notation.

The basic duality of interest is the following:
\be
SQED + N_f=1,\, W_{\cN=4} = \phi q \tilde q \quad \longleftrightarrow \quad (\cX, \cY, \cZ, \cS),\, W_{XYZ'} = \cX \cY \cZ+ \cS \cX \nonumber
\ee
On the LHS, in \emph{theory A}, we have a charged hyper and its cubic coupling to the complex scalar in the vector multiplet. 

On the RHS, \emph{theory B}, we have four neutral chirals, such that only the hyper $(\cY, \cZ)$ survives in the IR. It might seem like overkill to keep all four chirals in this theory, given that two are massive. However, the real case of interest, to be discussed in Section \ref{Sec:montsu2}, will be a theory with exactly the same field content, but a modified superpotential. It is therefore instructive to keep this information here.

Let us examine the  symmetries of the problem:

Theory $A$ has a $U(1)_G$ gauge symmetry, and a $U(1)_H \times U(1)_C \subset SU(2)_H \times SU(2)_C$ symmetry, such that $U(1)_{H+C}$ is an R-symmetry, and $U(1)_{H-C}$ is an axial symmetry. Finally, there is a topological $U(1)_T$ symmetry that shifts the dual photon, and therefore acts as a phase on the monopole operators $W_\pm$. Table \ref{tab:chsqedn1} summarizes the charges.

\begin{table} [h!]
$$
\renewcommand{\arraystretch}{1.5}
\begin{array}{c|c|c|c|c}
& U(1)_G &  U(1)_{H+C} & U(1)_{H-C} & U(1)_T \\ \hline
\phi & 0  & 1 & -1 & 0 \\ \hline
q & 1  & 1/2 & 1/2 & 0 \\ \hline
\tilde q & -1  & 1/2 & 1/2 & 0  \\ \hline
W_\pm & 0 & 1/2 & -1/2 & \pm 1 \\ \hline
W_{\cN=4} & 0 & 2 & 0 & 0
\end{array}
$$
\caption{Gauge and global symmetry charges for $\cN=4$ SQED with one flavor.} \label{tab:chsqedn1}
\end{table}

Theory $B$ similarly has a $U(1)_{\hat H} \times U(1)_{\hat C} \subset SU(2)_{\hat H} \times SU(2)_{\hat C}$. Instead of a topological symmetry, however, it has a $U(1)_F$ flavor symmetry acting on the pair $(\cY, \cZ)$. The charges are summarized in Table \ref{tab:chn4xyz}.
\begin{table} [h!]
$$
\renewcommand{\arraystretch}{1.5}
\begin{array}{c|c|c|c}
 &  U(1)_{\hat H+\hat C} & U(1)_{\hat H-\hat C} & U(1)_F \\ \hline

\cX  & 1 & -1 & 0 \\ \hline
\cY  & 1 /2& 1/2 &  1 \\ \hline
\cZ  & 1/2 & 1/2 & - 1 \\ \hline
\cS   & 1 & 1 & 0 \\ \hline
W_{XYZ'}  & 2 & 0 & 0
\end{array}
$$
\caption{Global symmetry charges for the $\cN=4$ SQED version of the XYZ-model.} \label{tab:chn4xyz}
\end{table}
From this table, we see that, under the mirror map
\be
W_+ \leftrightarrow \cY \,, W_- \leftrightarrow \cZ \,, q \tilde q \leftrightarrow \cX \,, \Phi \leftrightarrow \cS
\ee
the global symmetries are identified as follows:
\be
U(1)_H \leftrightarrow U(1)_{\hat C}\,, U(1)_C \leftrightarrow U(1)_{\hat H}\,, U(1)_T \leftrightarrow U(1)_{F}
\ee

Let us now see how this correspondence works at the level of the partition function on the squashed sphere. We define the squashed sphere of radius $1$ as the following hypersurface:
\be
b^2 |z_1|^2+ b^{-2} |z_2|^2 = 1 \subset \mathbb{C}^2\,,
\ee
where $b$ is a real number such that for $b=1$ we have the round three-sphere. A useful quantity to define is $Q = b+1/b$. In order to compute it, we need to define the putative R-charge for the IR theory.
In $\cN=2$ language, the R-charge is in general a linear combination of U(1) symmetries $R_{IR} = \sum_{i=0} c^i R_i$, with $c^0=1$, and $R_0 = H+C$, the Cartan of the original $\cN=4$ $SU(2)_H \times SU(2)_C$ R-symmetry.

A twisted real mass is a vev for the scalar component of a background vector multiplet associated to a global $U(1)$ symmetry.
Given a chiral multiplet $\chi$ with charge $\mathfrak{q}^i$ under the $i$-th $U(1)$ symmetry $R_i$, its real twisted mass will be $\mathfrak{q}_i m^i$, where $m_i$ is the background scalar vev. This $U(1)$ will contribute $\mathfrak{q}_i c^i$ to the R-charge of $\chi$. 

Then, the contribution to the partition function of a chiral multiplet with charges $\{\mathfrak{q}_i\}$ under the various $U(1)'s$ will be the following
\be
Z_\chi = s_b\left(i \tfrac{Q}{2}-\tilde m_\chi\right)\:,
\ee
where we defined the total complex twisted mass of $\chi$ as follows 
\be \tilde m_\chi = \sum_i \mathfrak{q}_i (m^i+i \tfrac{Q}{2} c^i)\,,
\ee
with $m^0=0$, and $s_b$ is the \emph{double sine} function
\be
s_b(x) = \prod_{m, n \in \mathbb{Z} \geq 0} \frac{m b+ n b^{-1} + \tfrac{Q}{2}- ix}{m b+ n b^{-1} + \tfrac{Q}{2}+ ix} \:.
\ee

 This prompts us to define the individual complex parameters
\be
\tilde m^i \equiv m^i+i\tfrac{Q}{2} c^i\,.
\ee
The reason these are useful is the discovery by Jafferis \cite{Jafferis}, that the partition function $Z(m^i, c^i)$ depends only on the holomorphic combinations $Z(\tilde m^i)$.

Let us begin with theory $A$, and assume that the IR R-symmetry is 
\be
R_\alpha \equiv H+C + \alpha(H-C)\,.
\ee
So we only allow mixing with the axial symmetry. In that case, we will have the following complex twisted masses:
\be
\tilde m_0 = i \tfrac{Q}{2}\,, \quad \tilde m_A = m_A+i \tfrac{Q}{2}\,, \quad \tilde m_T = -\xi\,,
\ee
whereby $A \equiv H-C$. Here, we have introduce a Fayet-Iliopoulos constant $\xi$, which can be regarded as a twisted mass for the topological $U(1)_T$, as explained in \cite{Aharony:1997bx}. 
The $\tilde m_0$ is real because we are not switching on a twisted mass for the R-symmetry. Our fields now have the following R-charges: 
\[
\begin{array}{c|ccc}
 &q & \tilde q & \phi \\ \hline
R_\alpha & \frac{1+\alpha}{2} & \frac{1+\alpha}{2} & 1-\alpha
\end{array}
\]
Let us write down the contributions from the various fields to the partition function:
$$
\begin{array}{c|ccc}
& \tilde m && s_b\left(i\tfrac{Q}{2}-\tilde{m}\right) \\ \hline
q & i \tfrac{Q}{4}+\tfrac{1}{2}\tilde m_A && s_b\left(i \tfrac{Q}{4}-u-\tfrac{1}{2} \tilde m_A\right) \\ 
\tilde q & i \tfrac{Q}{4}+\tfrac{1}{2}\tilde m_A && s_b\left(i \tfrac{Q}{4}+u-\tfrac{1}{2} \tilde m_A\right) \\ 
\phi & i \tfrac{Q}{2}-\tilde m_A&& s_b(\tilde m_A)
\end{array}
$$
Here, $u$ corresponds to a vev for the scalar in the gauge multiplet. Now we can write down the partition function for $\cN=4$ SQED with one flavor:
\be
Z_{\rm SQED_1 \cN=4}(\tilde m_A, \xi) = s_b(\tilde m_A) \int_{-\infty}^{\infty} e^{-2 \pi i \xi} s_b\left(i \tfrac{Q}{4}-u-\tfrac{1}{2} \tilde m_A\right)s_b\left(i \tfrac{Q}{4}+u-\tfrac{1}{2} \tilde m_A\right)
\ee
This partition function depends on the parameter $\alpha$, which defines the mixing of the UV R-symmetry with the axial symmetry through the relation $\tilde m_A = m_A + i \alpha \tfrac{Q}{2}$.

We now analyze theory B, the $\cN=4$ version of the XYZ-model. In that case, under the mirror map, the putative IR R-symmetry is given by 
\be
R_\alpha = \hat H + \hat C- \alpha (\hat H - \hat C)
\ee
The twisted complex masses are the following:
\be \tilde m_0 = i \tfrac{Q}{2},\qquad \tilde m_{\hat A} = m_{\hat A} - i \alpha \tfrac{Q}{2},\qquad \tilde m_F = m_F\,.
\ee
Here, we assume that the $U(1)_F$ will not mix with the IR R-symmetry, hence $\tilde m_F$ is real.
Our fields now have the following R-charges: 
\[
\begin{array}{c|cccc}
 &\cX & \cY & \cZ & \cS \\ \hline
R_\alpha & 1+\alpha & \frac{1-\alpha}{2} & \frac{1-\alpha}{2} & 1-\alpha
\end{array}
\]
the partition function is the following:
\begin{eqnarray}
Z_{XYZ'}(m_{\hat A}, m_F) &=& s_b(\tilde m_A) s_b\left(i \tfrac{Q}{4}-\tfrac{1}{2} \tilde m_{\hat A}-\tilde m_F\right) s_b\left(i \tfrac{Q}{4}-\tfrac{1}{2} \tilde m_{\hat A}+\tilde m_F\right) s_b(-\tilde m_A) \nonumber \\
&& =s_b\left(i \tfrac{Q}{4}-\tfrac{1}{2} \tilde m_{\hat A}-\tilde m_F\right) s_b\left(i \tfrac{Q}{4}-\tfrac{1}{2} \tilde m_{\hat A}+\tilde m_F\right)\,.
\end{eqnarray}
The last equality follows from the fact that $s_b(-x) = 1/s_b(x)$ as can be seen from the definition of the double-sine function. This makes sense, it corresponds to the two massive chirals dropping out of the theory in the IR.

We now want to confront these two partition functions. Under the mirror map, we expect $\tilde m_A \leftrightarrow -\tilde m_A$, and $\xi \leftrightarrow m_F$. So we should have the following equality:
\be
Z_{\rm SQED_1, \cN=4}(\tilde m_A, \xi) = Z_{XYZ'}(-\tilde m_A, \xi)
\ee
In other words, we expect the following to hold:
\be
s_b(\tilde m_A) \int_{-\infty}^{\infty} e^{-2 \pi i \xi} s_b\left(i \tfrac{Q}{4}-u-\tfrac{1}{2} \tilde m_A\right)s_b\left(i \tfrac{Q}{4}+u-\tfrac{1}{2} \tilde m_A\right) = s_b\left(i \tfrac{Q}{4}+\tfrac{1}{2} \tilde m_{A}-\xi\right) s_b\left(i \tfrac{Q}{4}+\tfrac{1}{2} \tilde m_{A}+\xi\right)
\ee
It so happens that this identity is known to hold for the double-sine function, and goes by the name of \emph{pentagon identity} \cite{Benvenuti:2011ga,Dimofte:2011ju,Benvenuti:2016wet}. We will record it here in generic notation for future convenience:
\be \label{pentagon}
s_b(y+z) s_b(y-z) = s_b\left(2 y- i\tfrac{Q}{2}\right)\int d x e^{-2 \pi i z x} s_b\left(i \tfrac{Q}{2}+x-y\right)s_b\left(i \tfrac{Q}{2}-x-y\right)\,.
\ee

\subsection{$\cN=4$ self-mirror symmetry of $T(SU(2))$} \label{dualqed}

In this section, we will study $\cN=4$ SQED with two flavors, also known as $T(SU(2))$ (see also Section \ref{Sec:U1monop}). The theory is known to be self-mirror. We will show this at the level of the partition function. This, of course, has been done before in the literature (see e.g. \cite{Benvenuti:2011ga,Dimofte:2011ju,Benvenuti:2016wet}), but with the assumption that the IR R-symmetry is fixed by the $\cN=4$ supersymmetry. We will rederive this mirror symmetry, but with unknown IR R-symmetry. Once this is done, we will be able to easily add our monopole deformation.

Let us first define the theory: It has one $\cN=4$ vector multiplet with complex scalar $\phi$, and two hypers $(q_i, \tilde q^i)$, with $i=1,2$, and superpotential $W_{\cN=4} = \phi \sum_i q_i \tilde q^i$.

The global symmetry is $SU(2)_H \times SU(2)_C \times SU(2)_F \times U(1)_T$. In $\cN=2$ language, the R-symmetry is given by
\be
R_{\alpha,\beta_T} = (H+C)+\alpha (H-C)+\beta_T T\,.
\ee 
Here, we have assumed again that the $U(1)_F \subset SU(2)_F$ will not mix with the R-symmetry. However, we will allow for the topological symmetry to mix via the parameter $\beta_T$.
Let us record the charges of all fields under the Cartan subalgebras:
$$\renewcommand{\arraystretch}{1.5}
\begin{array}{c|c|c|c|c|c}
& U(1)_{H+C} & U(1)_{H-C} & U(1)_F & U(1)_T & R_{\alpha,\beta_T}\\ \hline
\Phi & 1 & -1 & 0 & 0 & 1-\alpha_A \\ \hline
q_i & 1/2 & 1/2 & \pm 1/2 & 0 & (1+\alpha)/2\\ \hline
\tilde q^i & 1/2 & 1/2 & \mp 1/2 & 0 & (1+\alpha)/2 \\ \hline
W_\pm & 1 & -1 & 0 & \pm 1 & 1-\alpha \pm \beta_T \\ \hline
W_{\cN=4} & 2 & 0 & 0 & 0 & 2
\end{array}
$$
Note, that $U(1)_F$ acts on $q_2$ with an opposite phase to $q_1$, which explains the $\pm$ symbol, and similarly with $\tilde q^i$.

From this we can define the following twisted complex masses:
\be
\tilde m_0 = i\tfrac{Q}{2},\,\,\,\,\, \tilde m_A = m_A+i \alpha \tfrac{Q}{2}, \,\,\,\,\, \tilde m_F = m_F, \,\,\,\,\, \tilde m_T = \xi+i \beta_T \tfrac{Q}{2}\:.
\ee
Recall that a FI parameter can be regarded as a twisted mass for $U(1)_T$. Let us write down the contributions of the various fields to the partition function:
\[
\renewcommand{\arraystretch}{1.5}
\begin{array}{c|ccc}
& \tilde m && s_b\left(i\tfrac{Q}{2}-\tilde{m}\right) \\ \hline
q_i & i \tfrac{Q}{4}+\tfrac{1}{2}(\tilde m_A\pm \tilde m_F) && s_b\left(i\tfrac{Q}{4}-u-\tfrac{1}{2}(\tilde m_A\pm \tilde m_F) \right)\\ 
\tilde q^i & i \tfrac{Q}{4}  +\tfrac{1}{2}(\tilde m_A\mp \tilde m_F) && s_b \left(i\tfrac{Q}{4}+u-\tfrac{1}{2}(\tilde m_A\mp \tilde m_F)\right)\\ 
\Phi & i \tfrac{Q}{2} - \tilde m_A && s_b (\tilde m_A)\\ 
\end{array}
\]

Let us now write down the partition function of $\mathcal{N}=4, d=3$ SQED with $N_f=2$:
\begin{eqnarray}
Z_{TSU(2)}(\tilde m_A, m_F, \tilde m_T) &=& s_b (\tilde m_A) \int_{-\infty}^{\infty} d u \,e^{-2 \pi i u \tilde m_T} \\
&&s_b\left(i\tfrac{Q}{4}-u-\tfrac{1}{2}(\tilde m_F+\tilde m_A) \right) s_b\left(i\tfrac{Q}{4}-u-\tfrac{1}{2}(- \tilde m_F+\tilde m_A) \right)\nonumber \\
&&s_b \left(i\tfrac{Q}{4}+u-\tfrac{1}{2}(- \tilde m_F+\tilde m_A)\right) s_b \left(i\tfrac{Q}{4}+u-\tfrac{1}{2}( \tilde m_F+\tilde m_A)\right) \:.\nonumber
\end{eqnarray}

We begin by demonstrating the self mirror-duality of $T(SU(2))$. For this, we must choose $\beta_T=0$, which implies $\tilde m_T = \xi$.
In order to demonstrate that this theory is self-mirror, we will make repeated use of the pentagon identity \eqref{pentagon}. The idea here is to make two pairs out of the four chiral contributions, and apply the identity on them. Notice that we have three ways of pairing them. Let us use the natural hyper pairings $(q_i, \tilde q^i)$. This yields the following:
\begin{eqnarray}
&&Z_{TSU(2)}(\tilde m_A, m_F, \xi) = s_b (\tilde m_A) s_b^2(-\tilde m_A) \int du1, dx_1 dx_2 \,e^{-2\pi i u (\xi +x_1+x_2)-\pi i \tilde m_F (x_1- x_2) } \nonumber \\
&&\qquad\qquad s_b\left(i\tfrac{Q}{4}+\tfrac{1}{2}\tilde m_A+x_1\right) s_b\left(i\tfrac{Q}{4}+\tfrac{1}{2}\tilde m_A-x_1\right) s_b\left(i\tfrac{Q}{4}+\tfrac{1}{2}\tilde m_A+x_2\right) s_b\left(i\tfrac{Q}{4}+\tfrac{1}{2}\tilde m_A-x_2\right)\,. \nonumber
\end{eqnarray}
The double sine funciton satisfies $s_b(x) s_b(-x)=1$ by definition. So we can simplify the factors outside the integral. The integral in $\sigma$ acts as a delta function $\delta(x_1+x_2+\xi)$. After integrating over $x_2$ and redefining the remaining variable $x \equiv x_1+\xi/2$, we arrive at the following expression
\begin{eqnarray}
&&Z_{TSU(2)}(\tilde m_A,  m_F, \xi) = s_b(-\tilde m_A) \int dx \,e^{-2 \pi i m_F x} s_b\left(i\tfrac{Q}{4}+x+\tfrac{1}{2} (\tilde m_A-\xi)\right)  \\  
&&\qquad \qquad\qquad
s_b\left(i\tfrac{Q}{4}-x+\tfrac{1}{2} (\tilde m_A+\xi)\right) s_b\left(i\tfrac{Q}{4}-x+\tfrac{1}{2} (\tilde m_A-\xi)\right) s_b\left(i\tfrac{Q}{4}+x+\tfrac{1}{2}(\tilde m_A+\xi)\right)\,. \nonumber
\end{eqnarray}
From this, we read off the following identity
\be \label{partitionmirror}
Z_{TSU(2)}(\tilde m_A,  m_F, \xi) = Z_{T(SU(2))}(-\tilde m_A, \xi, m_F )
\ee
This is consistent with mirror symmetry, which exchanges $U(1)_H \leftrightarrow U(1)_C$, and therefore $U(1)_A \leftrightarrow U(1)_{-A}$, and exchanges the real mass parameter $m_F$ with the FI term. Note, that we have proven this identity without assuming any value for $\alpha$.

\subsection{$T(SU(2))$ deformed by a monopole operator} \label{Sec:montsu2}
Let us now deform the $\mathcal{N}=4$ superpotential with a monopole operator $\Delta W = W_+$. How should we implement this at the level of the partition function? We find two consequences to this deformation:
\begin{enumerate}
\item Now we need to change the R-charge assignments in order to make the superpotential still have charge 2. Given that $W_+$ has R-charge $1-\alpha+\beta_T$, this fixes the choice $\beta_T = 1+\alpha$. In other words, our putative R-symmetry now has only one free parameter:
\be
R_\alpha = (H+C)+\alpha (H-C)+(1+\alpha) T
\ee
Jafferis conjectured that the partition function should be holomorphic in all twisted masses, including those concerning topological symmetries \cite{Jafferis}. This is why we defined the complex twisted mass $\tilde m_T = \xi+i \beta_T \tfrac{Q}{2}$. So, allowing the topological symmetry to mix with the R-symmetry manifests itself as a complexified FI parameter in the partition function. 

\item The first point is necessary in order to allow for the superpotential deformation, but one could in principle have these R-charge assignments without it. If the deformation is turned on, however, one explicitly breaks the $U(1)_T$ symmetry. This means that we can no longer simply switch on a twisted mass for it. Only the diagonal subgroup of $U(1)_{H-C} \times U(1)_T$ survives. Concretely, this means we must set $m_A = \xi$.
\end{enumerate}
Implementing these two points, the partition function now reads:
\begin{eqnarray}\label{FSQEDMonEffTh}
&&Z_{TSU(2)}^{W_+}\left(\tilde m_A = \xi+i  \tfrac{Q}{2}\alpha, m_F, \xi\right) = s_b (\tilde m_A) \int_{-\infty}^{\infty} d \sigma e^{-2 \pi i \xi+ \pi Q (1+\alpha)}\\
&& \qquad\qquad s_b\left(i\tfrac{Q}{4} -\sigma-\tfrac{1}{2}(\tilde m_F+\tilde m_A) \right) s_b\left(i\tfrac{Q}{4}-\sigma-\tfrac{1}{2}(- \tilde m_F+\tilde m_A) \right)\nonumber \\
&&\qquad\qquad s_b \left(i\tfrac{Q}{4}+\sigma-\tfrac{1}{2}(- \tilde m_F+\tilde m_A)\right) s_b \left(i\tfrac{Q}{4}+\sigma-\tfrac{1}{2}( \tilde m_F+\tilde m_A) \right) \:. \nonumber
\end{eqnarray}
This can be regarded as an analytic continuation of the partition function, where we let $\xi \rightarrow \xi+i \tfrac{Q}{2} (1+\alpha)$. In other words, the partition function of the monopole deformed theory can be written as the partition function of the undeformed theory with the complexified FI parameter:
\be
Z_{TSU(2)}^{W_+}(\tilde m_A, m_F, \xi) = Z_{TSU(2)}\left(\tilde m_A, m_F, \xi+i \tfrac{Q}{2}(1+\alpha)\right)\,.
\ee
If this analytic continuation makes sense, then we expect the mirror relation \eqref{partitionmirror} to also hold:
\be \label{intermmirror1}
Z_{TSU(2)}\left(\tilde m_A, m_F, \xi+i \tfrac{Q}{2}(1+\alpha)\right) = Z_{TSU(2)}\left(-\tilde m_A, \xi+i \tfrac{Q}{2}(1+\alpha), m_F\right)\:.
\ee
Let us now compute this explicitly, imposing the condition $\tilde m_A = \xi+i \alpha \tfrac{Q}{2}$. After the dust settles, we find the following expression:
\begin{eqnarray}
&&Z_{TSU(2)}\left(-\xi-i \alpha \tfrac{Q}{2},\, \xi+i \tfrac{Q}{2}(1+\alpha),\, m_F\right) = 
s_b \left(-\xi-i \alpha \tfrac{Q}{2}\right) \int d u e^{-2 \pi i m_F u} 
 s_b\left(-u \right) s_b\left(u \right) \nonumber
\\&&  
\qquad\qquad\qquad\qquad\qquad\qquad\qquad s_b \left(i\tfrac{Q}{2}  (1+\alpha)+u+\xi \right) s_b \left(i\tfrac{Q}{2} (1+\alpha)- u+ \xi)\right)
\\&& \qquad =  s_b \left(-\xi-i \alpha \tfrac{Q}{2}\right) \int d u e^{-2 \pi i m_F u} 
s_b \left(i\tfrac{Q}{2} (1+\alpha)+u+\xi \right) s_b \left(i\tfrac{Q}{2} (1+\alpha)- u+ \xi)\right)\nonumber 
\end{eqnarray}
Let us pause to interpret what just happened in terms of our method from section \ref{Sec:U1monop}. The first mirror symmetry we applied in \eqref{intermmirror1} turned the monopole operator deformation into an off-diagonal mass deformation, as we have studied earlier in this paper (see Section~\ref{Sec:U1monop}). We can readily see this in the contribution $s(u) s(-u)$, which corresponds to two oppositely charged chirals $(P, \tilde Q)$, each of R-charge $1$. This is consistent with a superpotential term of the form $\Delta W = P \tilde Q$. The fact that both can be integrated out is implemented by the cancellation of their contributions to $Z$. 

Having performed this elimination, we do another mirror symmetry to get the effective theory we sought after. This is implemented by another use of the pentagon identity, which eliminates the integral and leaves us with the following expression:

\begin{eqnarray}
&&Z_{TSU(2)}\left(-\xi-i\tfrac{Q}{2} \alpha ,\, \xi+i\tfrac{Q}{2} (1+\alpha),\, m_F\right) = \\
&& \qquad\qquad\qquad  s_b\left(-i\tfrac{Q}{2} \alpha -\xi\right) s_b\left(i\tfrac{Q}{2}  (1+2 \alpha)-2 \xi\right) s_b\left(-i\tfrac{Q}{2} \alpha -m_F-\xi\right) s_b\left(-i\tfrac{Q}{2} \alpha +m_F-\xi\right) \nonumber
\end{eqnarray}
This corresponds to three chirals of dimension: $(1+\alpha)$ and one chiral of dimension $-2 \alpha$.

Let $\cX, \cY, \cZ, \cS$ have dimensions $(-2 \alpha,\,1+\alpha,\, 1+\alpha,\, 1+\alpha)$, respectively. Then this is consistent with the superpotential of the \emph{modified XYZ-model}
\be\label{supXYZmod}
W = \cX \cY \cZ + \cS^2 \cX\,.
\ee
This completes the verification of our claim in Section \ref{Sec:U1monop}, i.e. that $\cN=4$ 3d SQED with two flavors deformed by a monopole superpotential is dual to the modified XYZ model with superpotential \eqref{supXYZmod}. In fact, after integrating out the massive fields\footnote{In Section \ref{Sec:U1monop}, the description of the mirror theory was a bit different. However, from equation \eqref{SQEDmirrMassDef} one sees that $s_1,s_2,\phi_\ell$ can be integrated out (in particular $s_1=-s_2\equiv \mathcal{S}$), leaving a superpotential that is the sum of the $\cN=4$ SQED superpotential, plus the mass deformation for $P$ and $\tilde{Q}$.}, this is the theory with superpotential \eqref{modpot} of Section \ref{Sec:U1monop} where $\tilde{\mathfrak{m}}= \left(\begin{array}{cc} { \mathcal{S}} &  \mathcal{Y} \\  \mathcal{Z} & -{ \mathcal{S}} \end{array}\right)$ and in which we set $m=1$ and neglect the term including $\Psi$ (in this section we have not gauged the $SU(2)$ flavor symmetry).

\subsection{Monopole deformed $U(2)$ with four flavors} \label{Sec:monu2}

Let us consider the theory of Section \ref{Sec:BBduality}, i.e. 3d $\cN=4$ $U(2)$ SQCD with four flavors $Q_i\tilde Q^j$ deformed by a monopole superpotential. We want now to prove that its partition function is equal to the partition function of the effective theory we discussed in Section \ref{U2monopDef}: the S-dual theory with gauge group $U(1)\times SU(2)$ (see Section \ref{Sec:BBduality}) with the $U(1)$ node replaced by the modified XYZ model. 

In Section \ref{dualqed} we matched the  partition functions of dual theories by analysing the self-mirror property of $\cN=4$ SQED with two flavors, which is also called $T(SU(2))$ in the literature. Analogously, in the present section we obtain the desired result by considering mirror symmetry for $T(SU(3))$, which is the 3d $\cN=4$ theory in Figure \ref{tsu3th}. The basic result we need is that it is self-mirror. 
\begin{center}
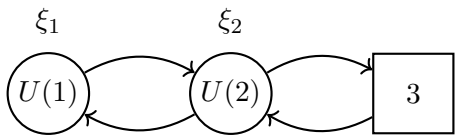
\begin{figure}[h!]
\centering
\begin{tikzpicture}[->,thick, scale=0.8]
  \node[circle, draw, inner sep= 2pt](L0) at (7,0){$U(1)$};
    \node[inner sep= 2pt](L01) at (7,1.2){$\xi_1$};
  \node[circle, draw, inner sep= 2pt](L1) at (10,0){$U(2)$};
      \node[inner sep= 2pt](L11) at (10,1.2){$\xi_2$};
  \node[draw, rectangle, minimum width=30pt, minimum height=30pt](L2) at (13,0){$3$};
 \path[every node/.style={font=\sffamily\small,
  		fill=white,inner sep=1pt}]
(L0) edge [bend left] node[above=2mm] {} (L1)
(L1) edge [bend left] node[below=2mm] {}(L0)
(L1) edge [bend left] node[above=2mm] {} (L2)
(L2) edge [bend left] node[below=2mm] {}(L1)
;
\end{tikzpicture}
\caption{Quiver of $T(SU(3))$ theory. $\xi_i$ is the FI parameter relative to the $i$-node.}\label{tsu3th}
\end{figure}
\end{center}

The partition function will be a function of the FI parameters $\xi_1,\xi_2$, associated with the $U(1)$ and $U(2)$ gauge groups respectively, and the real masses for the $SU(3)_F$ symmetry $m_1,m_2,m_3$ (subject to constraint $m_1+m_2+m_3=0$). The sphere partition function was computed in \cite{Benvenuti:2011ga} and is manifestly invariant under permutations of $m_i$. Since the theory is self-mirror, we expect a similar relation also for the FI parameters and indeed one can show that the partition function is also invariant under the transformation $\xi_1\leftrightarrow -\xi_2$, $\xi_2\leftrightarrow -\xi_1$. 
Said differently, we have the identity 
\begin{equation}\label{RelTSU3}
 Z_{TSU(3)}(\xi_1,\xi_2) =  Z_{TSU(3)}(-\xi_2,-\xi_1)\:.
\end{equation}
As in the previous section, we can promote the FI parameters to complex variables and the partition function is a holomorphic function of $z_i\equiv \xi_i+\tfrac{i}{2} \beta_{T,i} Q$, where $\beta_{T,i}$ is the mixing parameter between the R-charge and the topological $U(1)_T$ relative to the node $i$. The relation \eqref{RelTSU3} then implies
\begin{equation}\label{RelTSU3hol}
 Z_{TSU(3)}(z_1,z_2) =  Z_{TSU(3)}(-z_2,-z_1)\:.
\end{equation}
Now, remember that the monopole operators with R-charge $H+C=1$ have trial R-charge $1-\alpha$ (where we mixed the $\cN=4$ R-charge with the axial generator $H-C$ as in Section (\ref{dualqed})). In order to proceed we choose
\be
 z_1 = i\tfrac{Q}{2}  (1-\alpha)  \qquad\mbox{and}\qquad z_2 = \xi
\ee
Plugging this into \eqref{RelTSU3hol}, we have 
\begin{equation}\label{RelTSU3holmon}
 Z_{TSU(3)}\left(i\tfrac{Q}{2}  (1-\alpha) ,\xi\right) =  Z_{TSU(3)}\left(-\xi,-i\tfrac{Q}{2}  (1-\alpha) \right) \:.
\end{equation}
The LHS is the partition function of $T(SU(3))$ with an FI-term for the $U(2)$ node and a superpotential deformation $\Delta W= W_+$, with $W_+$ a monopole operator (with $\cN=4$ R-charge equal to one) relative to the $U(1)$ node (the mixing parameter $\beta_{T,1}=1+\alpha$ is the one we need to have $W_+$ with R-charge equal to $2$). 
The RHS instead is the partition function of $T(SU(3))$ with an FI-term for the $U(1)$ node and a superpotential deformation $\Delta W= V_-$, with $V_-$ a monopole operator (with $\cN=4$ R-charge equal to one) relative to the $U(1)$ node.

We now integrate \eqref{RelTSU3holmon} over $\xi$:
\begin{equation}\label{RelTSU3holmonIntegr}
 \int_{-\infty}^{+\infty} d\xi \,  Z_{TSU(3)}\left(i\tfrac{Q}{2}  (1-\alpha) ,\xi\right) =  \int_{-\infty}^{+\infty} d\xi \, Z_{TSU(3)}\left(-\xi,-i\tfrac{Q}{2}  (1-\alpha) \right) \:.
\end{equation}

At the level of the partition function, gauging a global $U(1)$ means introducing a real mass for it (i.e. a background gauge field is introduced for the global $U(1)$) and integrating over it (i.e. the $U(1)$ is now a gauge symmetry). Hence, the LHS of \eqref{RelTSU3holmonIntegr} is the partition function of the (monopole deformed) $T(SU(3))$ theory where we gauged the topological $U(1)_T$ relative to the $U(2)$ node (an FI parameter is a real mass for the topological symmetry). Since gauging a topological $U(1)$ is equivalent to ungauging the central $U(1)$ factor,\footnote{At the level of the sphere partition function for a $U(N)$ gauge theory turning on a FI term means multiplying the integrand by $\text{Exp}(-2\pi i\xi\sum_iu_i)$, where $\xi$ is the FI parameter and $u_i$ represent the $U(N)$ Cartan coordinates over which we integrate. The integral over $\xi$ produces a delta function $\delta(\sum_iu_i)$ thus reducing the partition function to that of a $SU(N)$ gauge theory} the LHS of \eqref{RelTSU3holmonIntegr} reduces to the partition function of the  $SU(2)\times U(1)$ theory discussed in Section \ref{Sec:BBduality}, with the superpotential deformation $\Delta W=W_+$ at the abelian node. On the other hand, on the RHS we are integrating over the FI parameter of the $U(1)$ node and the net effect is to ungauge it. Consequently, the RHS becomes the partition function of the $U(2)$ theory with four flavors deformed by $\Delta W = V_-$,  i.e.
\begin{equation}\label{PartFnctSduality}
   Z_{U(2)}^{V_-} (\alpha) =  Z_{U(1)\times SU(2)}^{W_+} (\alpha)\:.
\end{equation}
This, in particular, prove that the monopole operators $V_\mp$ are mapped to $W_\pm$, as claimed in Section \ref{U2monopDef}. 

Let us write down the RHS of \eqref{PartFnctSduality}explicitly:
\begin{eqnarray}
Z_{U(1)\times SU(2)}^{W_+} (\alpha) &=& \int du_2 \,\,\sinh(2 b\pi u_2) \,\,\sinh(2 b^{-1}\pi u_2)\,  \\
&&\overbrace{{s_b}^6\left(i\tfrac{Q}{4}  (1-\alpha)+ u_2\right) {s_b}^6\left(i\tfrac{Q}{4}  (1-\alpha)-u_2\right)  }^{q_i,\tilde{q}^i} \, \nonumber 
\overbrace{s_b\left(i\tfrac{Q}{2}  \alpha \right) s_b\left(i\tfrac{Q}{2} \alpha +2 u_2\right) s_b\left(i\tfrac{Q}{2}  \alpha -2 u_2\right) }^\Phi
\nonumber \\ \nonumber\\
   &&    \underbrace{s_b\left(i\tfrac{Q}{2}  \alpha \right)}_\phi \int du_1  \underbrace{e^{\pi Q(1+\alpha)u_1}}_{\Delta W=W_+}  \underbrace{\prod_{\epsilon_1=\pm} \prod_{\epsilon_2=\pm}
		s_b\left(i\tfrac{Q}{4}  (1-\alpha)+ \epsilon_1 u_1+\epsilon_2 u_2\right)
 }_{v,\tilde{v}} \nonumber
\end{eqnarray}
We now notice that the second line is exactly $Z_{TSU(2)}^{W_+}(\tfrac{i}{2} \alpha Q, 2 u_2, 0)$, that we have computed in \eqref{FSQEDMonEffTh}. Hence, using $\eqref{PartFnctSduality}$ we reach the desired conclusion:
\begin{eqnarray}
Z_{U(2)}^{V_-} (\alpha) &=& \int du_2 \,\sinh(2 b\pi u_2) \,\,\sinh(2 b^{-1}\pi u_2)\,  \\
&&\overbrace{{s_b}^6\left(i\tfrac{Q}{4}  (1-\alpha)+ u_2\right) {s_b}^6\left(i\tfrac{Q}{4}  (1-\alpha)-u_2\right)  }^{q_i,\tilde{q}^i} \, \nonumber 
\overbrace{s_b\left(i\tfrac{Q}{2}  \alpha \right) s_b\left(i\tfrac{Q}{2}  \alpha +2 u_2\right) s_b\left(i\tfrac{Q}{2}  \alpha -2 u_2\right) }^\Phi
\nonumber \\  \nonumber \\
   &&    \underbrace{s_b\left(-i\tfrac{Q}{2} \alpha \right) s_b\left(i\tfrac{Q}{2}  (1+2 \alpha)\right) s_b\left(-i\tfrac{Q}{2} \alpha -2 u_2\right) s_b\left(-i\tfrac{Q}{2} \alpha +2 u_2\right) }_{\mathcal{X},\tilde{\mathfrak{m}}}
   \nonumber
\end{eqnarray}
The RHS of the above equation is the partition function of the effective theory discussed in Section \ref{U2monopDef}.

\subsection{Monopole deformations of $\cN=4$ $U(N)$ SQCD with $2N$ flavors} \label{Sec:un2n}

Let us comment briefly on higher rank theories. In this case one duality frame involves a nonlagrangian theory, so we cannot directly compare the sphere partition functions. Nevertheless, since the $S^3$ partition function can be extracted from the superconformal index of the four dimensional theory, we conclude that the partition functions of $SU(N)$ SQCD with $2N$ flavors and the S-dual theory necessarily match: the index of class $\mathcal{S}$ theories is known to depend only on the data of the UV curve \cite{Gaiotto:2012xa} and S-dual theories are described by the same UV curve. Since the baryon number of SQCD is identified with the $U(1)$ symmetry acting on the $SU(2)$ doublet in the S-dual theory, the corresponding real mass parameters in three dimensions are identified. The equality of the partition functions for the dual pair we discussed in the previous sections then follow just by integrating over the real mass (i.e. gauging the $U(1)$ symmetry) and by turning on an ``imaginary'' FI parameter to introduce the mixing of the R-symmetry with the topological symmetry.

\section{Concluding remarks} 

In this paper we studied ADE quiver gauge theories in three dimensions with monopole superpotential terms, which describe the worldvolume theory of D2 branes probing T-brane backgrounds. These superpotential terms affect the moduli space of the theory in a subtle way and as we have seen it is often possible to identify a dual description of the theory which makes their effect manifest. We always find that the geometric branch (i.e. the ADE singularity) stays undeformed but the resolution of the singularity is obstructed, in agreement with the analysis of \cite{Anderson:2013rka}. 

The main tool we exploited in our analysis is (the 3d version of) Argyres-Seiberg duality. This is essential to reduce the problem to analyzing a $U(1)$ theory, which in turn can be handled using $\mathcal{N}=2$ abelian mirror symmetry as we did in \cite{Collinucci:2016hpz}. In this way we find a large new class of $\mathcal{N}=2$ theories whose Higgs branch coincides with that of the parent theory with eight supercharges. The key step is to turn our attention to a dual non-Lagrangian description of the theory: we lose in part the simplicity of the ``conventional'' description but we gain a simpler representation of the deformation we need to understand. This is a perfect example of the power of dualities: in every duality frame some observables are easy to compute and in order to achieve a complete understanding of a theory it is often necessary to consider simultaneously several dual descriptions. 

As we have already discussed, T-branes can also be understood as nilpotent mass terms in the mirror of the ADE quiver theories. Our method can be straightforwardly applied to study all T-branes for $D_N$ or $E_N$ singularities corresponding to mass matrices which square to zero. More general cases require a generalization of our method: When we apply our duality to handle the monopole superpotential at one node, we affect nontrivially neighboring nodes as well, since they are now coupled to non-Lagrangian matter. Hence, when we turn on a monopole superpotential at all the gauge nodes in a subquiver, it is convenient to look for a suitable dual description of the subquiver as a whole. When the problem can be reduced to discussing linear subquivers we expect class $\mathcal{S}$ dualities to provide the right duality frame. In the general case (such as a mass matrix in the principal nilpotent orbit) we probably need analogs of the Argyres-Seiberg duality for quivers of D or E type, which are not known at present. Similar considerations apply also to the duality studied in Section \ref{Sec:dualnew}: We can understand the equivalence of the new quivers for a subclass of monopole superpotentials. The general case necessarily requires exploiting dualities for more complicated subquivers.   

We believe our method can be generalized further and represents an essential starting point both for the study of monopole operators in supersymmetric theories and for a more thorough study of T-brane backgrounds from the perspective of the probe brane. 

It would finally be interesting to provide a proper description of how the Coulomb branch is modified by the monopole deformation. From the mirror side, we know that some directions of the moduli space should be lifted. Perhaps there might be a way to use for this purpose the Hilbert series constructions of \cite{Cremonesi:2013lqa,Cremonesi:2014kwa,Cremonesi:2014uva,Cremonesi:2015dja,Hanany:2015via,Cremonesi:2016nbo} that proved so successful in constructing moduli spaces.

\section*{Acknowledgements}
We are grateful to Raffaele Savelli for his early collaboration in our previous work, and for numerous insightful discussions. We also thank Sergio Benvenuti, Amihay Hanany and Sara Pasquetti for useful discussions.

A.C. is a Research Associate of the Fonds de la Recherche Scientifique F.N.R.S. (Belgium). The work of A.C. is partially supported by IISN - Belgium (convention 4.4503.15). 

The work of R.V. is supported by the Programme ``Rita Levi Montalcini for young researchers'' (incentivisation of mobility of foreign and Italian researchers resident abroad). 

\appendix

\section{The Higgs branch of ADE quivers} 
\label{app:higgsade}


\subsection{D-type quivers}\label{D-typeHB}
 \begin{center}
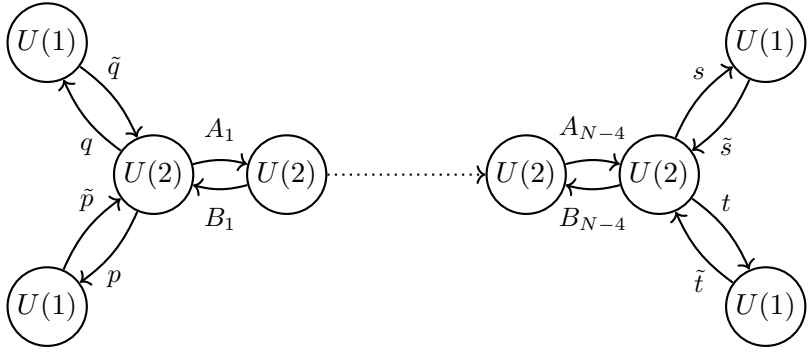
\begin{figure}[ht!] 
\centering
\begin{tikzpicture}[->, every node/.style={circle,draw},thick, scale=0.7]
  \node[inner sep=1.7](L1) at (-10,2.5){$U(1)$};
  \node[inner sep=1.7](L2) at (-10,-2.5){$U(1)$};
  \node[inner sep=1.7](V1) at (-8,0){$U(2)$};
  \node[inner sep=1.7](V2) at (-5.5,0){$U(2)$};
  \node[inner sep=1.7](VN) at (-1,0){$U(2)$};
  \node[inner sep=1.7](VN+1) at (1.5,0){$U(2)$};
  \node[inner sep=1.7](L3) at (3.5,-2.5) {$U(1)$};
    \node[inner sep=1.7](L4) at (3.5,2.5) {$U(1)$};

 \path[every node/.style={font=\sffamily\small,
  		fill=white,inner sep=1pt}]
(L1) edge [bend left=15] node[above=2mm] {$\tilde q$} (V1)
(V1) edge [bend left=15] node[below=2mm] {$q$}(L1)
(L2) edge [bend left=15] node[above=2mm] {$\tilde p$} (V1)
(V1) edge [bend left=15] node[below=2mm] {$p$}(L2)
(V1) edge [bend left=15] node[above=2mm] {$A_1$} (V2)
(V2) edge [bend left=15] node[below=2mm] {$B_1$} (V1)
(V2) edge [dotted] (VN)
(VN) edge [bend left=15] node[above=2mm] {$A_{N-4}$} (VN+1)
(VN+1) edge [bend left=15] node[below=2mm] {$B_{N-4}$} (VN)
(L3) edge [bend left=15] node[below=2mm] {$\tilde t$} (VN+1)
(VN+1) edge [bend left=15] node[above=2mm] {$t$}(L3)
(L4) edge [bend left=15] node[below=2mm] {$\tilde s$} (VN+1)
(VN+1) edge [bend left=15] node[above=2mm] {$s$}(L4)
;
\end{tikzpicture}
\caption{D$_N$ quiver.}\label{DNquiverapp}
\end{figure}
\end{center}
The worldvolume theory of a D-brane probing an $D_N$ singularity is a quiver gauge theory in the shape of the affine $D_N$ 
Dynkin diagram (see Figure  \ref{DNquiverapp}).
The Higgs branch chiral operators (that are traces of products of bifundamental fields) can be expressed in terms of three generators constrained by one relation, that is the same as the algebraic equation defining the D-type singularity (see
\cite{Lindstrom:1999pz} for the explicit derivation). The relations that express all the chiral operators in terms of the three generators and that constrain the three generators come from the F-term equations derived from differentiating the superpotential \eqref{WDN} with respect to the vector multiplet scalars $\Psi_i,\phi_q,\phi_p,\phi_t\phi_s$.

Let us sketch the derivation of the Higgs branch. When $N$ is even, the three invariants that generate the chiral ring are
\begin{eqnarray}\label{DngeneratorsHB}
 Z &\equiv& - \tilde{q}p\tilde{p}q\:, \nonumber \\
 Y &\equiv& 2 \tilde{p}A_1\cdots A_{N-4}s \tilde{s}B_{N-4}\cdots B_1 p + (-z)^{N/2-1}\:, \\
 X &\equiv& 2 \tilde{q}A_1\cdots A_{N-4}s \tilde{s}B_{N-4}\cdots B_1 p \tilde{p}q \:. \nonumber
\end{eqnarray}
They are subject to the relation 
\begin{equation}\label{DNsingHBBapp}
  X^2+ Z Y^2 = Z^{N-1} \:.
\end{equation}

All the other possible gauge invariants either vanish or can be written in terms of the generators by using the F-terms of $\Psi_i,\phi_q,\phi_p,\phi_t\phi_s$. For instace:
\begin{eqnarray}
\tilde{s}t\tilde{t}s &=& \mbox{tr}( t\tilde{t} + s\tilde{s} )^2 = \mbox{tr}( A_{N-4}B_{N-4} )^2 =  \mbox{tr}( A_{1}B_{1} )^2 = \tilde{q}p\tilde{p}q = -Z \nonumber \\
\tilde{p}A_1\cdots A_{k} B_{k}\cdots B_1 p &=&  \tilde{p} (A_1 B_1)^k p= \tilde{p} (p\tilde{p}+q\tilde{q} )^k p = \delta_1^k Z^2 \qquad k\leq N-4 \nonumber \\
\tilde{s}B_{N-4}\cdots B_{N-4-h} A_{N-4-h}\cdots A_{N-4} s &=&  \tilde{s} (B_{N-4} A_{N-4})^k s= \tilde{s} (s\tilde{s}+t\tilde{t} )^k s = \delta_1^h Z^2 \qquad h\leq N-4 \nonumber \\
\end{eqnarray}

For $N$ odd, analogous computations can be done (see \cite{Lindstrom:1999pz} for detail).

\subsection{E-type quivers} 
\label{app:equivers}

\begin{figure}[ht!]
\begin{center}
\begin{tikzpicture}[->, thick]
\node[shape=circle, draw] (18) at (0,0) {\tiny{$1$}};
\node[shape=circle, draw] (19) [right= .2cm of 18] {\tiny{$2$}};
\node[shape=circle, draw] (20) [right= .2cm of 19] {\tiny{$3$}};
\node[shape=circle, draw] (21) [right= .2cm of 20] {\tiny{$2$}};
\node[shape=circle, draw] (22) [right= .2cm of 21] {\tiny{$1$}};
\node[shape=circle, draw] (23) [above= .2cm of 20] {\tiny{$2$}};
\node[shape=circle, draw] (24) [above= .2cm of 23] {\tiny{$1$}};
\node[shape=circle, draw] (10) [right= .5cm of 22] {\tiny{$1$}};
\node[shape=circle, draw] (11) [right= .2cm of 10] {\tiny{$2$}};
\node[shape=circle, draw] (12) [right= .2cm of 11] {\tiny{$3$}};
\node[shape=circle, draw] (13) [right= .2cm of 12] {\tiny{$4$}};
\node[shape=circle, draw] (17) [above= .2cm of 13] {\tiny{$2$}};
\node[shape=circle, draw] (14) [right= .2cm of 13] {\tiny{$3$}};
\node[shape=circle, draw] (15) [right= .2cm of 14] {\tiny{$2$}};
\node[shape=circle, draw] (16) [right= .2cm of 15] {\tiny{$1$}};
\node[shape=circle, draw] (1) [right= .5cm of 16] {\tiny{$1$}};
\node[shape=circle, draw] (2) [right= .2cm of 1] {\tiny{$2$}};
\node[shape=circle, draw] (3) [right= .2cm of 2] {\tiny{$3$}};
\node[shape=circle, draw] (4) [right= .2cm of 3] {\tiny{$4$}};
\node[shape=circle, draw] (5) [right= .2cm of 4] {\tiny{$5$}};
\node[shape=circle, draw] (6) [right= .2cm of 5] {\tiny{$6$}}; 
\node[shape=circle, draw] (7) [above= .2cm of 6] {\tiny{$3$}}; 
\node[shape=circle, draw] (8) [right= .2cm of 6] {\tiny{$4$}}; 
\node[shape=circle, draw] (9) [right= .2cm of 8] {\tiny{$2$}};

\draw[-] (1) -- (2);
\draw[-] (2) -- (3);
\draw[-] (3) -- (4);
\draw[-] (4) -- (5);
\draw[-] (5) -- (6); 
\draw[-] (6) -- (7); 
\draw[-] (6) -- (8); 
\draw[-] (8) -- (9); 
\draw[-] (10) -- (11);
\draw[-] (11) -- (12);
\draw[-] (12) -- (13);
\draw[-] (13) -- (14);
\draw[-] (14) -- (15);
\draw[-] (15) -- (16);
\draw[-] (13) -- (17);
\draw[-] (18) -- (19);
\draw[-] (19) -- (20);
\draw[-] (20) -- (21);
\draw[-] (21) -- (22);
\draw[-] (20) -- (23);
\draw[-] (23) -- (24);
;

\end{tikzpicture}
\caption{From left to right the quivers of $E_6$, $E_7$ and $E_8$ type}
\label{eccezz}
\end{center} 
\end{figure}
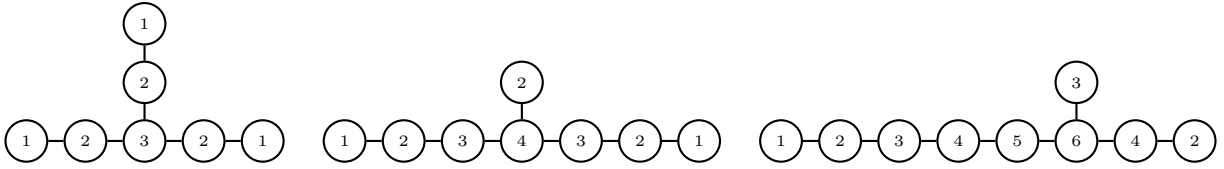

The worldvolume theory of a D-brane probing an $E_N$ singularity is a quiver gauge theory in the shape of the affine $E_N$ 
Dynkin diagram. These theories have a central gauge node coupled to three linear tails of unitary groups, as displayed in Figure \ref{eccezz}.
All the Higgs branch chiral operators of these models (i.e. 
traces of products of bifundamental multiplets) can be expressed in terms of three fundamental generators satisfying a chiral 
ring relation, which is the same as the defining equation of the corresponding E-type singularity. This was shown explicitly in 
\cite{Lindstrom:1999pz} and is essentially a consequence of F-term equations derived from the superpotential of the theory. We will now sketch the derivation, referring to \cite{Lindstrom:1999pz} for details. A first important observation is the following: we can construct three mesons (let us call them $M_{1,2,3}$) quadratic in the bifundamental fields, which transform in the adjoint representation of the central $U(n)$ gauge node (and neutral under the other gauge groups) and they satisfy the chiral ring relations:
\be\label{relcent}M_i^{l_i}=0;\quad M_1+M_2+M_3=0,\ee 
where $l_{1,2,3}$ denote the length of the linear tails (including the central node) and the last equation is simply a consequence of the F-term equation for the central gauge node. 

It is possible to show that all nonvanishing Higgs branch chirals can be written in terms of $M_{1,2,3}$. Using this description and (\ref{relcent}), one can identify the three basic generators of the Higgs branch. For example in the $E_6$ case (the quiver has three identical legs of length three) the three generators are (in the notation of \cite{Lindstrom:1999pz})
\be U=\tr(M_1^2M_2^2M_3^2);\quad W=\tr(M_1^2M_2);\quad V=\tr(M_1^2M_2^2).\ee 
All other gauge invariants are polynomials in $U$, $V$ and $W$. The task now is to extract the $E_6$ singularity from these three objects and the key tool for this is the Schouten identity discussed in \cite{Lindstrom:1999pz}. In the $E_6$ case one only needs the identity for one-by-one matrices, which simply states that such matrices commute. This fact is helpful because the invariants $U$ and $V$ defined above can also be written as products of $U(1)\times U(1)$ bifundamentals (these arise by multiplying together bifundamental fields, starting and ending at $U(1)$ nodes). These are one-by-one matrices and hence can be freely commuted. Using this fact one can find suitable combinations $X$, $Y$ and $Z$ of the three generators which satisfy the relation $X^2+Y^3=Z^4$, which is the desired result. 

The logic for the other two cases is similar; the only difference is that we need the two-dimensional Schouten identity which reads: $$\tr(\{A,B\}C)=\tr(AB)(\tr C)+\tr(AC)(\tr B)+\tr(BC)(\tr A)-(\tr A)(\tr B)(\tr C)$$ for any $2\times 2$ matrices $A$, $B$ and $C$. This identity can be applied to extract the desired chiral ring relation if we can rewrite the basic generators as the trace of products of two-by-two matrices. In \cite{Lindstrom:1999pz} this is done by introducing certain ``chains'' of the bifundamental chirals in the quiver which transform in the adjoint representation of one $U(2)$ node in the quiver, and showing that the Higgs branch generators can be written in terms of them. 

In the $E_7$ quiver, which has two tails of length four and one of length two, the three generators are (we call $M_3$ the meson of the tail of length two) 
\be X=\tr(M_1^2M_2^3M_1^3M_3);\quad Y=-\tr(M_1^3M_2^3);\quad Z=\tr(M_1^3M_3).\ee 
Using the Schouten identity at the $U(2)$ node of one of the long tails, one can prove that $X^2+Y^3+YZ^3=0$. 

Finally, in the $E_8$ case the three generators are 
\be X=\tr(M_1^5M_2^2M_1M_2^2M_1^3M_2^2);\quad Y=\tr(M_1^5M_2^2M_1M_2^2);\quad Z=\tr(M_1^5M_2),\ee 
where $M_1$ is associated to the tail of length six and $M_2$ to the tail of length three. In order to prove that $X^2+Y^3+Z^5=0$, we need to use the Schouten identity at the $U(2)$ node of the length-three tail. 

In summary, in order to extract the $E_N$ singularities, we need to check that all the Higgs branch operators can be written in terms of three matrices satisfying (\ref{relcent}) and that they can be written in terms of $U(2)$ adjoints as in \cite{Lindstrom:1999pz}. The last step is needed in order to use the Schouten identity.

\section{Monopole deformation along a $U(1)$ node of $D_N$}\label{app:DNU1extnode}

\noindent Consider a $D_N$ quiver gauge theory (see Figure  \ref{DNquiverapp}).
The arrows of the quiver represent bifundamental chirals, as the diagram shows.\footnote{We use the convention where the arrows that go from a non-Abelian node to an Abelian one represent column vectors.}
The four external are associated with Abelian vector multiplets, which have each a complex scalar fields. Starting from the upper left in clockwise orientation, these are $\phi_q, \phi_s, \phi_t, \phi_p$. 
Similarly, each non-Abelian node in the middle horizontal line has an adjoint complex scalar field $\Psi_1\,, \ldots \Psi_{N-3}$.

The $\cN=4$ theory has the following superpotential
\begin{eqnarray}\label{WDN}
W &=& {\rm tr}\big[\left(\Psi_1-\mathbb{1} \phi_q \right) q \tilde q+\left(\Psi_{N-3}-\mathbb{1} \phi_s \right) s \tilde s+\left(\Psi_{N-3}-\mathbb{1} \phi_t \right) t \tilde t +\left(\Psi_1-\mathbb{1} \phi_p \right) p \tilde p \nonumber\\
&+& \sum_{i=1}^{N-4} \left(B_i \Psi_i A_i-A_i \Psi_{i+1} B_i \right) \big]
\end{eqnarray}

The Higgs branch (HB) is described by gauge invariant combinations of the quark fields subject to relations coming from the F-terms for the fields $\Psi_i$ and $\phi_{p,q,s,t}$ \cite{Lindstrom:1999pz,Borokhov:2003yu} (see Appendix \ref{D-typeHB}). 
All the gauge invariants can be written in terms of three generators $x$, $y$ and $z$ satisfying the equation defining the $D_N$ singularity (see Appendix \ref{D-typeHB}):
\begin{equation}\label{DNsingHBB}
  x^2+ z y^2 = z^{N-1} \:.
\end{equation}

We now want to deform the superpotential \eqref{WDN} by adding the coupling (that is our definition of a T-brane along the corresponding Jordan block)
\begin{equation}\label{WmonopWq}
  \Delta W = m W_{q,+} \:,
\end{equation}
where $W_{q,+}$ is the monopole operator relative to the $U(1)_q$ node, i.e. the one that has R-charge equal to one and sits in the same $\cN=4$ supermultiplet as the conserved topological current relative to the given photon.
We then proceed as outlined above. We ungauged the nearby $U(2)$ node, obtaining a $U(1)$ gauge theory with two flavors and coupled to the complex scalar field $\Psi_1$ like in \eqref{WBloc} (where now $\phi_\ell \rightarrow \phi_q$ and $\Psi \rightarrow \Psi_1$). 
As seen above, the monopole deformation produces 
a local theory with no gauge fields and with superpotential
\be \label{WeffDN}
W_{\rm loc}^{\rm eff}= \mbox{tr}(\Psi_1\,  \mathfrak{m})-\frac{\mathcal{X}}{m}\text{det} \mathfrak{m} \:,
\ee
with $ \mathfrak{m}$ a $2\times 2$ traceless complex  matrix.
We finally have to glue again our theory to the $U(2)$ gauge node. Since the gauge group has now disappeared, our quiver has lost one Abelian tail and has now the shape of a D$_N$ (not affine) Dynkin diagram. The previously trivalent vertex now has two adjoint chiral multiplets and two neutral chirals ($\phi$ and $\mathcal{X}$) coupled to them. The rest of the quiver and superpotential terms are unaltered. 

\begin{figure}[ht!]
\begin{center}
\begin{tikzpicture}[->, every node/.style={circle,draw},thick, scale=0.8]
  \node(L2) at (-10,-2.5){$\phi_p$};
  \node(V1) at (-8,0){$\Psi_1$};
  \node(V2) at (-5.5,0){$\Psi_2$};
  \node[inner sep=.7](VN+1) at (-1,0){$\Psi_{N+1}$};
  \node(L3) at (1,-2.5) {$\phi_y$};
    \node(L4) at (1,2.5) {$\phi_x$};

 \path[every node/.style={font=\sffamily\small,
  		fill=white,inner sep=1pt}]
(L2) edge [bend left=15] node[above=2mm] {$\tilde p$} (V1)
(V1) edge [bend left=15] node[below=2mm] {$p$}(L2)
(V1) edge [bend left=15] node[above=2mm] {$A_1$} (V2)
(V2) edge [bend left=15] node[below=2mm] {$B_1$} (V1)
(V2) edge [dotted] (VN+1)
(L3) edge [bend left=15] node[below=2mm] {$\tilde t$} (VN+1)
(VN+1) edge [bend left=15] node[above=2mm] {$t$}(L3)
(L4) edge [bend left=15] node[below=2mm] {$\tilde s$} (VN+1)
(VN+1) edge [bend left=15] node[above=2mm] {$s$}(L4)
(V1) edge[loop, out=100, in=160, looseness=8] node[above=2mm]{$M$} (V1)
;

\end{tikzpicture}
\caption{Effective quiver for deformed D$_{N+4}$}
\label{fig:dn->reduced}
\end{center}
\end{figure}
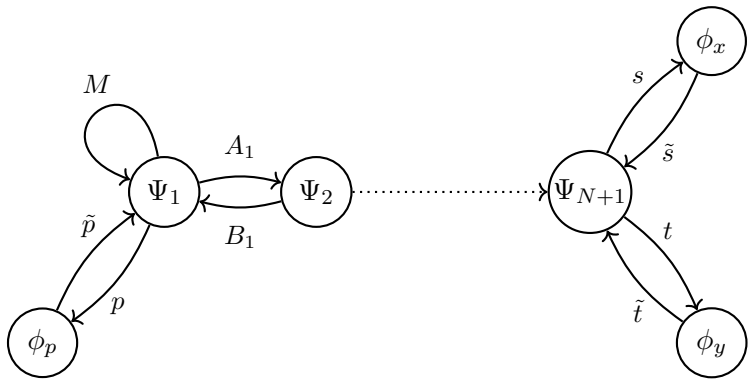

The Higgs branch is not modified by the monopole deformation \eqref{WmonopWq}. 
The Higgs branch of the $\mathcal{N}=4$ theory is the singularity of type D$_N$: one constructs suitable gauge invariant operators out of the bifundamentals (for $N$ even, they are given by \eqref{DngeneratorsHB}) and using the F-term constraints one proves 
that they satisfy the desired relation \cite{Lindstrom:1999pz}. All the gauge invariants considered in extracting the 
singularity are constructed using the meson matrix built out of these bifundamentals.
The theory we obtain after the deformation 
differs from the original more supersymmetric model only in 
one aspect: the bifundamentals fields between one $U(1)$ node and its adjacent $U(2)$ node are replaced by the meson matrix $\mathfrak{m}$. The F-terms of the deformed theory says that $\mathfrak{m}$ is traceless and square to zero. 
We can hence apply the results at the end of Section \ref{Sec:GoalStrategy} and conclude that $\mathfrak{m}$  satisfies all possible relation satisfied by the meson matrix in the original theory.


\begin{thebibliography}{10}

\bibitem{Gomez:2000zm}
T.~Gomez and E.~R. Sharpe, \emph{{D-branes and scheme theory}},
  \href{http://arxiv.org/abs/hep-th/0008150}{{\tt hep-th/0008150}}.

\bibitem{Donagi:2003hh}
R.~Donagi, S.~Katz and E.~Sharpe, \emph{{Spectra of D-branes with higgs vevs}},
  \href{http://dx.doi.org/10.4310/ATMP.2004.v8.n5.a3}{\emph{Adv. Theor. Math.
  Phys.} {\bf 8} (2004) 813--859},
  [\href{http://arxiv.org/abs/hep-th/0309270}{{\tt hep-th/0309270}}].

\bibitem{Donagi:2011jy}
R.~Donagi and M.~Wijnholt, \emph{{Gluing Branes, I}},
  \href{http://dx.doi.org/10.1007/JHEP05(2013)068}{\emph{JHEP} {\bf 05} (2013)
  068}, [\href{http://arxiv.org/abs/1104.2610}{{\tt 1104.2610}}].

\bibitem{Cecotti:2010bp}
S.~Cecotti, C.~Cordova, J.~J. Heckman and C.~Vafa, \emph{{T-Branes and
  Monodromy}}, \href{http://dx.doi.org/10.1007/JHEP07(2011)030}{\emph{JHEP}
  {\bf 07} (2011) 030}, [\href{http://arxiv.org/abs/1010.5780}{{\tt
  1010.5780}}].

\bibitem{Anderson:2013rka}
L.~B. Anderson, J.~J. Heckman and S.~Katz, \emph{{T-Branes and Geometry}},
  \href{http://dx.doi.org/10.1007/JHEP05(2014)080}{\emph{JHEP} {\bf 05} (2014)
  080}, [\href{http://arxiv.org/abs/1310.1931}{{\tt 1310.1931}}].

\bibitem{Collinucci:2014taa}
A.~Collinucci and R.~Savelli, \emph{{F-theory on singular spaces}},
  \href{http://dx.doi.org/10.1007/JHEP09(2015)100}{\emph{JHEP} {\bf 09} (2015)
  100}, [\href{http://arxiv.org/abs/1410.4867}{{\tt 1410.4867}}].

\bibitem{Anderson:2017rpr}
L.~B. Anderson, J.~J. Heckman, S.~Katz and L.~Schaposnik, \emph{{T-Branes at
  the Limits of Geometry}},  \href{http://arxiv.org/abs/1702.06137}{{\tt
  1702.06137}}.

\bibitem{Collinucci:2014qfa}
A.~Collinucci and R.~Savelli, \emph{{T-branes as branes within branes}},
  \href{http://dx.doi.org/10.1007/JHEP09(2015)161}{\emph{JHEP} {\bf 09} (2015)
  161}, [\href{http://arxiv.org/abs/1410.4178}{{\tt 1410.4178}}].

\bibitem{Marchesano:2015dfa}
F.~Marchesano, D.~Regalado and G.~Zoccarato, \emph{{Yukawa hierarchies at the
  point of E$_{8}$ in F-theory}},
  \href{http://dx.doi.org/10.1007/JHEP04(2015)179}{\emph{JHEP} {\bf 04} (2015)
  179}, [\href{http://arxiv.org/abs/1503.02683}{{\tt 1503.02683}}].

\bibitem{Cicoli:2015ylx}
M.~Cicoli, F.~Quevedo and R.~Valandro, \emph{{De Sitter from T-branes}},
  \href{http://dx.doi.org/10.1007/JHEP03(2016)141}{\emph{JHEP} {\bf 03} (2016)
  141}, [\href{http://arxiv.org/abs/1512.04558}{{\tt 1512.04558}}].

\bibitem{Bena:2016oqr}
I.~Bena, J.~Bl{\aa}b{\"a}ck, R.~Minasian and R.~Savelli, \emph{{There and back
  again: A T-brane's tale}},
  \href{http://dx.doi.org/10.1007/JHEP11(2016)179}{\emph{JHEP} {\bf 11} (2016)
  179}, [\href{http://arxiv.org/abs/1608.01221}{{\tt 1608.01221}}].

\bibitem{Marchesano:2016cqg}
F.~Marchesano and S.~Schwieger, \emph{{T-branes and $\alpha'$-corrections}},
  \href{http://dx.doi.org/10.1007/JHEP11(2016)123}{\emph{JHEP} {\bf 11} (2016)
  123}, [\href{http://arxiv.org/abs/1609.02799}{{\tt 1609.02799}}].

\bibitem{Ashfaque:2017iog}
J.~M. Ashfaque, \emph{{Monodromic T-Branes And The $SO(10)_{GUT}$}},
  \href{http://arxiv.org/abs/1701.05896}{{\tt 1701.05896}}.

\bibitem{Bena:2017jhm}
I.~Bena, J.~Bl{\aa}b{\"a}ck and R.~Savelli, \emph{{T-branes and Matrix
  Models}},  \href{http://arxiv.org/abs/1703.06106}{{\tt 1703.06106}}.

\bibitem{Collinucci:2016hpz}
A.~Collinucci, S.~Giacomelli, R.~Savelli and R.~Valandro, \emph{{T-branes
  through 3d mirror symmetry}},
  \href{http://dx.doi.org/10.1007/JHEP07(2016)093}{\emph{JHEP} {\bf 07} (2016)
  093}, [\href{http://arxiv.org/abs/1603.00062}{{\tt 1603.00062}}].

\bibitem{Benini:2009qs}
F.~Benini, C.~Closset and S.~Cremonesi, \emph{{Chiral flavors and M2-branes at
  toric CY4 singularities}},
  \href{http://dx.doi.org/10.1007/JHEP02(2010)036}{\emph{JHEP} {\bf 02} (2010)
  036}, [\href{http://arxiv.org/abs/0911.4127}{{\tt 0911.4127}}].

\bibitem{Intriligator:1996ex}
K.~A. Intriligator and N.~Seiberg, \emph{{Mirror symmetry in three-dimensional
  gauge theories}},
  \href{http://dx.doi.org/10.1016/0370-2693(96)01088-X}{\emph{Phys. Lett.} {\bf
  B387} (1996) 513--519}, [\href{http://arxiv.org/abs/hep-th/9607207}{{\tt
  hep-th/9607207}}].

\bibitem{Hanany:1996ie}
A.~Hanany and E.~Witten, \emph{{Type IIB superstrings, BPS monopoles, and
  three-dimensional gauge dynamics}},
  \href{http://dx.doi.org/10.1016/S0550-3213(97)00157-0,
  10.1016/S0550-3213(97)80030-2}{\emph{Nucl. Phys.} {\bf B492} (1997)
  152--190}, [\href{http://arxiv.org/abs/hep-th/9611230}{{\tt
  hep-th/9611230}}].

\bibitem{deBoer:1996mp}
J.~de~Boer, K.~Hori, H.~Ooguri and Y.~Oz, \emph{{Mirror symmetry in
  three-dimensional gauge theories, quivers and D-branes}},
  \href{http://dx.doi.org/10.1016/S0550-3213(97)00125-9}{\emph{Nucl. Phys.}
  {\bf B493} (1997) 101--147}, [\href{http://arxiv.org/abs/hep-th/9611063}{{\tt
  hep-th/9611063}}].

\bibitem{Porrati:1996xi}
M.~Porrati and A.~Zaffaroni, \emph{{M theory origin of mirror symmetry in
  three-dimensional gauge theories}},
  \href{http://dx.doi.org/10.1016/S0550-3213(97)00061-8}{\emph{Nucl. Phys.}
  {\bf B490} (1997) 107--120}, [\href{http://arxiv.org/abs/hep-th/9611201}{{\tt
  hep-th/9611201}}].

\bibitem{deBoer:1996ck}
J.~de~Boer, K.~Hori, H.~Ooguri, Y.~Oz and Z.~Yin, \emph{{Mirror symmetry in
  three-dimensional theories, SL(2,Z) and D-brane moduli spaces}},
  \href{http://dx.doi.org/10.1016/S0550-3213(97)00115-6}{\emph{Nucl. Phys.}
  {\bf B493} (1997) 148--176}, [\href{http://arxiv.org/abs/hep-th/9612131}{{\tt
  hep-th/9612131}}].

\bibitem{deBoer:1997ka}
J.~de~Boer, K.~Hori, Y.~Oz and Z.~Yin, \emph{{Branes and mirror symmetry in N=2
  supersymmetric gauge theories in three-dimensions}},
  \href{http://dx.doi.org/10.1016/S0550-3213(97)00444-6}{\emph{Nucl. Phys.}
  {\bf B502} (1997) 107--124}, [\href{http://arxiv.org/abs/hep-th/9702154}{{\tt
  hep-th/9702154}}].

\bibitem{deBoer:1997kr}
J.~de~Boer, K.~Hori and Y.~Oz, \emph{{Dynamics of N=2 supersymmetric gauge
  theories in three-dimensions}},
  \href{http://dx.doi.org/10.1016/S0550-3213(97)00328-3}{\emph{Nucl. Phys.}
  {\bf B500} (1997) 163--191}, [\href{http://arxiv.org/abs/hep-th/9703100}{{\tt
  hep-th/9703100}}].

\bibitem{Aharony:1997bx}
O.~Aharony, A.~Hanany, K.~A. Intriligator, N.~Seiberg and M.~J. Strassler,
  \emph{{Aspects of N=2 supersymmetric gauge theories in three-dimensions}},
  \href{http://dx.doi.org/10.1016/S0550-3213(97)00323-4}{\emph{Nucl. Phys.}
  {\bf B499} (1997) 67--99}, [\href{http://arxiv.org/abs/hep-th/9703110}{{\tt
  hep-th/9703110}}].

\bibitem{Kapustin:1999ha}
A.~Kapustin and M.~J. Strassler, \emph{{On mirror symmetry in three-dimensional
  Abelian gauge theories}},
  \href{http://dx.doi.org/10.1088/1126-6708}{\emph{JHEP} {\bf 04} (1999) 021},
  [\href{http://arxiv.org/abs/hep-th/9902033}{{\tt hep-th/9902033}}].

\bibitem{Borokhov:2002ib}
V.~Borokhov, A.~Kapustin and X.-k. Wu, \emph{{Topological disorder operators in
  three-dimensional conformal field theory}},
  \href{http://dx.doi.org/10.1088/1126-6708}{\emph{JHEP} {\bf 11} (2002) 049},
  [\href{http://arxiv.org/abs/hep-th/0206054}{{\tt hep-th/0206054}}].

\bibitem{Borokhov:2002cg}
V.~Borokhov, A.~Kapustin and X.-k. Wu, \emph{{Monopole operators and mirror
  symmetry in three-dimensions}},
  \href{http://dx.doi.org/10.1088/1126-6708}{\emph{JHEP} {\bf 12} (2002) 044},
  [\href{http://arxiv.org/abs/hep-th/0207074}{{\tt hep-th/0207074}}].

\bibitem{Benvenuti:2016wet}
S.~Benvenuti and S.~Pasquetti, \emph{{3d $ \mathcal{N} $ = 2 mirror symmetry,
  pq-webs and monopole superpotentials}},
  \href{http://dx.doi.org/10.1007/JHEP08(2016)136}{\emph{JHEP} {\bf 08} (2016)
  136}, [\href{http://arxiv.org/abs/1605.02675}{{\tt 1605.02675}}].

\bibitem{Aharony:2013dha}
O.~Aharony, S.~S. Razamat, N.~Seiberg and B.~Willett, \emph{{3d dualities from
  4d dualities}}, \href{http://dx.doi.org/10.1007/JHEP07(2013)149}{\emph{JHEP}
  {\bf 07} (2013) 149}, [\href{http://arxiv.org/abs/1305.3924}{{\tt
  1305.3924}}].

\bibitem{Heckman:2010qv}
J.~J. Heckman, Y.~Tachikawa, C.~Vafa and B.~Wecht, \emph{{N = 1 SCFTs from
  Brane Monodromy}},
  \href{http://dx.doi.org/10.1007/JHEP11(2010)132}{\emph{JHEP} {\bf 11} (2010)
  132}, [\href{http://arxiv.org/abs/1009.0017}{{\tt 1009.0017}}].

\bibitem{Chacaltana:2010ks}
O.~Chacaltana and J.~Distler, \emph{{Tinkertoys for Gaiotto Duality}},
  \href{http://dx.doi.org/10.1007/JHEP11(2010)099}{\emph{JHEP} {\bf 11} (2010)
  099}, [\href{http://arxiv.org/abs/1008.5203}{{\tt 1008.5203}}].

\bibitem{Argyres:2007cn}
P.~C. Argyres and N.~Seiberg, \emph{{S-duality in N=2 supersymmetric gauge
  theories}},
  \href{http://dx.doi.org/10.1088/1126-6708/2007/12/088}{\emph{JHEP} {\bf 12}
  (2007) 088}, [\href{http://arxiv.org/abs/0711.0054}{{\tt 0711.0054}}].

\bibitem{Gaiotto:2009we}
D.~Gaiotto, \emph{{N=2 dualities}},
  \href{http://dx.doi.org/10.1007/JHEP08(2012)034}{\emph{JHEP} {\bf 08} (2012)
  034}, [\href{http://arxiv.org/abs/0904.2715}{{\tt 0904.2715}}].

\bibitem{Benini:2010uu}
F.~Benini, Y.~Tachikawa and D.~Xie, \emph{{Mirrors of 3d Sicilian theories}},
  \href{http://dx.doi.org/10.1007/JHEP09(2010)063}{\emph{JHEP} {\bf 09} (2010)
  063}, [\href{http://arxiv.org/abs/1007.0992}{{\tt 1007.0992}}].

\bibitem{Gaiotto:2008ak}
D.~Gaiotto and E.~Witten, \emph{{S-Duality of Boundary Conditions In N=4 Super
  Yang-Mills Theory}},
  \href{http://dx.doi.org/10.4310/ATMP.2009.v13.n3.a5}{\emph{Adv. Theor. Math.
  Phys.} {\bf 13} (2009) 721--896}, [\href{http://arxiv.org/abs/0807.3720}{{\tt
  0807.3720}}].

\bibitem{Benini:2017dud}
F.~Benini, S.~Benvenuti and S.~Pasquetti, \emph{{SUSY monopole potentials in
  2+1 dimensions}},  \href{http://arxiv.org/abs/1703.08460}{{\tt 1703.08460}}.

\bibitem{Minahan:1996fg}
J.~A. Minahan and D.~Nemeschansky, \emph{{An N=2 superconformal fixed point
  with E(6) global symmetry}},
  \href{http://dx.doi.org/10.1016/S0550-3213(96)00552-4}{\emph{Nucl. Phys.}
  {\bf B482} (1996) 142--152}, [\href{http://arxiv.org/abs/hep-th/9608047}{{\tt
  hep-th/9608047}}].

\bibitem{Gaiotto:2008nz}
D.~Gaiotto, A.~Neitzke and Y.~Tachikawa, \emph{{Argyres-Seiberg duality and the
  Higgs branch}},
  \href{http://dx.doi.org/10.1007/s00220-009-0938-6}{\emph{Commun. Math. Phys.}
  {\bf 294} (2010) 389--410}, [\href{http://arxiv.org/abs/0810.4541}{{\tt
  0810.4541}}].

\bibitem{Maruyoshi:2013hja}
K.~Maruyoshi, Y.~Tachikawa, W.~Yan and K.~Yonekura, \emph{{N=1 dynamics with
  $T_N$ theory}}, \href{http://dx.doi.org/10.1007/JHEP10(2013)010}{\emph{JHEP}
  {\bf 10} (2013) 010}, [\href{http://arxiv.org/abs/1305.5250}{{\tt
  1305.5250}}].

\bibitem{Tachikawa:2015bga}
Y.~Tachikawa, \emph{{A review of the $T_N$ theory and its cousins}},
  \href{http://dx.doi.org/10.1093/ptep/ptv098}{\emph{PTEP} {\bf 2015} (2015)
  11B102}, [\href{http://arxiv.org/abs/1504.01481}{{\tt 1504.01481}}].

\bibitem{Gadde:2013fma}
A.~Gadde, K.~Maruyoshi, Y.~Tachikawa and W.~Yan, \emph{{New N=1 Dualities}},
  \href{http://dx.doi.org/10.1007/JHEP06(2013)056}{\emph{JHEP} {\bf 06} (2013)
  056}, [\href{http://arxiv.org/abs/1303.0836}{{\tt 1303.0836}}].

\bibitem{Cremonesi:2013lqa}
S.~Cremonesi, A.~Hanany and A.~Zaffaroni, \emph{{Monopole operators and Hilbert
  series of Coulomb branches of $3d$ $\mathcal{N} = 4$ gauge theories}},
  \href{http://dx.doi.org/10.1007/JHEP01(2014)005}{\emph{JHEP} {\bf 01} (2014)
  005}, [\href{http://arxiv.org/abs/1309.2657}{{\tt 1309.2657}}].

\bibitem{Cremonesi:2014kwa}
S.~Cremonesi, A.~Hanany, N.~Mekareeya and A.~Zaffaroni, \emph{{Coulomb branch
  Hilbert series and Hall-Littlewood polynomials}},
  \href{http://dx.doi.org/10.1007/JHEP09(2014)178}{\emph{JHEP} {\bf 09} (2014)
  178}, [\href{http://arxiv.org/abs/1403.0585}{{\tt 1403.0585}}].

\bibitem{Cremonesi:2014uva}
S.~Cremonesi, A.~Hanany, N.~Mekareeya and A.~Zaffaroni,
  \emph{{T$_{\rho}^{\sigma}$ (G) theories and their Hilbert series}},
  \href{http://dx.doi.org/10.1007/JHEP01(2015)150}{\emph{JHEP} {\bf 01} (2015)
  150}, [\href{http://arxiv.org/abs/1410.1548}{{\tt 1410.1548}}].

\bibitem{Cremonesi:2015dja}
S.~Cremonesi, \emph{{The Hilbert series of 3d ${\boldsymbol{\mathcal{N}}}=2$
  Yang--Mills theories with vectorlike matter}},
  \href{http://dx.doi.org/10.1088/1751-8113}{\emph{J. Phys.} {\bf A48} (2015)
  455401}, [\href{http://arxiv.org/abs/1505.02409}{{\tt 1505.02409}}].

\bibitem{Hanany:2015via}
A.~Hanany, C.~Hwang, H.~Kim, J.~Park and R.-K. Seong, \emph{{Hilbert Series for
  Theories with Aharony Duals}},
  \href{http://dx.doi.org/10.1007/JHEP11(2015)132,
  10.1007/JHEP04(2016)064}{\emph{JHEP} {\bf 11} (2015) 132},
  [\href{http://arxiv.org/abs/1505.02160}{{\tt 1505.02160}}].

\bibitem{Cremonesi:2016nbo}
S.~Cremonesi, N.~Mekareeya and A.~Zaffaroni, \emph{{The moduli spaces of 3d $
  \mathcal{N}\ge 2 $ Chern-Simons gauge theories and their Hilbert series}},
  \href{http://dx.doi.org/10.1007/JHEP10(2016)046}{\emph{JHEP} {\bf 10} (2016)
  046}, [\href{http://arxiv.org/abs/1607.05728}{{\tt 1607.05728}}].

\bibitem{Lindstrom:1999pz}
U.~Lindstrom, M.~Rocek and R.~von Unge, \emph{{HyperKahler quotients and
  algebraic curves}},
  \href{http://dx.doi.org/10.1088/1126-6708/2000/01/022}{\emph{JHEP} {\bf 01}
  (2000) 022}, [\href{http://arxiv.org/abs/hep-th/9908082}{{\tt
  hep-th/9908082}}].

\bibitem{Borokhov:2003yu}
V.~Borokhov, \emph{{Monopole operators in three-dimensional N=4 SYM and mirror
  symmetry}},
  \href{http://dx.doi.org/10.1088/1126-6708/2004/03/008}{\emph{JHEP} {\bf 03}
  (2004) 008}, [\href{http://arxiv.org/abs/hep-th/0310254}{{\tt
  hep-th/0310254}}].




\bibitem{Benvenuti:2011ga}
S.~Benvenuti and S.~Pasquetti, 
\emph{{3D-partition functions on the sphere: exact evaluation and mirror symmetry}},
  \href{http://dx.doi.org/10.1007/JHEP05(2012)099}{\emph{JHEP} {\bf 05}
  (2012) 099}, [\href{http://arxiv.org/abs/1105.2551}{{\tt
  1105.2551}}].
  
\bibitem{Dimofte:2011ju}
  T.~Dimofte, D.~Gaiotto and S.~Gukov,
  \emph{{Gauge Theories Labelled by Three-Manifolds}},
   \href{http://dx.doi.org/10.1007/s00220-013-1863-2}{\emph{Math.Phys} {\bf 325}
 (2014) 367},[\href{http://arxiv.org/abs/1108.4389}{{\tt
  1108.4389}}].

\bibitem{Jafferis}
  D.~L.~Jafferis,
  \emph{{The Exact Superconformal R-Symmetry Extremizes Z}},
  \href{http://dx.doi.org/10.1007/JHEP05(2012)159}{\emph{JHEP} {\bf 05}
  (2012) 159}, [\href{http://arxiv.org/abs/1012.3210}{{\tt
  1012.3210}}].
 
\bibitem{Hama:2011ea} 
  N.~Hama, K.~Hosomichi and S.~Lee,
  \emph{{SUSY Gauge Theories on Squashed Three-Spheres}},
  \emph{JHEP} {\bf 1105}, 014 (2011)
  doi:10.1007/JHEP05(2011)014
  [\href{http://arXiv:1102.4716}{\tt 1102.4716}].

\bibitem{Hama:2010av} 
  N.~Hama, K.~Hosomichi and S.~Lee,
  \emph{{Notes on SUSY Gauge Theories on Three-Sphere}},
  JHEP {\bf 1103}, 127 (2011)
  doi:10.1007/JHEP03(2011)127
  [\href{arXiv:1012.3512}{\tt 1012.3512}].
 
\bibitem{Gaiotto:2012xa}
  D.~Gaiotto, L.~Rastelli and S.~S.~Razamat,
  \emph{{Bootstrapping the superconformal index with surface defects}},
  \href{http://dx.doi.org/10.1007/JHEP01(2013)022}{\emph{JHEP} {\bf 01}
  (2013) 022}, [\href{http://arxiv.org/abs/1207.3577}{{\tt
  1207.3577}}].
 

  
\end{thebibliography}

\providecommand{\href}[2]{#2}\begingroup\raggedright\endgroup

\end{document}